\documentclass[twocolumn,showpacs,preprintnumbers,amsmath,amssymb]{revtex4}
\usepackage{amsmath,epsfig,subfigure}
\usepackage{graphicx}
\usepackage{dcolumn}
\usepackage{bm}

\newcount\EquaNo \EquaNo=0
\def\newEq#1{\advance\EquaNo by 1 #1=\EquaNo}
\newcount\TablNo \TablNo=0
\def\newTabl#1{\advance\TablNo by 1 #1=\TablNo}
\newcount\FigNo \FigNo=0
\def\newFig#1{\advance\FigNo by 1 #1=\FigNo}
\newcount\ChapNo \ChapNo=0
\def\newCh#1{\advance\ChapNo by 1 #1=\ChapNo}
\def\expec#1{\big\langle{#1}\big\rangle}
\begin {document}

\title { Transient behavior in Single-File Systems}

\author{S.V. Nedea}
\altaffiliation{ Department of Mathematics and Computing Science, Eindhoven University of Technology, P.O. Box 513, 5600 MB Eindhoven, The Netherlands.}
\email{silvia@win.tue.nl}
\author{A.P.J. Jansen}
\altaffiliation{
 Department of Chemical Engineering, Eindhoven University of Technology, P.O. Box 513, 5600 MB Eindhoven, The Netherlands.
}
\author{J.J. Lukkien}
\altaffiliation{ Department of Mathematics and Computer Science,
Eindhoven University of Technology, P.O. Box 513, 5600 MB
Eindhoven, The Netherlands.}
\author{P.A.J. Hilbers}
\altaffiliation{ Department of Biomedical Engineering, Eindhoven
University of Technology, P.O. Box 513, 5600 MB Eindhoven, The
Netherlands.}




\date{\today}

\begin {abstract}
   We have used Monte-Carlo methods and analytical techniques
to investigate the influence of the characteristics, such as pipe
length, diffusion, adsorption, desorption and reaction rates
on the transient properties of {\it Single-File Systems}. The transient
or the relaxation regime is the period in which the system is
evolving to equilibrium. We have studied the system when all the sites are
reactive and when only some of them are reactive. Comparisons
between Mean-Field predictions, Cluster Approximation predictions,
and Monte Carlo simulations for the relaxation time of the system
are shown. We outline the cases where Mean-Field analysis gives good results
compared to Dynamic Monte-Carlo results. For some specific cases we can analytically
derive the relaxation time. Occupancy profiles for different
distribution of the sites both for Mean-Field and simulations are
compared. Different results for slow and fast reaction systems and
different distribution of reactive sites are discussed.
\end{abstract}


\maketitle

\section { Introduction }

 Although systems in nature evolve by obeying
physical laws, it is in most cases difficult or not feasible to describe the
system properties accurately since details of the microscopic dynamics are
not fully known.
 Therefore we usually deal with simplified models for these systems of which
stochastic models are an example.
They are thus described by a reduced set of dynamic variables.
 Although many exact solutions have been found,~\cite{Baxter82, DerridaEvans97,
Alcaraz94, DDM92, SchuetzDomany93, DEHP93} the vast majority of
stochastic models cannot be solved exactly. Many results
for equilibrium systems~\cite{tsikoyannis, rodenbeck, okino, coppens2,
coppens1} have been classified.
 In nature, however, equilibrium is rather an exception than a rule. In most
cases the temporal evolution starts from an initial state
which is far away from equilibrium. The relaxation of such a
system towards its stationary state depends on the specific
dynamical properties and cannot be described within the framework
of equilibrium statistical mechanics. Instead it is necessary to
set up a model for the microscopic dynamics of the system.
Assuming certain transition probabilities, the time-dependent
probability  distribution $P_{t}(s)$ to find the system in
configuration $s$ has to be derived from the Master Equation (ME).
Solving the ME is usually a difficult task. Therefore,
the theoretical understanding of non-equilibrium processes is
still at its beginning. Better understanding of these phenomena
would be an important step as non-equilibrium systems exhibit a
richer behavior than equilibrium systems.~\cite{Privman97,MarroDickman98,
DickmanJensen91, GLB89, DickmanBurschka88}

 We investigate non-equilibrium processes for a {\it Single-File System} (SFS) with conversion.
 In~\cite{silvia} we have already elaborated on the special properties of
porous structures such as zeolites. The one-dimensional nature of
the zeolite channel leads to extraordinary effects on the kinetic
properties of these materials. These structures are modelled by
one-dimensional systems called Single-File Systems where particles
are not able to pass each other.
 In~\cite{silvia} we have focused on the steady-state properties of
a SFS with conversion. The process of diffusion in SFS has different
characteristics as ordinary diffusion, which affects the nature
of both transport and conversion by chemical reactions. We
are investigating the kinetic properties of this system, and,
more precisely, we are interested in the properties of the system
before reaching equilibrium (the transient or relaxation regime).

  Different methods and techniques have been described in the literature to solve
the ME exactly.~\cite{rodenbeck, Privman97, DerridaEvans99}
  In spite of the remarkable progress in the field of exactly solvable
non-equilibrium processes, the majority of reaction-diffusion
models cannot be solved exactly.~\cite{BenNaimKrapivsky94,DHZ96}
  It is therefore necessary to use approximation techniques in order to
describe their essential properties (e.g. Mean-Field Approximation, Cluster
Approximation).~\cite{kuzovkov2}
 Also, as was already realized by Smoluchowski,~\cite{smoluchowski} fluctuations and correlations may be
extremely important in low-dimensional systems where the diffusive
mixing is not strong. Therefore, these approximation techniques can give results
that deviate strongly from the system behavior.
 Dynamic Monte Carlo methods are used to simulate the system according to the ME.

  Few researchers have concentrated on the properties of the system in the
transient regime and only studied the reactivity of the system in this
regime. Moreover, few research has been done for an open system
where adsorption/desorption is present at the marginal sites. The
reason that many of the analytical approaches fail is because of
the asymmetry of the system.

 In the present work we focus on the non-equilibrium phase properties
of SFS with conversion. We study the relaxation time of the system
(time evolution of the system properties, starting with no
particles) for different sets of kinetic parameters, lengths of
the pipe and distributions of the reactive sites.
 In the transient regime we observe that MF results are close to the 
DMC results both for slow and fast reaction systems. We outline the cases
for which the differences are significant.
 We compare with the steady-state situation in which the MF was not describing the
Single-File effects properly.~\cite{silvia}
 We analyze the situations when analytical results can be derived,
and we compare these results with Mean-Field (MF) and Dynamic Monte Carlo
(DMC) results. We look at the relaxation of the total loading, loading with different
 components, occupancies of individual sites for various
 parameters and different distributions of the reactive sites.
 As MF is a coarse approximation, for the analysis of profile occupancies
we introduce a better approximation (Cluster Approximation).
  We analyse the results using different analytic methods such as pair
and MF approximation. Pair Approximation and MF Approximation tend
to give similar results due to the high-order correlations in the system.
We investigate the effect of various model assumptions made about
diffusion, adsorption, desorption and reaction on the kinetic
behavior of the system. 

 In section~\ref{sec:lev0} we specify our mathematical model
together with the theoretical background for analytical and
simulation results.
  In section~\ref{sec:lev1} we present the various results for
transient regimes for the simplified model without conversion. We
solve numerically and analytically the Master Equation in
order to get the relaxation time of the system. In section
~\ref{sec:lev12} we use MF theory to simplify the rate equations
~\cite{silvia} of the system for the case when all the sites have the same
activity towards conversion.
  We present the results obtained using DMC simulations for the case with 
conversion when all the sites are reactive in
section~\ref{sec:lev3}, and when only some of the sites are reactive in section
~\ref{sec:lev4}. For all of these cases we compare the DMC results with MF
and Pair Approximation results. The influence of the position of the reactive 
sites is also outlined. The last section summarizes our main conclusions.

\section {\label{sec:lev0} Theory}

  In this section we give the theoretical background for our analytical
and simulation results. First we specify our model and then we show
that the defined system obeys a Master Equation (ME).
 We derive then a set of exact rate equations from the Master
Equation of the system describing the reaction kinetics.
 We look at the properties of the system in the transient regime
by solving these rate equations. In order to do this we have
to use approximation techniques.
 We use MF analysis and we derive a set of equations
that we can solve numerically. 
 Because MF is a strong simplification and neglects all spatial
correlations in the system, we introduce another approximation called
the Cluster Approximation. 
 We have also simulated the system governed by the ME using DMC simulations.

\subsection {The Model}

    We model a Single-File System by a one-dimensional array of
sites, each possibly occupied by an adsorbate. This is the model of
diffusion and reaction in a one-dimensional arrangements of particles with
hard-core interaction. The sites are numbered $1,2,\ldots ,S$. A particle can
 only move to the right or to the left if an adjacent site is vacant. 
 The sites could be reactive and unreactive and we note
with $N_{\rm prot}$ the number of reactive sites. A reactive site is the only place
where a conversion may take place.

\begin{figure*}
\centering
\subfigure {\epsfig {figure=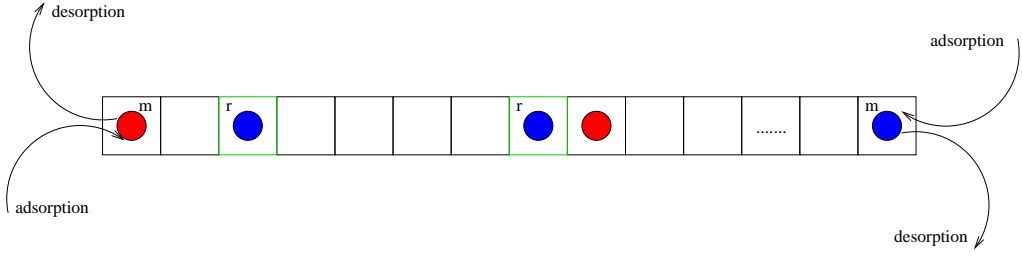, width=13.5cm} }
\caption {Picture of a Single-File System with two types of adsorbed particles}
\end {figure*}


       We consider two types of adsorbates, $\rm A$ and $\rm B$, in our model and we denote
with $\rm Y$ the site occupation of a site, $\rm Y$=($*$, $\rm A$, $\rm B$), which stands for a
vacant site, a site occupied by $\rm A$, or a site occupied by a $\rm B$, respectively.
   We restrict ourselves to the following mono- and bi-molecular transitions.
\\
\\
a) Adsorption and desorption
\\
\\
    Adsorption and desorption take place only at the two marginal sites
i.e., the left and rightmost sites at the ends of the system.
\\
\begin{center}
    ${\rm A}({\rm gas})$ + $\rm *_{\it m}$ $\longrightarrow$ ${\rm A}_{\it m}$
\\
    ${\rm A}_{\it m}$   $\longrightarrow$   ${\rm A}({\rm gas})$ + ${\rm *}_{\it m}$
\\
    ${\rm B}_{\it m}$ $\longrightarrow$ ${\rm B}({\rm gas})$ + $*_{\it m},$
\\
\end{center}
where subscripts $\it m$ denotes a marginal site. Note that there is no $\rm B$ adsorption.
$\rm B$'s can only be formed  by conversion.
\\
\\
b) Diffusion
\\
\\
    In the pipe, particles are allowed to diffuse via hopping to
    vacant nearest neighbor sites.
\\
\begin{center}
   ${\rm A}_{\it n}$ + ${\rm *}_{\it n+1}$ $\longleftrightarrow$ ${\rm *}_{\it n}$ + ${\rm A}_{\it n+1}$
\\
   ${\rm B}_{\it n}$ + ${\rm *}_{\it n+1}$ $\longleftrightarrow$ ${\rm *}_{\it n}$ + ${\rm B}_{\it n+1}$
\\
\end{center}
where the subscripts are site indices: $n$= 1, 2, $\ldots$, $S$-1.
\\
\\
c) Conversion
\\
\\
   An $\rm A$ can transform into a $\rm B$ at a reactive site.
\\
\begin {center}
    $\rm A_{\it r}$ $\longrightarrow$ $\rm B_{\it r}$.
\end{center}

       In the initial state of the system all the sites are vacant (no particles 
in the pipe) as we are interested in the behavior of the system towards equilibrium.

\subsection { Master Equation}

  Reaction kinetics is described by a stochastic process. Every reaction has a microscopic
rate constant associated with it that is the probability per unit time that
the reaction occurs.
  Stochastic models of physical systems can be described by a Master
Equation.~\cite{kampen}

 By $\alpha$, $\beta$, we will indicate a particular configuration of the
system i.e., a particular way to distribute adsorbates over all the sites.
 $P_\alpha(t)$ will indicate the probability of finding the system in
configuration $\alpha$ at time $t$ and $W_{\alpha\beta}$ is the rate
constant of the reaction changing configuration $\beta$ to configuration
$\alpha$.

    The probability of the system being in configuration $\alpha$ at time
$t+dt$ can be expressed  as the sum of two terms. The first term is the
probability to find the system already in configuration $\alpha$ at time $t$
multiplied by the probability to stay in this configuration during $dt$.
 The second term is the probability to find the system in some other
configuration $\beta$ at time $t$ multiplied by the probability to go from
$\beta$ to $\alpha$ during $dt$.

\begin{equation}
P_{\alpha}(t+dt)=(1-dt\sum_{\beta} W_{\beta\alpha})P_{\alpha}(t) +
         dt\sum_{\beta}W_{\alpha\beta}P_{\beta}(t)
\end {equation}

By taking the limit $dt \to 0$ this equation reduces to a Master Equation:

\begin{equation}
  {dP_\alpha(t)\over dt}
  =\sum_{\beta}
   \left[W_{\alpha\beta}P_{\beta}(t)-W_{\beta\alpha}P_\alpha(t)\right].
\end{equation}

Analytical results can be derived as follows. The value of a property $X$ is a weighted average over the values
$X_{\alpha}$ which is the value of $X$ in configuration $\alpha$:

\begin{equation}
   \langle X \rangle=\sum_{\alpha}P_{\alpha}X_{\alpha}.
\end{equation}

  From this follows the rate equation
\begin{equation}
\begin{split}
 {{d\langle X \rangle}\over{dt}}
  & =\sum_{\alpha}{dP_{\alpha}\over{dt}}X_{\alpha}\cr
  &=\sum_{\alpha\beta}[W_{\alpha\beta}P_{\beta}-W_{\beta\alpha}P_{\alpha}]X_{\alpha}\cr
  &=\sum_{\alpha\beta}W_{\alpha\beta}P_{\beta}(X_{\alpha}-X_{\beta}).
\end{split}
\end{equation}

\subsection {Analytical methods}

\subsubsection{Rate equations}

   Starting from the Master Equation (2) and using expression (4)
the rate equations of the system are derived.~\cite{silvia}
 We denote by $W_{\rm ads}$, $W_{\rm des}$, $W_{\rm diff}$ 
the rate constants of adsorption, desorption and diffusion
respectively.
 For simplicity we assume that the rate constants of ${\rm A}$ and ${\rm B}$
desorption are equal, and also the rate constants of ${\rm A}$ and ${\rm B}$
diffusion are equal.
 We denote by $\langle {\rm Y}_n \rangle$ the probability that a particle of
type $\rm Y$ is on site $n$, and with $\langle {\rm Y}_n {\rm Y}_{n+1} \rangle$
the probability that a particle of type ${\rm Y}$ is at site $n$ and one at
site $n+1$, where $\rm Y$=($\rm *$,$\rm A$,$\rm B$).
 The coefficients ${\Delta}_n$, where $n$=1, 2,$\ldots$, $S$, are 1 if site
$n$ is reactive and 0 otherwise.
   The rate equations for a non-marginal site are
\begin{equation}
\begin{split}
 { d\langle {\rm A}_n \rangle\over{dt}}
&=W_{\rm diff}[-\langle {\rm A}_n *_{n+1}\rangle-\langle *_{n-1} {\rm A}_n\rangle+\langle {\rm A}_{n-1}*_n\rangle \cr
&+\langle *_n {\rm A}_{n+1}\rangle] - {\Delta}_n W_{\rm rx}\langle {\rm A}_n\rangle.
\end{split}
\end{equation}
 For $\langle {\rm B}_n\rangle$ we get similarly
\begin{equation}
\begin{split}
 { d\langle {\rm B}_n\rangle\over{dt}}
&=W_{\rm diff}[-\langle {\rm B}_n *_{n+1}\rangle-\langle *_{n-1} {\rm B}_n\rangle+\langle {\rm B}_{n-1}*_n\rangle \cr
&+\langle *_n {\rm B}_{n+1}\rangle] + {\Delta}_n W_{\rm rx}\langle {\rm A}_n\rangle.
\end{split}
\end{equation}
 The marginal sites also have adsorption and desorption. They can be dealt
with similary as the conversion. The rate equations for $\rm A$ are
\begin{equation}
\begin{split}
{d\langle {\rm A}_1\rangle\over{dt}}
&=W_{\rm diff}[-\langle {\rm A}_1 *_{2}\rangle+\langle *_1
 {\rm A}_{2}\rangle]+W_{\rm ads}\langle *_1\rangle \cr
&-W_{\rm des}\langle{\rm A}_1\rangle -{\Delta}_1 W_{\rm rx}\langle {\rm A}_1\rangle,\cr
{d\langle {\rm A}_S\rangle\over{dt}}
& =W_{\rm diff}[-\langle *_{S-1}{\rm A}_S \rangle+\langle *_S {\rm A}_{S-1}
 \rangle]+W_{\rm ads}\langle *_S\rangle \cr
&-W_{\rm des}\langle {\rm A}_S\rangle - {\Delta}_S W_{\rm rx}\langle
{\rm A}_S\rangle,\cr
\end{split}
\end{equation}
and the rate equations for $\rm B$
\begin{equation}
\begin{split}
 {d\langle{\rm B}_1\rangle\over{dt}}
&=W_{\rm diff}[-\langle {\rm B}_1 *_{2}\rangle+\langle *_1 {\rm B}_{2}\rangle]-W_{\rm des}\langle {\rm B}_1\rangle+
{\Delta}_1W_{\rm rx}\langle{\rm A}_1\rangle,\cr
{d\langle {\rm B}_S\rangle\over{dt}}
&=W_{\rm diff}[-\langle *_{S-1}{\rm A}_S \rangle+\langle {\rm A}_{S-1} *_S \rangle]-W_{\rm des}\langle {\rm B}_1\rangle\cr
&+{\Delta}_S W_{\rm rx}\langle
{\rm A}_S\rangle.
\end{split}
\end{equation}

 Note that these coupled sets of differential equations are exact, but not
closed.

\subsubsection {Mean-Field}

 To solve this coupled set of differential equations, we need to make an
approximation for the two-site probabilities such as
 $\langle {\rm A}_n *_{n+1} \rangle$,$\langle{\rm B}_n *_{n+1}\rangle$, etc.
The closure relation
\begin{equation}
\langle X \rangle=\sum_Y {\langle XY \rangle}
\end{equation}
should hold for any approximation for these two-site probabilities.
We denote with $X$ the occupation of site $n$ and with $Y$ the occupation of
site $n+1$.
 The simplest approximation is
\begin{equation}
\langle XY \rangle=\langle X \rangle \langle Y \rangle,
\end{equation}
i.e. neighboring sites are considered independent. 

 The two-site probabilities then become $\langle {\rm A}_n
*_{n+1}\rangle$=$\langle {\rm A}_n\rangle \langle*_{n+1}\rangle$,
$\langle {\rm B}_n *_{n+1}\rangle$=$\langle{\rm B}_n\rangle\langle
*_{n+1}\rangle$.~\cite{kuzovkov2} This approximation is called
the Mean-Field Approximation and gives us a coupled set of differential equations 
that we can solve numerically.

\subsubsection {Cluster Approximation}

 The MF Approximation is a strong simplification because it neglects
all spatial correlations in the system. Because the system we
analyze is one-dimensional, the correlations might be significant
and important.
 The obvious possibility to eliminate the weakness of the MF approach is to
introduce another approximation. 

 Instead of using MF Approximation for the two-site probabilities, we 
write down their rate equations (see appendix). These equations have three-site
probabilities which we approximate. This leads to a so-called Cluster
Approximation. The closure relation
\begin{equation}
  \langle XY \rangle=\sum_Z{\langle XYZ \rangle}
\end{equation}
should now hold for any approximation.
We denote here with $X$ the occupation of site $n$, with $Y$ the occupation
of site $n+1$, and with $Z$ the occupation of site $n+2$.

 There are various decoupling scheme used in the literature~\cite{DerridaEvans99, DerridaEvans97, mamda, DDM92, DEHP93} as 
approximations for the $n$-site probabilities. For many of these decoupling
schemes the closure relation (11) no longer holds.
 For simplicity we will use the simplest Cluster Approximation, called Pair Approximation 
for which only the correlations between pairs of nearest neighbors (NN) are
considered. 

 The decoupling scheme for our pair Approximation is
\begin{equation}
\langle XYZ \rangle={{\langle XY \rangle \langle YZ
\rangle}\over{\langle Y \rangle}}.
\end{equation}
It is straightforward to see that the closure relation (11) holds.
\begin{equation}
\sum_Z{\langle XYZ \rangle}
={{\langle XY \rangle}\over {\langle Y \rangle}}\sum_Z{\langle YZ \rangle}
={1\over{\langle Y \rangle}} {\langle XY \rangle} {\langle Y \rangle} 
=\langle XY \rangle.
\end{equation}
 After decoupling, this system of coupled sets of differential equations,
consisting of the rate equations for one-site and two-site probabilities,
becomes closed and can be solved numerically.

\subsection {Dynamic Monte Carlo}

  We have seen that we can derive approximate analytical solutions to the 
Master Equation.
 DMC methods allow us to simulate the system governed by the
Master Equation over time.
  We simplify the notation of the Master Equation by defining a matrix $\bf W$
containing the rate constants $W_{\alpha\beta}$, and a diagonal matrix $\bf R$ by
${R}_{{\alpha}{\beta}}\equiv \sum_{\gamma}W_{\gamma\beta}$, if ${\alpha}={\beta}$,
and 0 otherwise.
 If we put the probabilities of the configurations $P_{\alpha}$ in a vector
$\bf P$, we can write the Master Equation as

\begin {equation}
 {d{\bf P}\over{dt}}=-({\bf R}-{\bf W}){\bf P}.
\end{equation}
where $\bf R$ and $\bf W$ are time independent.
We also introduce a new matrix $\bf Q$, ${\bf Q}(t) \equiv \exp[-{\bf R}t].$
This matrix is time dependent by definition and we can rewrite the Master
Equation in the integral form

\begin{equation}
{\bf P}(t)={\bf Q}(t){\bf P}(0)+\int_0^tdt^{\prime}{\bf Q}(t-t^{\prime}){\bf W}{\bf P}(t^{\prime}).
\end{equation}
By substitution we get from the right-hand-side  for $P(t^{\prime})$

\begin{equation}
\begin{split}
{\bf P}(t)
  & =[{\bf Q}(t)\cr
  & +
     \int_0^t dt^{\prime}{\bf Q}(t-t^{\prime}){\bf W}{\bf Q}(t^{\prime})\cr
  & +
    \int_0^tdt^{\prime}\int_0^{t^{\prime}}dt^{\prime\prime}{\bf Q}(t-t^{\prime}){\bf W}{\bf Q}(t^{\prime}-t^{\prime\prime})
    {\bf W}{\bf Q}(t^{\prime\prime})\cr
  & +\ldots]{\bf P}(0).
\end{split}
\end{equation}

 Suppose at $t=0$ the system is in configuration $\alpha$ with probability
$P_{\alpha}(0)$. The probability that, at time $t$, the system is still in
configuration $\alpha$ is given by
$Q_{\alpha\alpha}(t)P_{\alpha}(0)=\exp(-R_{\alpha\alpha}t)P_{\alpha}(0)$.
 This shows that the first term represents the contribution to the
probabilities when no reaction takes place up to time $t$. The matrix $\bf W$
determines how the probabilities change when a reaction takes place. The
second term represents the contribution  to the probabilities when no
reaction takes place between times $0$ and $t^{\prime}$, some reaction takes
place at time $t^{\prime}$, and then no reaction takes place between
$t^{\prime}$ and $t$. The subsequent terms represent contributions when two,
three, four, etc. reactions take place.
 The idea of the DMC method is not to compute probabilities
$P_{\alpha}(t)$ explicitly, but to start with some particular configuration,
representative for the initial state of the experiment one wants to
simulate, and then generate a sequence of other configurations with the
correct probability.
 The method generates a time $t^{\prime}$ when the first reaction occurs
according to the probability distribution $1-\exp[-R_{\alpha\alpha}t]$.
At time $t^{\prime}$ a reaction takes place such that a new configuration
${\alpha}^{\prime}$ is generated by picking it out of all possible new
configurations $\beta$ with a probability proportional to
$W_{{\alpha}^{\prime}{\alpha}}$. At this point we can proceed by repeating
the previous steps, drawing again a time for a new reaction and a new
configuration.~\cite{lukkien, gelten_all}

\section { Results and Discussion }

 In~\cite{silvia} various results for the system with conversion ($W_{\rm rx}
\neq 0$) and without conversion($W_{\rm rx}=0$) were reported.
In case $W_{\rm rx}=0$ we have only $\rm A$ particles in the system. The total
loading ($Q$) of the system is defined as the average number of particles
per site.
In case $W_{\rm rx}\neq 0$ we have ${\rm B}$'s as well in the system.
In this case, the total loading ($Q$), is the sum of the loading with $\rm A$
particles ($Q_{\rm A}$) and loading with $\rm B$ particles ($Q_{\rm B}$)

\begin{equation}
Q={1\over {S}} \sum_{n=1}^S{\langle {\rm A}_n\rangle} + {1\over {S}} \sum_{n=1}^S{\langle {\rm B}_n\rangle}.
\end{equation}
Note that the total loading of the pipe for the model with conversion is the
same as for the model without conversion.~\cite{silvia}
We study the relaxation time of the system without conversion and of the
system with conversion.

\subsection {\label{sec:lev1} No conversion}

  We are interested in the relaxation time of the system (transients). We
start with the evolution of the total loading over time starting
from a system with no particles at all.
 As the total loading is the same for the case with and without conversion, we
will consider for simplicity the case with no conversion first.

 As we can derive a finite set of exact rate equations(5,6,7,8) it's not necessary to
work with the Master Equation in this case. With $\expec{{\rm X}_n}$ the
probability that site $n$ is occupied we have

\begin{equation}
\begin{split}
  {d\expec{{\rm X}_n}\over dt} 
&=W_{\rm diff} [-\expec{{\rm X}_n*_{n+1}}-\expec{*_{n-1}{\rm X}_n}\cr
&+\expec{{\rm X}_{n-1}*_n}+\expec{*_n{\rm X}_{n+1}}]
\end{split}
\end{equation}
when $n$ is not a marginal site.
  The two-site probabilities can be eliminated by using closure relations,
\begin{equation}
  \expec{{\rm X}_n{\rm X}_{n+1}}+\expec{{\rm X}_n*_{n+1}}
  =\expec{{\rm X}_n},
\end{equation}
that hold in this specific case.
 The probabilities with particles on both neighboring sites cancel and
the result is
\begin{equation}
  {d\expec{{\rm X}_n}\over dt}=W_{\rm diff}
  \left[\expec{{\rm X}_{n-1}}-2\expec{{\rm X}_n}
        +\expec{{\rm X}_{n+1}}\right].
\end{equation}
 For the marginal sites we get
\begin{equation}
\begin{split}
  {d\expec{{\rm X}_1}\over dt}&=
  W_{\rm ads}\left[1-\expec{{\rm X}_1}\right]
  -W_{\rm des}\expec{{\rm X}_1}\cr
  &+W_{\rm diff}
  \left[\expec{{\rm X}_2}-\expec{{\rm X}_1}\right]\cr
\noalign{\hbox{and}}
  {d\expec{{\rm X}_S}\over dt}&=
  W_{\rm ads}\left[1-\expec{{\rm X}_S}\right]
  -W_{\rm des}\expec{{\rm X}_S}\cr
  &+W_{\rm diff}
  \left[\expec{{\rm X}_{S-1}}-\expec{{\rm X}_S}\right].\cr
\end{split}
\end{equation}

 These equations are used for the derivation of the relaxation time.
The rate equations are going to be simplified to a point where they are a set of
homogeneous linear ODE's with only one parameter apart from the system
size. Dividing by $W_{\rm ads}+W_{\rm des}$ and introducing the
dimensionless parameters

\begin{equation}
\begin{split}
  \expec{{\rm X}_{\rm eq}}
  &\equiv{W_{\rm ads}\over W_{\rm ads}+W_{\rm des}},\cr
  \lambda
  &\equiv{W_{\rm diff}\over W_{\rm ads}+W_{\rm des}},\cr
  \tau
  &\equiv(W_{\rm ads}+W_{\rm des})t,\cr
\end{split}
\end{equation}
we get
\begin{equation}
\begin{split}
  {d\expec{{\rm X}_1}\over d\tau}&=
  \expec{{\rm X}_{\rm eq}}-\expec{{\rm X}_1}
  +\lambda
  \left[\expec{{\rm X}_2}-\expec{{\rm X}_1}\right],\cr
  {d\expec{{\rm X}_n}\over d\tau}&=\lambda
  \left[\expec{{\rm X}_{n-1}}-2\expec{{\rm X}_n}
        +\expec{{\rm X}_{n+1}}\right],\cr
  {d\expec{{\rm X}_S}\over d\tau}&=
  \expec{{\rm X}_{\rm eq}}-\expec{{\rm X}_S}
  +\lambda
  \left[\expec{{\rm X}_{S-1}}-\expec{{\rm X}_S}\right].\cr
\end{split}
\end{equation}

 We can write these equations in matrix-vector notation as
 \begin{equation}
{{d\expec{{\bf X}}\over d\tau}}= - {\bf M}\expec{\bf X} + {\bf v},
 \end{equation}
where $\expec{\bf X}$ is a vector containing the occupancy
probabilities, ${\bf M}$ is the matrix of coefficients having the form

\begin{equation}
  \begin{pmatrix}
  1+\lambda & -\lambda & 0 & 0 & \ldots & & & & \cr
  -\lambda & 2\lambda & -\lambda & 0 & \ldots & & & & \cr
  0 & -\lambda & 2\lambda & -\lambda & & & & & \cr
  0 & 0 & -\lambda & 2\lambda & \ddots & & & & \cr
  \vdots & \vdots & & \ddots & \ddots & \ddots & & \vdots
    & \vdots \cr
  & & & & \ddots & 2\lambda & -\lambda & 0 & 0\cr
  & & & & & -\lambda & 2\lambda & -\lambda & 0\cr
  & & & & \ldots & 0 & -\lambda & 2\lambda & -\lambda \cr
  & & & & \ldots & 0 & 0 & -\lambda & 1+\lambda \cr
  \end{pmatrix}
\end{equation}
 and $\bf v$ is the vector that makes the system non-homogeneous.
The parameter $\expec{{\rm X}_{\rm eq}}$ is the equilibrium coverage. 

 Finally we can make the set of equations homogeneous by working with
probabilities for vacancies: i.e., with
\begin{equation}
  {\bf y} \equiv 1- {\expec{\bf X}\over{\expec{{\rm X}_{\rm eq}}}}
\end{equation}
we get
\begin{equation}
\begin{split}
  -{d{\bf y}\over d\tau} ={\bf M} {\bf y}
\end{split}
\end{equation}

  In order to solve equation (27), we try the substitution 
\begin{equation}
  {\bf y}={\bf a} e^{-\omega\tau}.
\end{equation}
 Taking out the exponential leaves us with
\begin{equation}
 {\bf M} {\bf a}=\omega {\bf a}.
\end{equation}
 Removing the time dependence yields relaxation times. We see that 
we have obtained an eigenvalue equation. The
eigenvalues $\omega$ are the reciprocal of relaxation times for
the corresponding eigenvectors. The relaxation time of the system
as a whole($t_{rel}$) is the reciprocal of the smallest eigenvalue. We can
get this time by simply numerically solving the eigenvalue
equation for given $S$ and $\lambda$.

\begin {center} {\it{ Solving the eigenvalue equation analytically.}}
\end{center}

 For some special cases analytical expressions for the eigenvalues can be
given. We consider the {\it ansatz}
\begin{equation}
  a_n=z^n.
\end{equation}
If we substitute $a_n$ from expression (29) into the equation for $n$ not a marginal site,
we get
\begin{equation}
  \lambda\left[-{1\over z}+2-z\right]=\omega.
\end{equation}
This equation has two solutions.

\begin{equation}
  z_\pm=1-{1\over 2}f\pm{1\over 2}\sqrt{f^2-4f}
\end{equation}
where
\begin{equation}
  f\equiv{\omega\over\lambda}.
\end{equation}

 There are two cases to be distinguished. If $f\ge 4$ then both solutions
are real. Because this is the same as $\omega\ge 4\lambda$, we will have
this for slow diffusion and for eigenvectors with fast relaxation
(large $\omega$). As we are interested in the slowest relaxation (small
$\omega$) we will look at the other case $f<4$ or $\omega<4\lambda$. The
two solutions are then each others complex conjugate. (Note for the
following that $f\ge 0$.)
\begin{center}{\it{ Fast diffusion or slow relaxation.}} \end{center}
 In this case we can write the solutions as
\begin{equation}
  z_\pm=re^{\pm i\varphi}
\end{equation}
where $r$ and $\varphi$ are both real. In fact the equation for $z$
shows that when $z$ is a solution, then so is $1/z$. This means that
$r=1$, or
\begin{equation}
  z_\pm=e^{\pm i\varphi}.
\end{equation}
Substitution in the equation for $z$ yields
\begin{equation}
  \cos\varphi=1-{1\over 2}f.
\end{equation}
This has indeed only solutions for $0\le f\le 4$.

Because there are two solutions for $z$, the solution for $a_n$ is a
linear combination of these two solutions: i.e.,
\begin{equation}
  a_n=c_1e^{in\varphi}+c_2e^{-in\varphi}.
\end{equation}

 From equation (28) we remark that $a_n$ should be always real.
This means that $c_1$ is the complex conjugate of $c_2$.
The coefficients will follow from the equations for the marginal
sites. There are two of these equations. The equation above for
$\cos\varphi$ is a third equation. We have four unknowns ($c_1$, $c_2$,
$\omega$, and $\varphi$), but, as only the ratio between the
coefficients can be determined, we should effectively be able to
determine all of them.

 Substitution of the expression for $a_n$ in the equations for the
marginal sites taking into account that $c_1$ and $c_2$ are complex
conjugate ($c_1={c_1}_{\rm R} + i {c_1}_{\rm I}$), leads to

\begin{widetext}
\begin{equation}
  \begin{pmatrix}
   2(1-\omega + \lambda) \cos(\varphi)-2\lambda \cos(2\varphi)
  &-2 \sin(\varphi)(1-\omega+\lambda)+2\lambda\sin(2\varphi) \cr
   2\cos(S{\varphi})(1-\omega +\lambda) -2\lambda \cos((S-1){\varphi})
  &-2(1-\omega+\lambda)\sin(S\varphi)+2\lambda\sin((S-1)\varphi)\cr
  \end{pmatrix}
  \begin {pmatrix}
     {c_1}_{\rm R}\cr {c_1}_{\rm I}\cr
  \end{pmatrix}
  =\begin{pmatrix} 0\cr 0\cr \end{pmatrix}.
\end{equation}
\end{widetext}
This equation only has non-trivial solutions (the trivial solution is
${c_1}_R={c_1}_I=0$) when the determinant of the matrix equals zero. This leads
to the following equation
\begin{equation}
\begin{split}
  8\lambda\left(-\omega+\lambda+1\right)\sin((S-2)\varphi)\cr
  -4\left(-\omega+\lambda+1\right)^2\sin((S-1)\varphi)\cr
   +4\lambda^2\sin((S-3)\varphi)=0.
\end{split}
\end{equation}
We can eliminate $\omega$ using $\cos\varphi=1-f/2$.

\begin {figure}[t]
\centering
\subfigure {\epsfig {figure=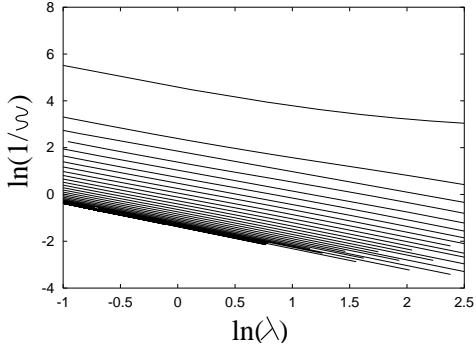, width=6.5cm} }
\caption {
      The general solution for $\ln(1/{\omega})$ as a function of $\ln(\lambda)$,
for $\varphi \in [0,\pi)$, and $S=30$.
         }
\label{relaxspec}
\end{figure}

 Using $\varphi$ as parameter in equation (39), $\varphi \in [0,\pi)$, we can
get the $\lambda$'s. Equations (35) and (32) gives us the $\omega$'s.
 We get in this way $\omega$ as a function of $\lambda$, coupled by the
parameter $\varphi$ (see figure~\ref{relaxspec}c).




\begin {center}{\it Solving the eigenvalue equation for the total
loading, in the case $\lambda\to\infty$}
\end{center}

 We solve the eigenvalue equation numerically in order to get the relaxation
time of a system. We want to describe how the relaxation time of the total
loading ($t_{rel}$) depends on system parameters like reaction, adsorption/desorption,
diffusion rate constants and system size $S$.
In figure~\ref{relaxspec} we show the influence of diffusion on the relaxation time for
a system of size $S$=30.
 Note that there are two regimes describing the dependence on diffusion of the
relaxation time $t_{rel}$. The first regime is for slow diffusion, when
$t_{rel}$ decreases rapidly with increasing diffusion, and the second for
fast diffusion, when $t_{rel}$ slowly decreases with diffusion to a
limiting value.

%

  Because diffusion is infinitely-fast all the sites have the same
probability to be occupied and the system is homogeneous.
 We can then analytically derive the limiting value of $t_{rel}$ for
infinitely-fast diffusion from the equation
\begin{equation}
{{d{\expec{{\rm X}_n}}} \over {dt}}
 = {{2W_{\rm ads}}\over {S}}\left[1- {\expec{{\rm X}_n}}\right]-
  {{2W_{\rm des}}\over {S}} {\expec{{\rm X}_n}}.
\end{equation}
 The first term is the probability of a particle to be
adsorbed at the two open ends, and the second is the probability
of a particle to be desorbed at the two open ends. The probability
of a particle to be adsorbed to one end equals the adsorption rate
constant ($W_{\rm ads}$) times the probability to have there a vacancy
(1-$\expec{{\rm X}_n}$), while the probability of a particle to be desorbed
equals the desorption rate constant ($W_{\rm des}$) times the
probability to have a particle ($\expec{{\rm X}_n}$).

 From the above equation we get the expression for $\expec{{\rm X}_n}$,
\begin{equation}
\expec{{\rm X}_n}={\expec{{\rm X}_{\rm eq}}}\left[ 1- \exp{\left[{{-2(W_{ads}+W_{des})\over S}t}\right]}\right].
\end{equation}

 The relaxation time is
\begin{equation}
t_{rel}={ S \over {2(W_{\rm ads}+W_{\rm des})} }
\end{equation}

\begin {figure}[th]
\centering
\subfigure {\epsfig {figure=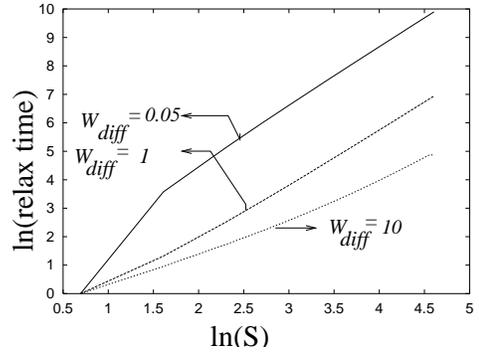, width=6.5cm} }
\caption {
          The logarithm of the relaxation time as a function of $\ln(S)$ for
          the parameters $W_{\rm ads}$=0.2, $W_{\rm des}$=0.8 and different
          $W_{\rm diff}$.
         }
\label{relaxS}
\end{figure}

 This limiting value is the same as the
one indicated by the convergence of the curves in figure~\ref{relaxspec}.
  The other parameter that influence the relaxation time $t_{rel}$ is the length
of the pipe $S$. For diffusion very fast, $t_{rel}$ increases linearly
with $S$ (see figure~\ref{relaxS}).



\subsection{ \label{sec:lev12} Conversion}

\begin {figure*}
\centering
\subfigure {\epsfig {figure=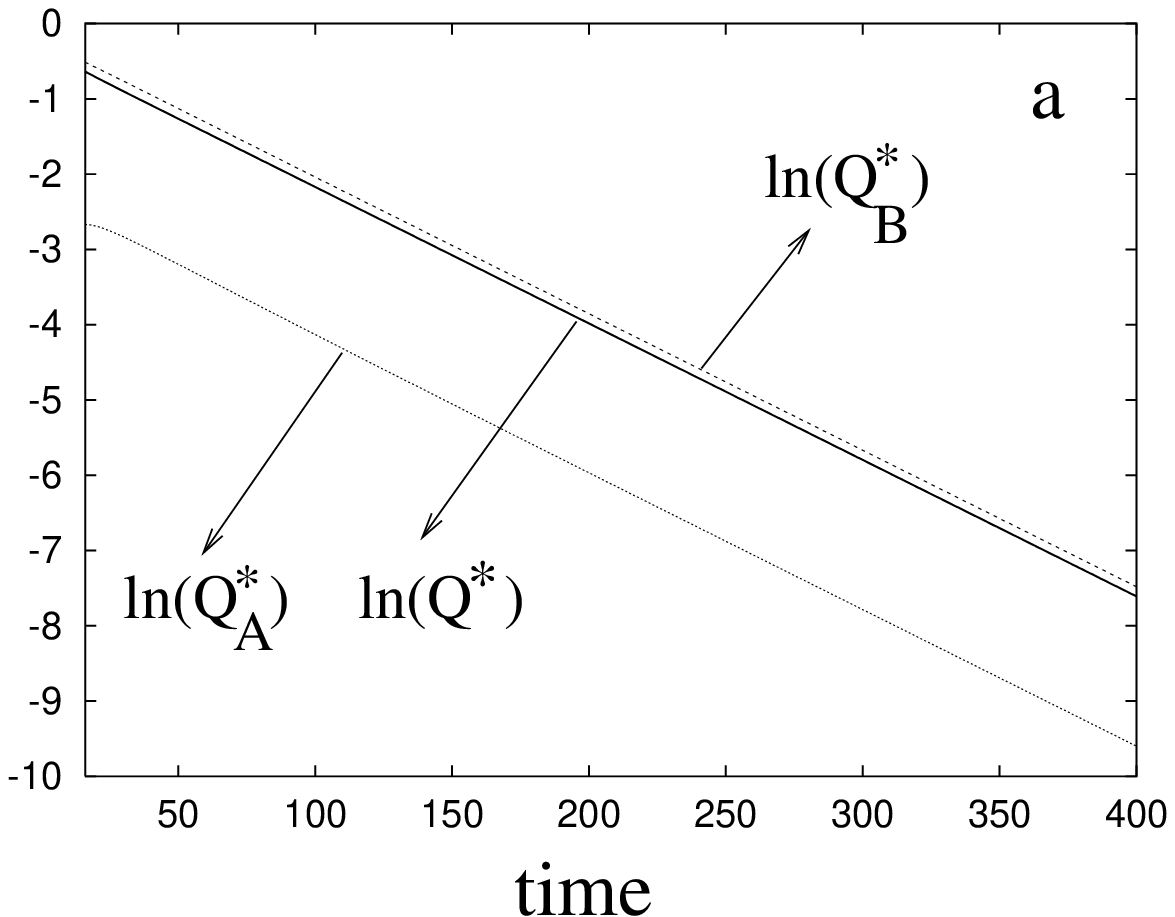, width=5.5cm} }
\subfigure {\epsfig {figure=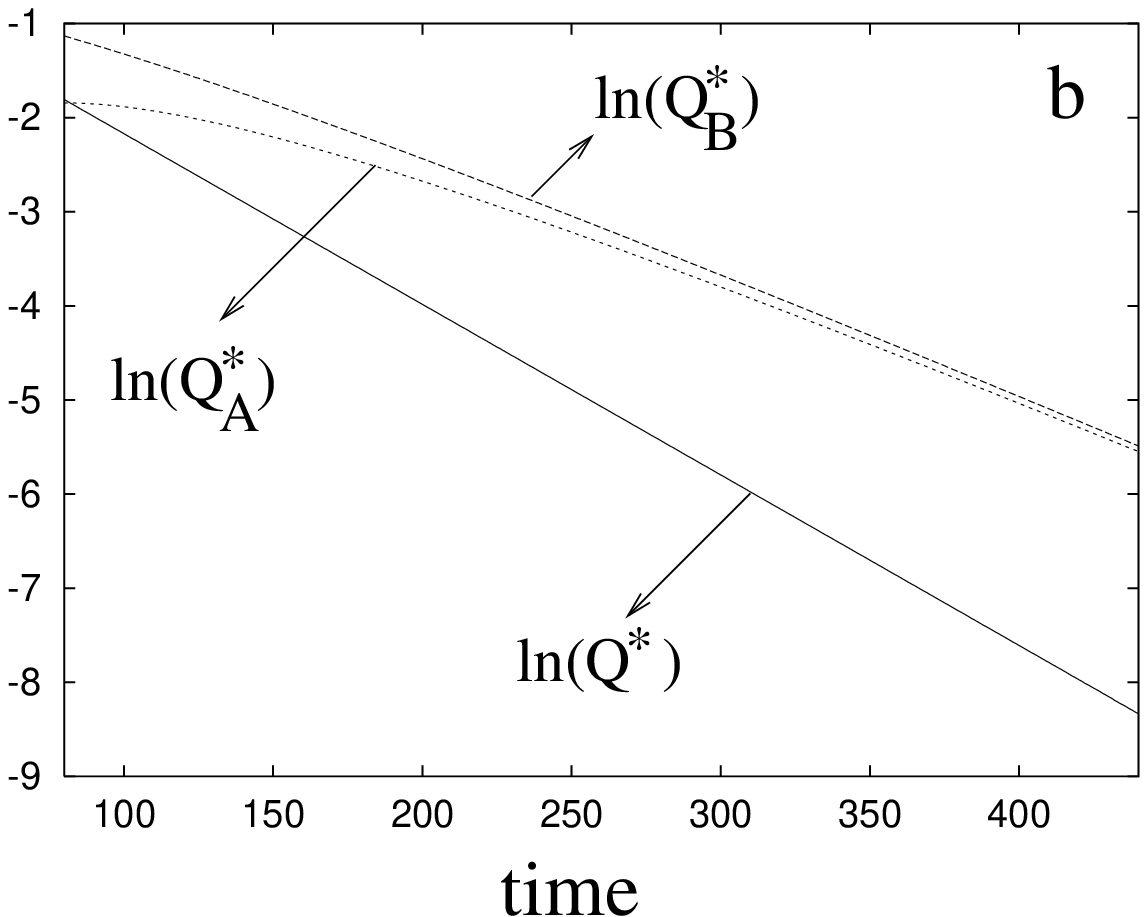, width=5.5cm} }
\subfigure {\epsfig {figure=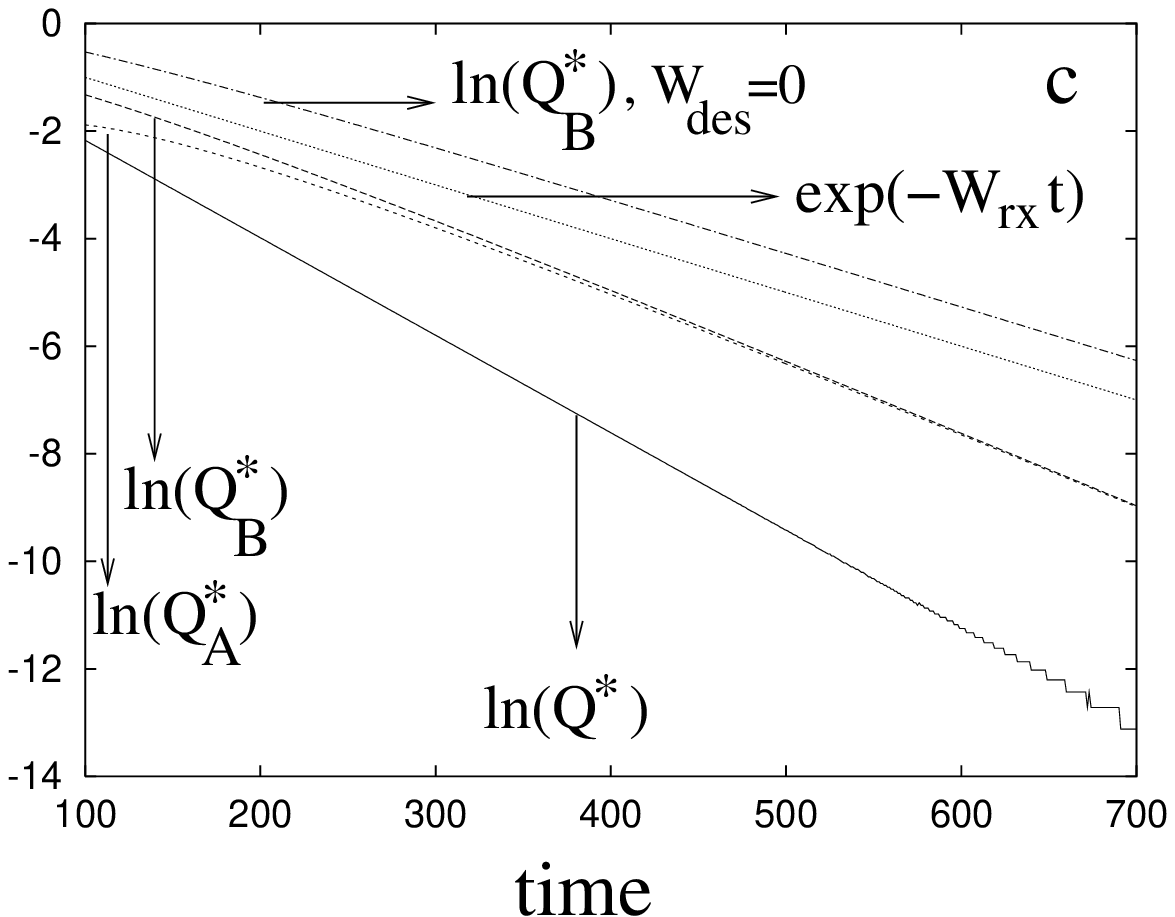, width=5.5cm} }
\caption {
          a) Time dependence of $\ln({Q}^{*})$, $\ln({Q_{\rm A}}^{*})$ and
           $\ln({Q_{\rm B}}^{*})$ 
           for fast reaction systems ($W_{\rm ads}=0.8$, $W_{\rm des}=0.2$,
           $W_{\rm diff}$=1; $W_{\rm rx}$=2). 
          b) Time dependence of the $\ln(Q^{*})$, $\ln({Q_{\rm A}}^{*})$ and
           $\ln({Q_{\rm B}}^{*})$ 
           for slow reaction systems ($W_{\rm ads}$=0.8, $W_{\rm des}=0.2$,
           $W_{\rm diff}$=2; $W_{\rm rx}$=0.01).
          c) Comparison between  time dependence of $\ln(Q^{*})$,
           $\ln({Q_{A}}^{*})$, $\ln({Q_{\rm B}}^{*})$ and the time 
           dependence of the function $\exp{(-W_{\rm rx}*t)}$. 
           For slow reaction systems, the slope does not depend only on
           $W_{\rm rx}$, but also on desorption. We have marked the time
           dependence of the $\ln({Q_{\rm B}}^{*})$ when $W_{\rm des}$=0.
           In a), b) and c) we have marked with * the absolute value of the 
           difference between the current value and the steady-state value of 
           the parameter.
         }
\label{tranz_sr_ssr}
\end{figure*}

  In the case with no conversion, we have derived a set of exact
equations and we have simplified the rate equations to
homogeneous linear ODE's.
  Including conversion in our model, the two-site probabilities can not be
eliminated and an approximation is needed. 
 We use the MF Approximation and we get a coupled set of differential
equations that we can solve numerically.
 In figure~\ref{tranz_sr_ssr}, from the MF results for the transients, we
observe that there are two different behaviors determined by conversion.

 For fast reaction systems (see figure~\ref{tranz_sr_ssr}a),
the relaxation time of the loading with $\rm A$'s ($t_{relA}$) and $\rm B$'s
($t_{relB}$) is equal to the relaxation time of the total loading ($t_{rel}$).
 This means that when $Q$ has reached equilibrium, $Q_{\rm A}$ and
$Q_{\rm B}$ have also reached equilibrium.

 For slow reaction systems (see figure~\ref{tranz_sr_ssr}b), the total loading $Q$
relaxes faster to equilibrium than the loading with $\rm A$'s ($Q_{\rm A}$) and
$\rm B$'s ($Q_{\rm B}$).
 The regime between $Q$ reaching equilibrium 
and $Q_{\rm A}$ and $Q_{\rm B}$ reaching equilibrium we call the reaction
limited regime.

 We remark that for slow reaction systems in the reaction limited regime, 
the vacancy probability can be replaced with the steady-state expression
 $\langle *_{n}\rangle$= $W_{\rm des}/({W_{\rm ads}+W_{\rm des}})$. 
The set of equations for the case when all the sites are reactive then becomes

\begin{equation}
\begin{split}
{d\langle {\rm A}_n\rangle\over{dt}}
  &={W_{\rm diff}W_{\rm des}\over W_{\rm ads}+W_{\rm des}}\left[
     \expec{{\rm A}_{n+1}}+\expec{{\rm A}_{n-1}}
    -2\expec{{\rm A}_n}\right]\cr
  & -W_{\rm rx}\expec{{\rm A}_n},\cr
{d\langle {\rm B}_n\rangle\over{dt}}
  &={W_{\rm diff}W_{\rm des}\over W_{\rm ads}+W_{\rm des}}\left[
     \expec{{\rm B}_{n+1}}+\expec{{\rm B}_{n-1}}
    -2\expec{{\rm B}_n}\right] \cr
  &+W_{\rm rx}\expec{{\rm A}_n},\cr
{d\langle {\rm A}_1\rangle\over{dt}}
  &={W_{\rm diff}W_{\rm des}\over W_{\rm ads}+W_{\rm des}}\left[
    \expec{{\rm A}_2}-\expec{{\rm A}_1}\right]
   -W_{\rm rx}\expec{{\rm A}_1}\cr
  & -W_{\rm des}\expec{{\rm A}_1}
    +{W_{\rm ads}W_{\rm des}\over W_{\rm ads}+W_{\rm des}},\cr
{d\langle {\rm B}_1\rangle\over{dt}}
  &={W_{\rm diff}W_{\rm des}\over W_{\rm ads}+W_{\rm des}}\left[
    \expec{{\rm B}_2}-\expec{{\rm B}_1}\right]
   +W_{\rm rx}\expec{{\rm A}_1}\cr
  & -W_{\rm des}\expec{{\rm B}_1},\cr
{d\langle {\rm A}_S\rangle\over{dt}}
  &={W_{\rm diff}W_{\rm des}\over W_{\rm ads}+W_{\rm des}}\left[
    \expec{{\rm A}_{S-1}}-\expec{{\rm A}_S}\right]
   -W_{\rm rx}\expec{{\rm A}_S}\cr
  &-W_{\rm des}\expec{{\rm A}_S}
   +{W_{\rm ads}W_{\rm des}\over W_{\rm ads}+W_{\rm des}},\cr
{d\langle {\rm B}_S\rangle\over{dt}}
  &={W_{\rm diff}W_{\rm des}\over W_{\rm ads}+W_{\rm des}}\left[
    \expec{{\rm B}_{S-1}}-\expec{{\rm B}_S}\right]
   +W_{\rm rx}\expec{{\rm A}_S}\cr
  &-W_{\rm des}\expec{{\rm B}_S}.\cr
\end{split}
\end{equation}

 We can use these approximate MF equations for the problem of relaxation of $\rm A$ and
$\rm B$ loadings for the case when the total loading has already reached
steady-state. The equations for $\rm A$'s can be written as
\begin{equation}
{{d \langle {\bf A} \rangle}\over{dt}}={\bf M^{\prime}}\langle {\bf A} \rangle +{\bf v^{\prime}},
\end{equation}
where $\langle{\bf A}\rangle$ is a vector containing the occupacy
probabilities with $\rm A$'s, $\bf M^{\prime}$ is the matrix of coefficients and $\bf v^{\prime}$
is the vector that makes this systems non-homogeneous having the non-zero
elements for indices 1 and $S$.
 We substitute $\langle {\bf A} \rangle $ = $\langle {\bf A} \rangle_{ss}$ +
$\bf c$ in the rate equations (47), where $\bf c$ is the vector with the deviations of 
the site occupancy probabilities with $\rm A$'s from the steady-state. The substitution
yields\

\begin{equation}
\begin{split}
{d{\bf c}\over{dt}} &= {\bf M^{\prime}} \left[ \langle{\rm A} \rangle_{ss} + {\bf c}\right] + {\bf v^{\prime}}
\cr &= {\bf M^{\prime}} {\bf c} + {\bf M^{\prime}} \langle{\bf A}\rangle_{ss} +
{\bf v^{\prime}} \cr &= {\bf M^{\prime}}{\bf c},
\end{split}
\end{equation}
because ${\bf M}^{\prime}{\langle{\bf A}\rangle}_{ss}+{\bf v}=0$ by definition.

 The equations in $\bf c$ are homogeneous and the matrix of coefficients is the
same as the one in the original rate equations.
 We can obtain the eigenvalue equation by making the substitution

\begin{equation}
{\bf c}={\bf c}^{\prime}e^{-{\omega}t}
\end{equation}
and taking out the exponential from the equations. The relaxation
time of the ${\rm A}$ loading($Q_{\rm A}$) is then the reciprocal of the
smallest eigenvalue. We can get this time by simply numerically solving the
eigenvalue equation.

\begin {figure*}[th]
\centering
\subfigure {\epsfig {figure=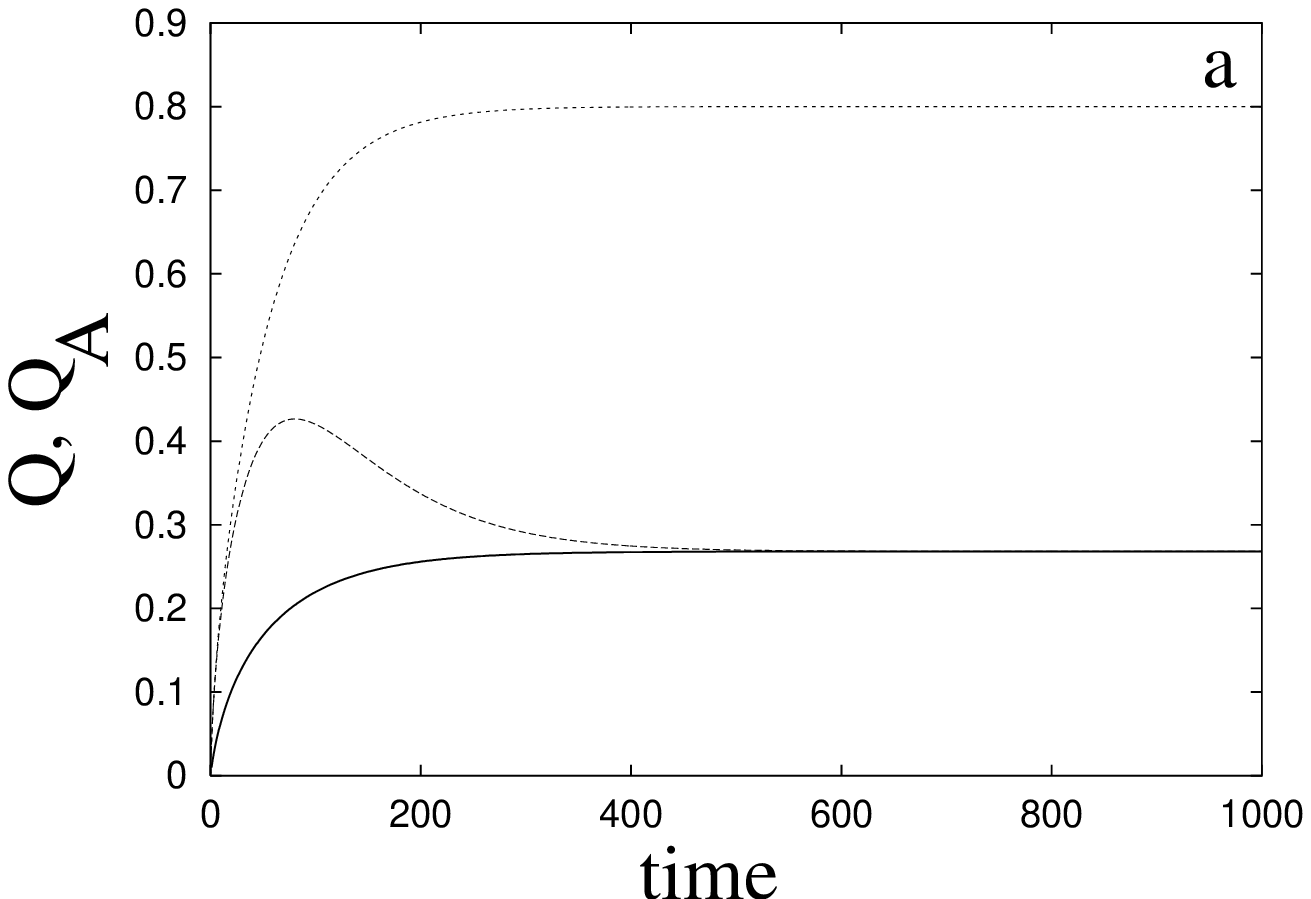, width=5.5cm} }
\subfigure {\epsfig {figure=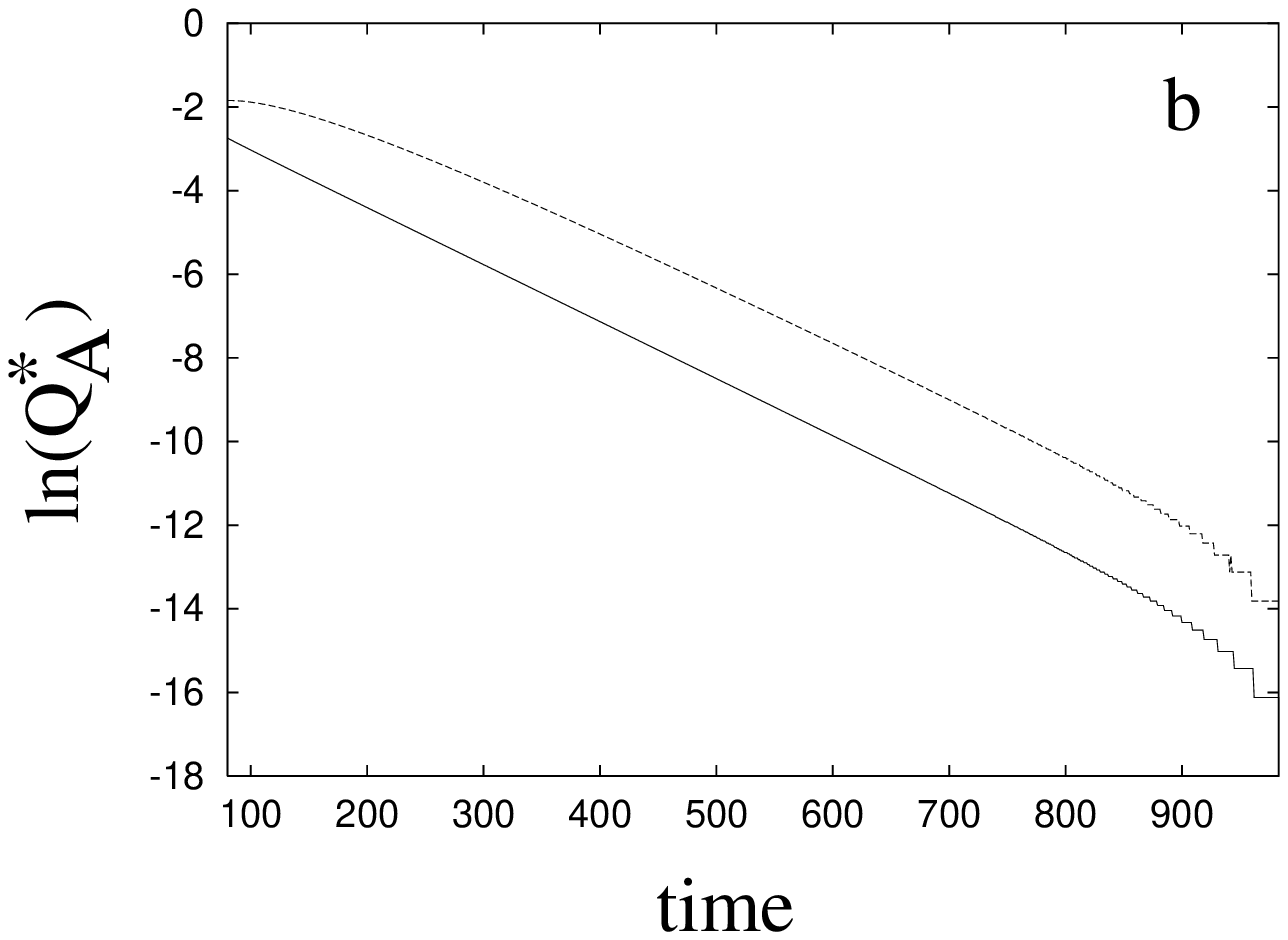, width=5.5cm} }
\caption {
          a) MF results (dashed line) and analytical results obtained
             solving the eigenvectors in the reaction limited regime(continuous
             line) for time dependence of $Q_{\rm A}$ for the parameters $W_{\rm ads}=0.8, W_{\rm des}=0.2, W_{\rm rx}=0.01,
             W_{\rm diff}=2.0, S=30$ and all the sites reactive. The
             upper dashed line represents the total loading ($Q$) in the system.
             The total loading $Q$ gives us the information when the
             reaction limited regime starts.
          b) MF(dashed) and analytical results obtained
             solving the eigenvectors in the reaction limited regim(continuous
             line) for the relaxation of $Q_{\rm A}$. We have marked with *
             the absolute value of the difference between the 
             current value and the steady-state value of $Q_{\rm A}$.
         }
\label{Arelaxare}
\end{figure*}


\begin{center} {\it {Solving the eigenvalue equation for the case
with conversion} }
\end{center}
 
\begin {figure*}[th]
\centering
\subfigure {\epsfig {figure=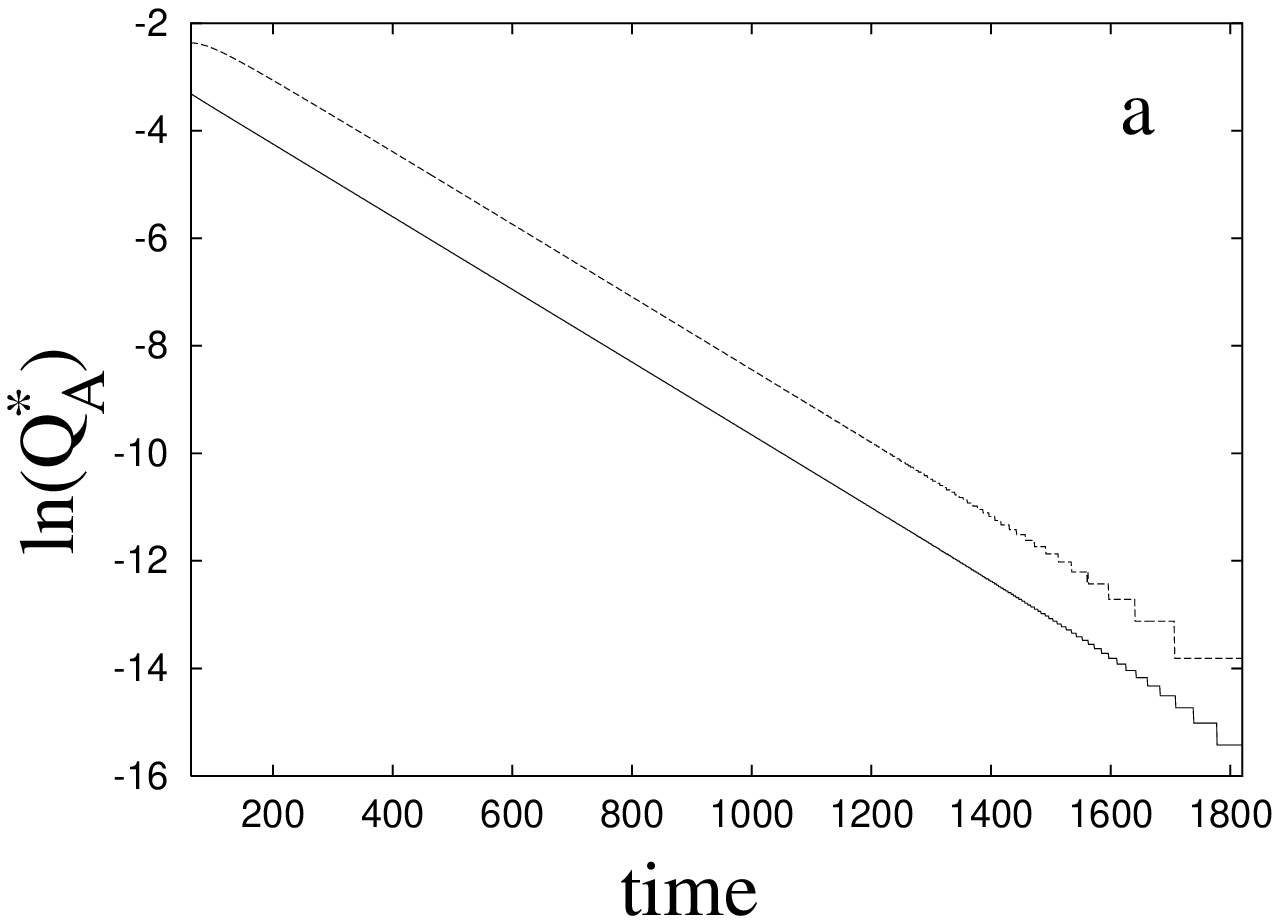, width=5.5cm} }
\subfigure {\epsfig {figure=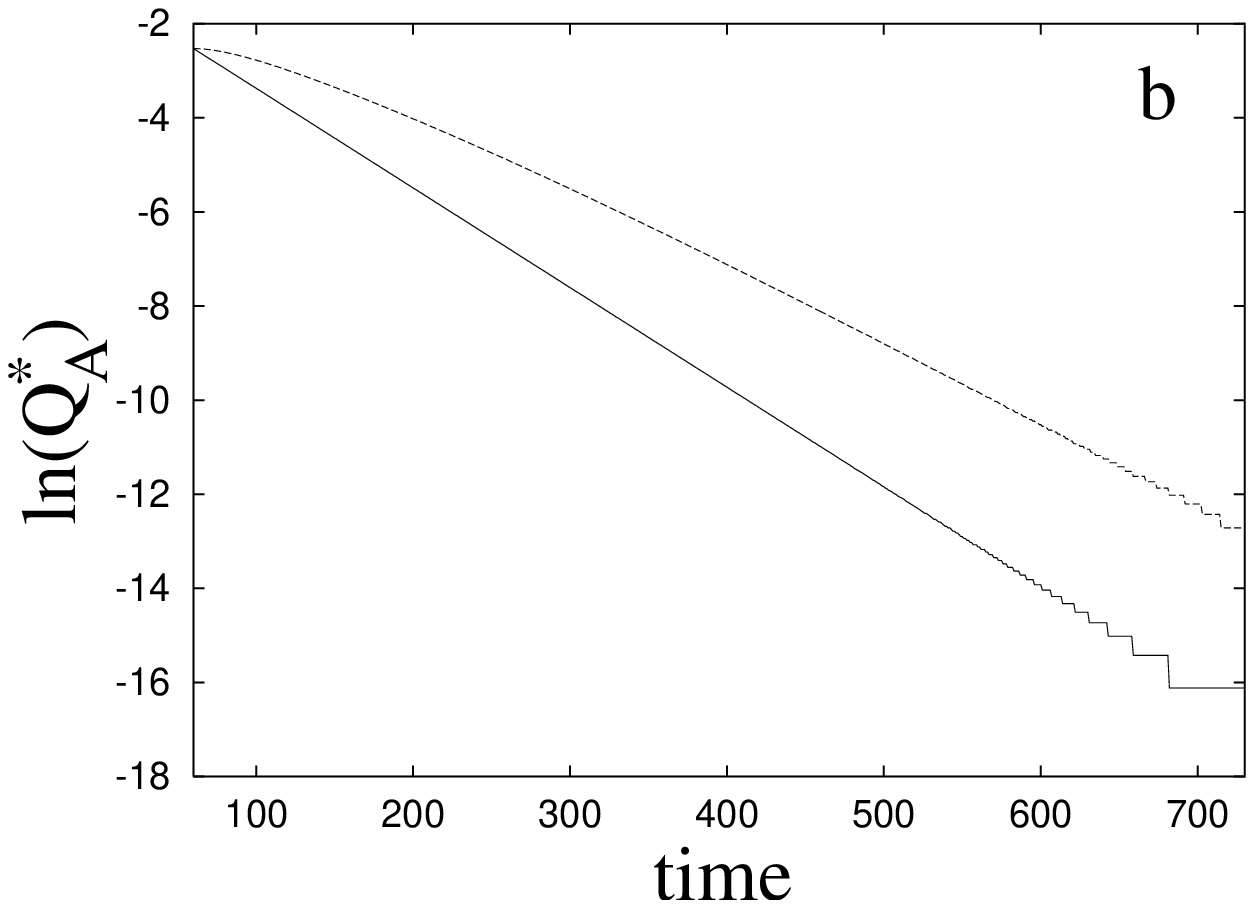, width=5.5cm} }
\subfigure {\epsfig {figure=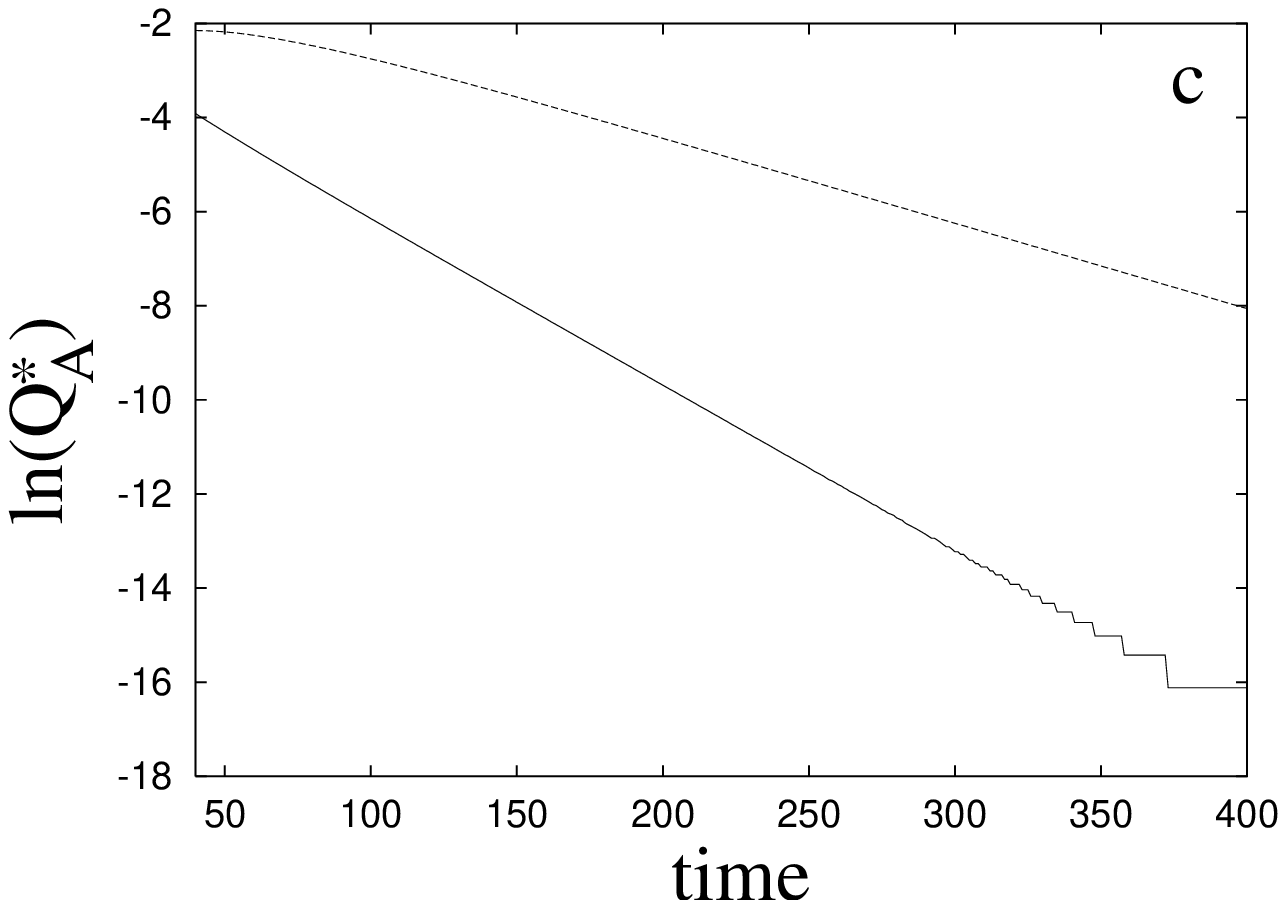, width=5.5cm} }
\caption {
          a) Time dependencies for $\ln({Q_{\rm A}}^{*})$ for a slow reaction
system ($W_{\rm ads}$=0.8, $W_{\rm des}$=0.2, $W_{\rm rx}$=0.1, $W_{\rm diff}=2$, $S$=30) in
the reaction limited regime when five left and right marginal sites are reactive.
         b)Time dependencies for $\ln({Q_{\rm A}}^{*})$ for a slow reaction
system ($W_{\rm ads}$=0.8, $W_{\rm des}$=0.2, $W_{\rm rx}$=0.1, $W_{\rm diff}=2$, $S$=30) in
the reaction limited regime when ten middle sites are reactive.
         c)Time dependencies for $\ln({Q_{\rm A}}^{*})$ for a slow reaction
system ($W_{\rm ads}$=0.8, $W_{\rm des}$=0.2, $W_{\rm rx}$=0.1, $W_{\rm diff}=2$, $S$=30) in
the reaction limited regime when a number of ten reactive sites are
homogeneously distributed in the system. 
  In a), b) and c) the continous line is for the numerical
results obtained solving the eigenvalue equation of the system in reaction
limited regime, and the dashed line is for the MF results. We have marked
with * the difference between the current value and the steady-state value
of the parameter.
          }
\label{LnA_h_homog_mid_marg}
\end{figure*}

  For fast reaction systems we have seen that using MF, the $\rm A$ and $\rm B$
loadings have the same relaxation, which is the relaxation of the total 
loading $Q$ (see figure~\ref{tranz_sr_ssr}a). The relaxation of $Q_{\rm A}$ 
can be thus derived from exact equations (13,14).
  For slow reaction systems we can derive analytically the relaxation of
$Q_{\rm A}$ and $Q_{\rm B}$ for the reaction limited regime.
 In this case we expect the relaxation of $Q_{\rm A}$ and $Q_{\rm B}$ to be 
determined only by reaction. From the MF
results we remark that desorption has also a strong influence on
the transients (see figure~\ref{tranz_sr_ssr}c). This is happening because,
in the case desorption is very high, the adsorbed particles at the marginal
sites will hardly diffuse into the pipe, most of them being desorbed
immediately. As a result, few particles will succeed in getting to the middle
sites and the residence time of the particles near the marginal sites will
decrease. The loading of the pipe with particles will converge than slower
to steady-state.
 In figure~\ref{Arelaxare} we can see that the analytical results for the
transients obtained solving the eigenvalue equation corresponds to the numerical
results obtained using the MF Approximation for the case all the sites are
reactive. In figure~\ref{LnA_h_homog_mid_marg} we have analytical results from the eigenvalue
equation and MF results for the system in reaction limited regime and for 
different distributions of the reactive sites. We observe that for marginal
sites reactive the results are similar but differ considerably for the middle sites
reactive and for the homogeneous distribution of the reactive sites.

\begin {figure}[th]
\centering
\subfigure {\epsfig {figure=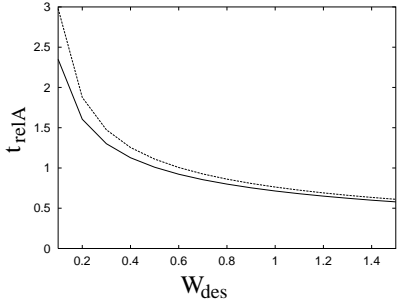, width=5.5cm} }
\caption {
          a) The analytical results for the relaxation of $Q_{\rm A}$ as a
             function on $W_{\rm des}$ for different reaction rates.
             The continuous line is for $W_{\rm rx}=0.1$ and the dashed line for
             $W_{\rm rx}=0.01$, when $W_{\rm des} \in [0, 2]$.
         }
\label{Arelaxare1}
\end{figure}

 Analytically, solving the eigenvalue equation for a very slow reaction
system, we find that the relaxation of $Q_{\rm A}$ ($t_{relA}$) as a function
of desorption varies with reaction for low desorption rates and converges to a 
limiting value for very high rates of desorption (see figure~\ref{Arelaxare1}-a,b). 
 The dependence of $\rm A$ loading on $W_{\rm des}$ has
two regimes, the first for low desorption rates when $\rm A$ loading
strongly decreases with desorption, and the second when the $\rm A$
loading is converging to a limiting value and the adsorption
process takes over the system behavior.
 
 We remark that the analytical results for $Q_{\rm A}$ obtained from the
eigenvalue equation don't give the MF peak of $\rm A$'s particles
accumulated in the transient regime for a slow reaction system.

\subsection { Comparisons with simulation results}

 We present now the results obtained for the transients using DMC methods for
different sets of parameters. We compare them with the Mean-Field and Pair Approximation 
results. As for very large pipes the computational effort is considerable,
we study a system of size $S=30$. We have considered separately the sets of
parameters in Table~\ref{tab:Table1}.

   The sets of parameters from a) to e) are for the cases of low loading and from f) to j)
for high loading. The parameters in the table describe the
following situations:
a) and f) for very slow reaction and slow diffusion;
b) and g) for slow reaction and slow diffusion;
c) and h) for slow reaction and fast diffusion;
d) and i) for fast reaction and slow diffusion;
e) and j) for fast reaction and fast diffusion.

 As the MF results for the total loading are exact, there are, as expected,
no differences between these results and those of the simulation for the
total loading.
 We compare the MF results with the simulation results for the case with
conversion in the case all sites are reactive and when only some of the sites
reactive.
\\
\begin {table}[hb]
\vspace*{-0.5cm}
\begin {center}
\begin{tabular}{|l|cccc|}
\hline case & $W_{\rm ads}$ & $W_{\rm des}$ & $W_{\rm diff}$ & $W_{\rm rx}$\\
\hline a)& 0.2 & 0.8 & 0.05 & 0.01\\
 b)& 0.2 & 0.8 & 0.05 & 0.1\\
 c)& 0.2 & 0.8 & 2 & 0.1\\
 d)& 0.2 & 0.8 & 1 & 2\\
 e)& 0.2 & 0.8 & 10 & 2\\
 f)& 0.8 & 0.2 & 0.05 & 0.01\\
 g)& 0.8 & 0.2 & 0.05 & 0.1\\
 h)& 0.8 & 0.2 & 2 & 0.1\\
 i)& 0.8 & 0.2 & 1 & 2\\
 j)& 0.8 & 0.2 & 10 & 2\\
\hline
\end{tabular}
\end{center}
\caption {\label{tab:Table1}Table containing the sets of parameters used for the simulations}
\end{table}

\subsubsection {\label{sec:lev3} All sites reactive}

\begin {figure*}
\centering
\subfigure {\epsfig {figure=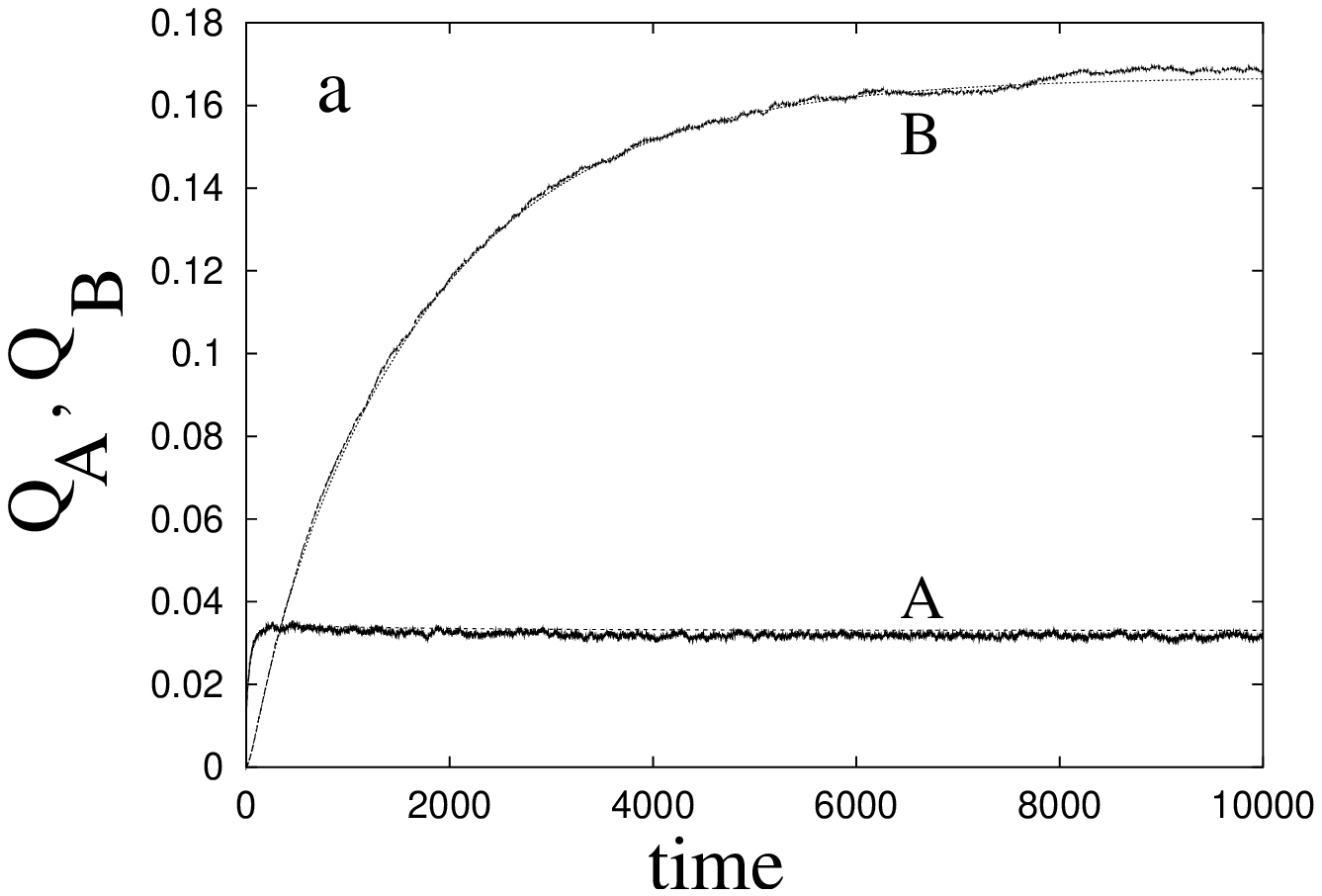, width=5.5cm} }
\subfigure {\epsfig {figure=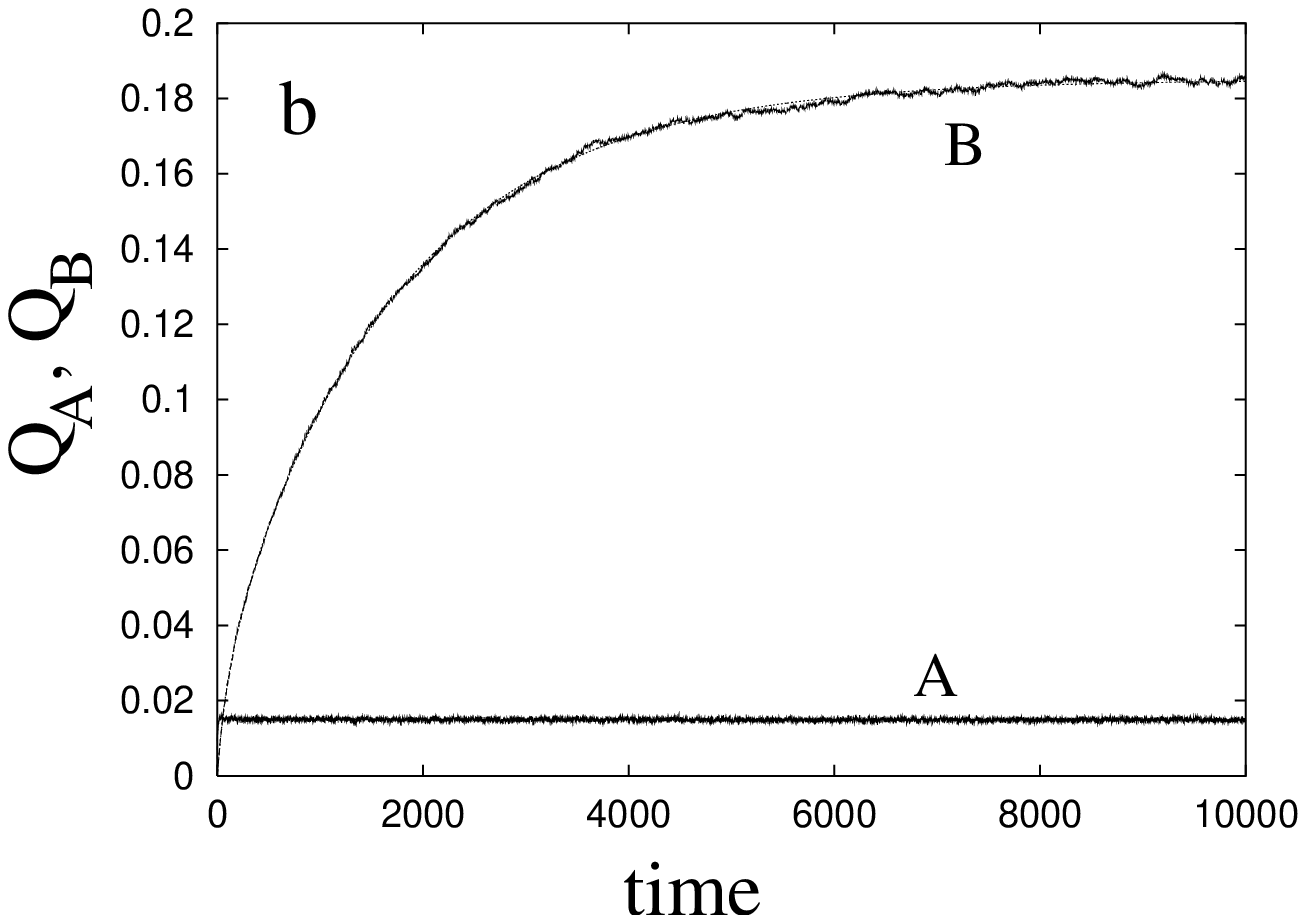, width=5.5cm} }
\subfigure {\epsfig {figure=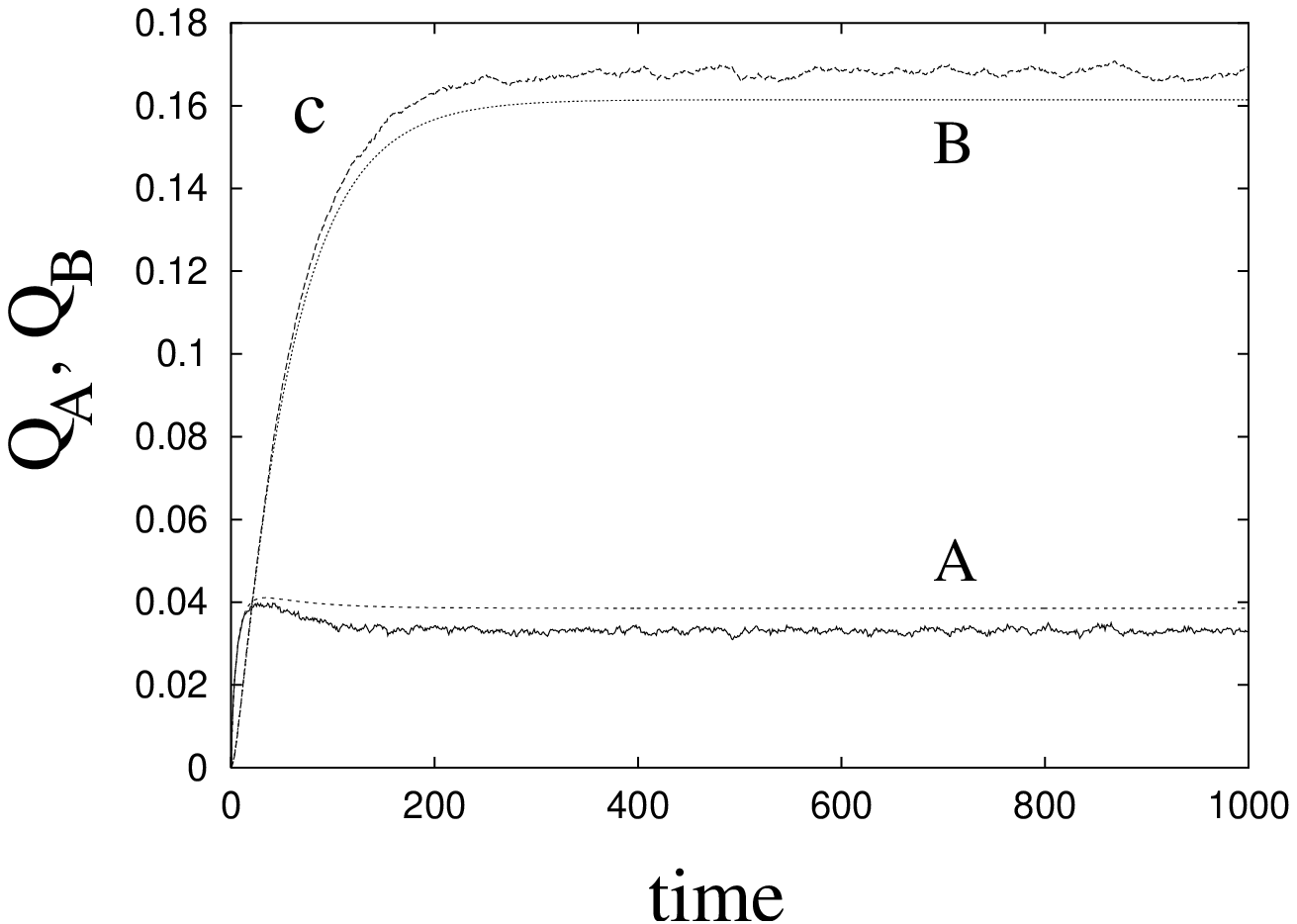, width=5.5cm} }
\subfigure {\epsfig {figure=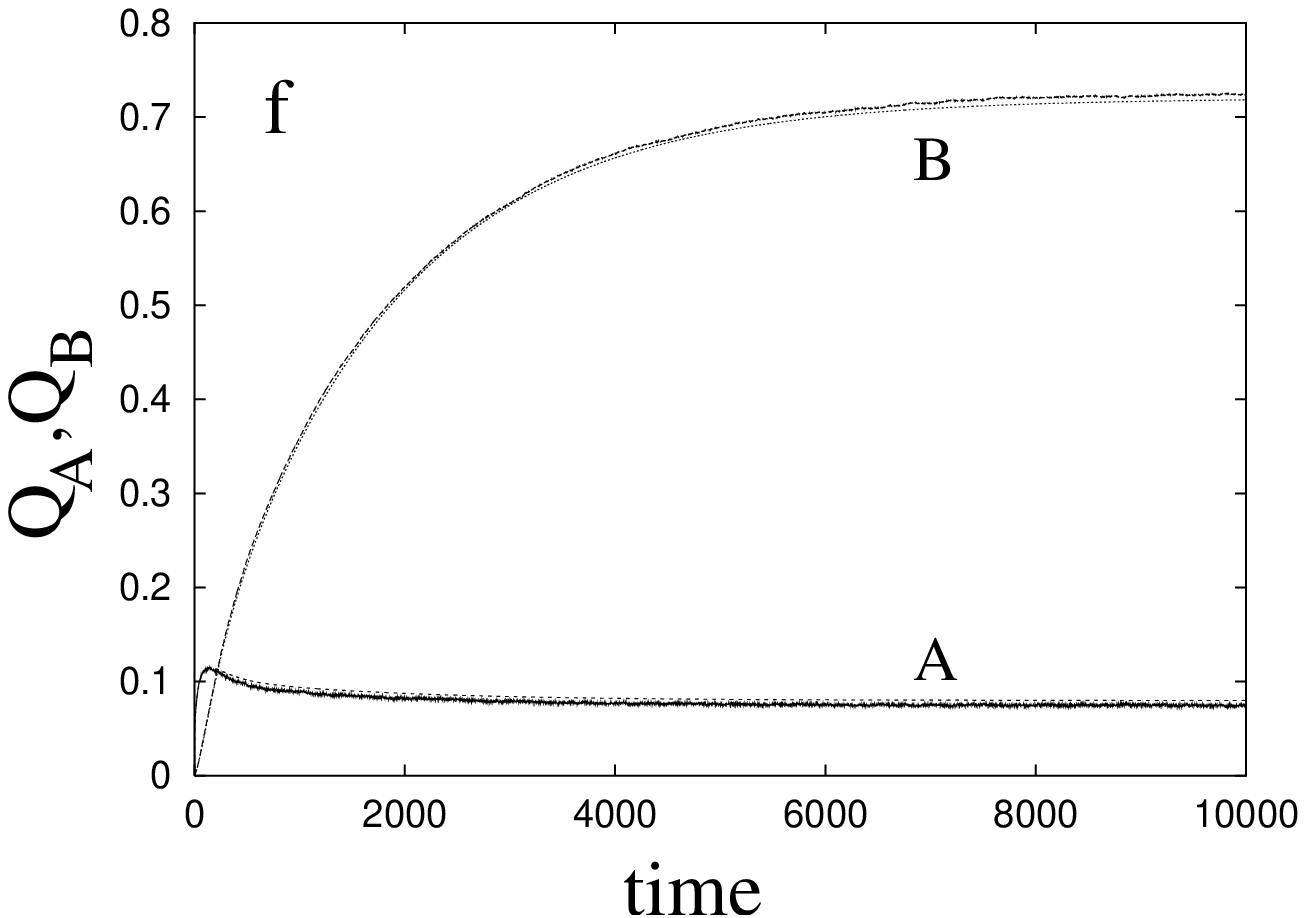, width=5.5cm} }
\subfigure {\epsfig {figure=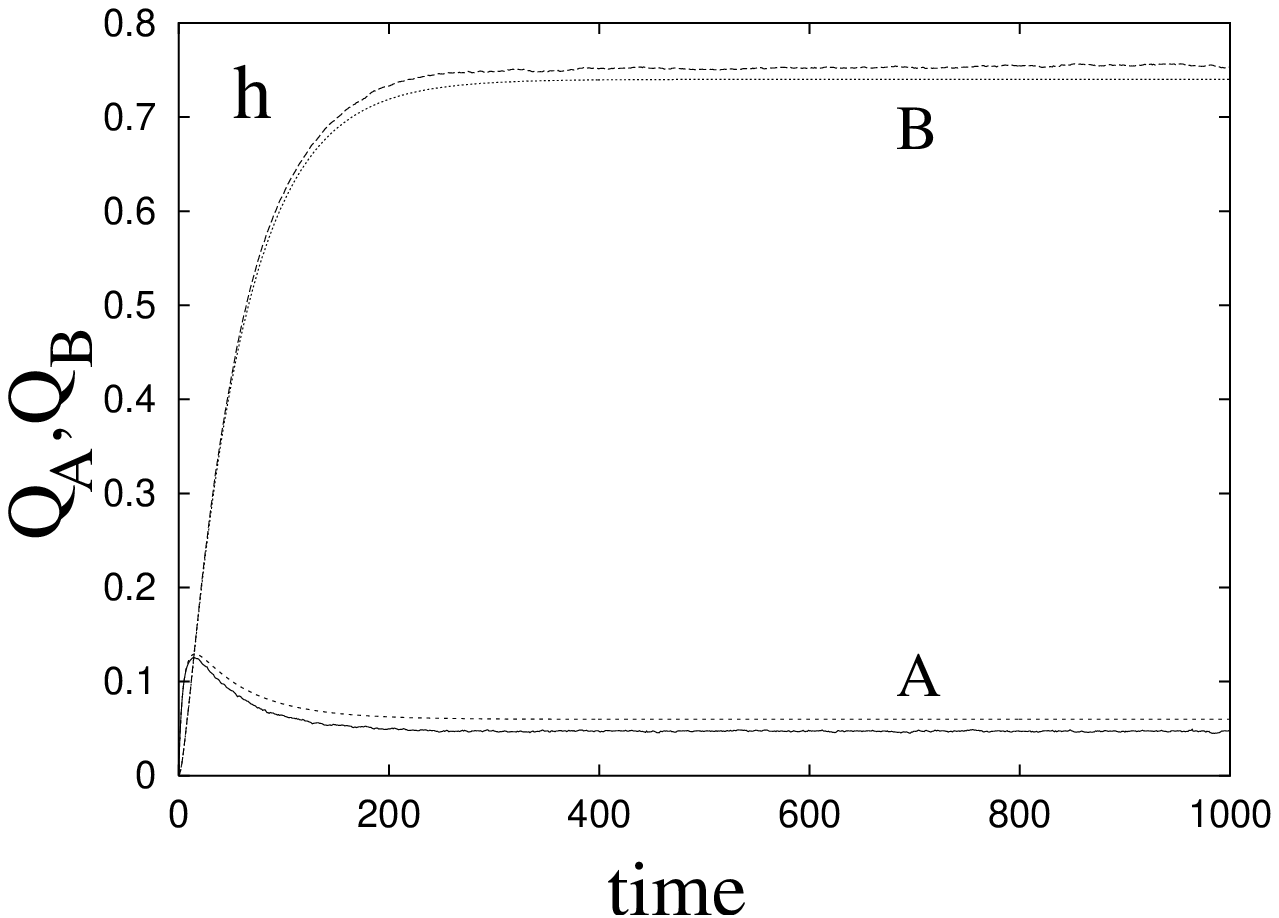, width=5.5cm} }
\caption { DMC and MF results for the time dependencies
          of $Q_{\rm A}$ and $Q_{\rm B}$ for the cases a, b, c, f, h in TABLE I
          when all the sites are reactive.
          The straight lines visible for cases c and h (fast
          diffusion-slow reaction) are the MF lines. In the other cases the
          MF and the DMC results are indistinguishable.
         }
\label{QA_QB_time_conv}
\end{figure*}

\begin {figure*}
\centering
\subfigure {\epsfig {figure=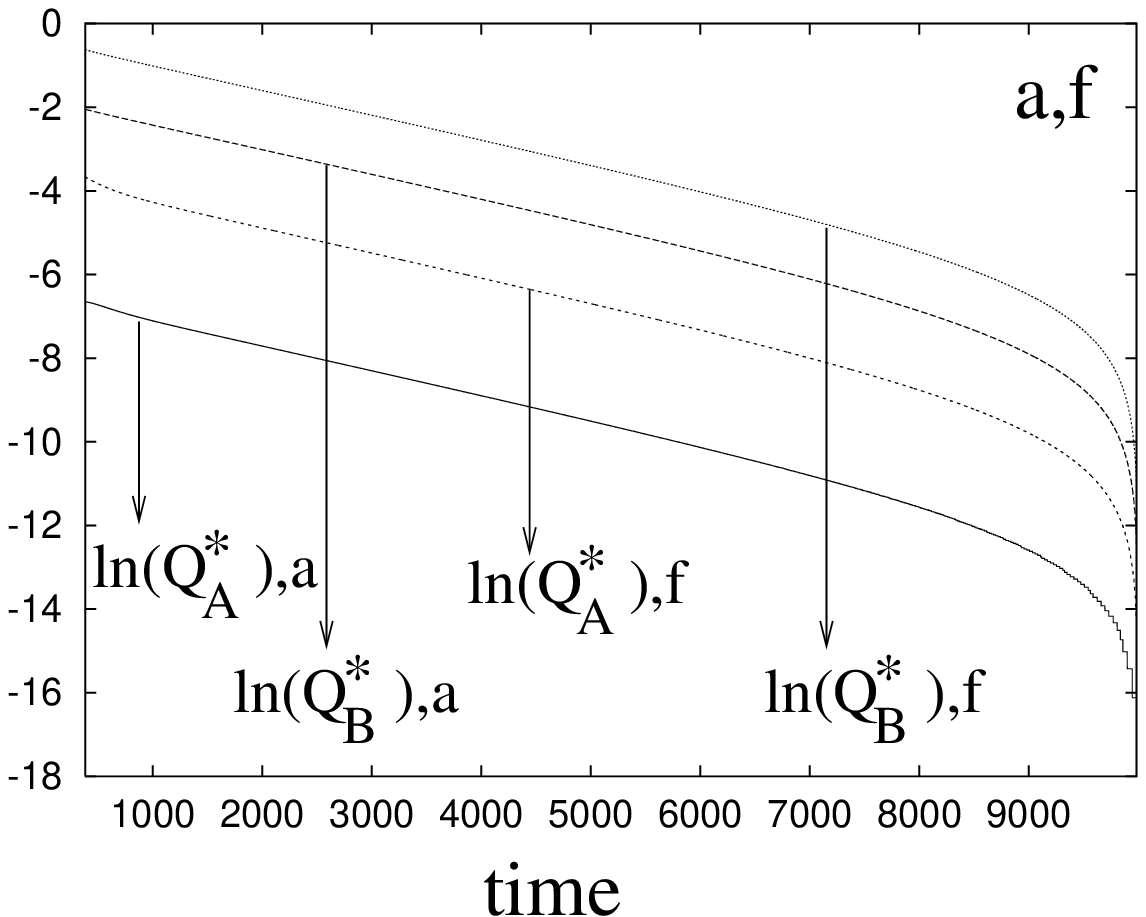, width=5.5cm} }
\subfigure {\epsfig {figure=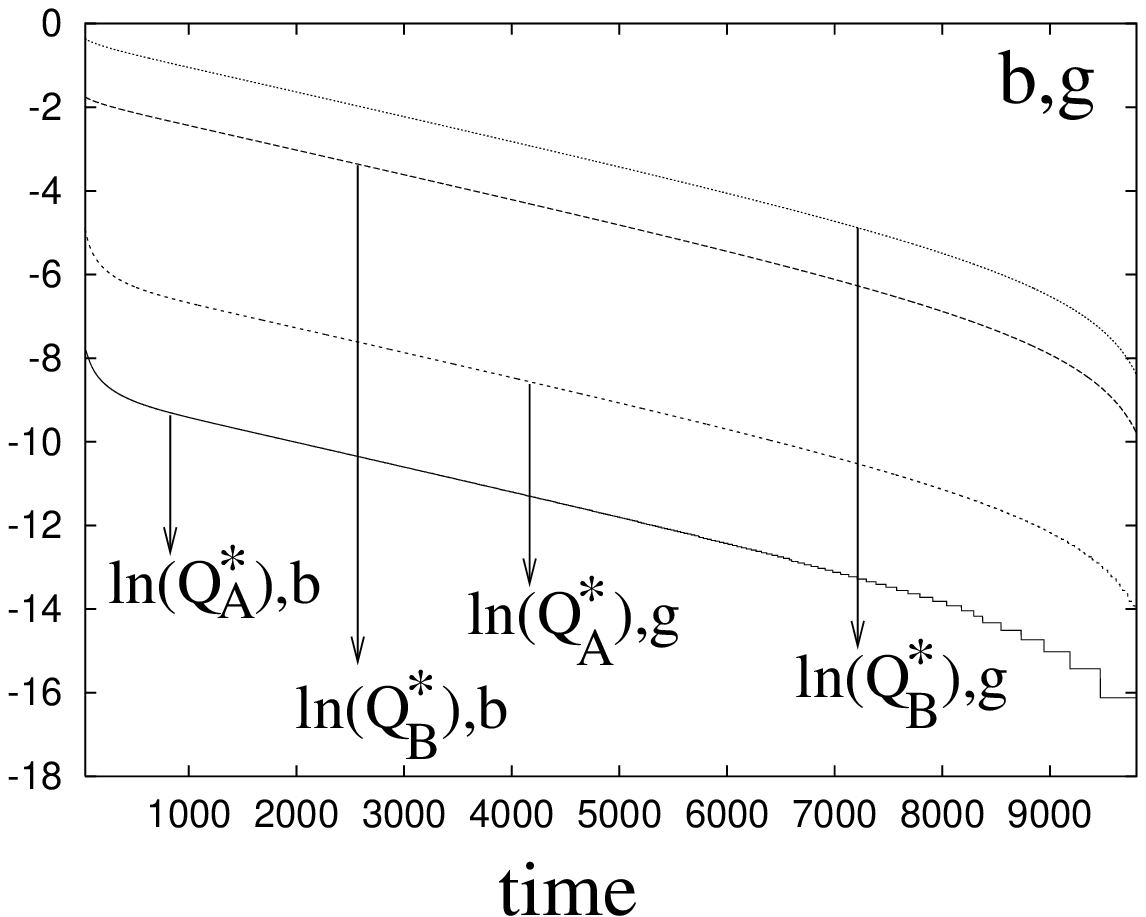, width=5.5cm} }
\subfigure {\epsfig {figure=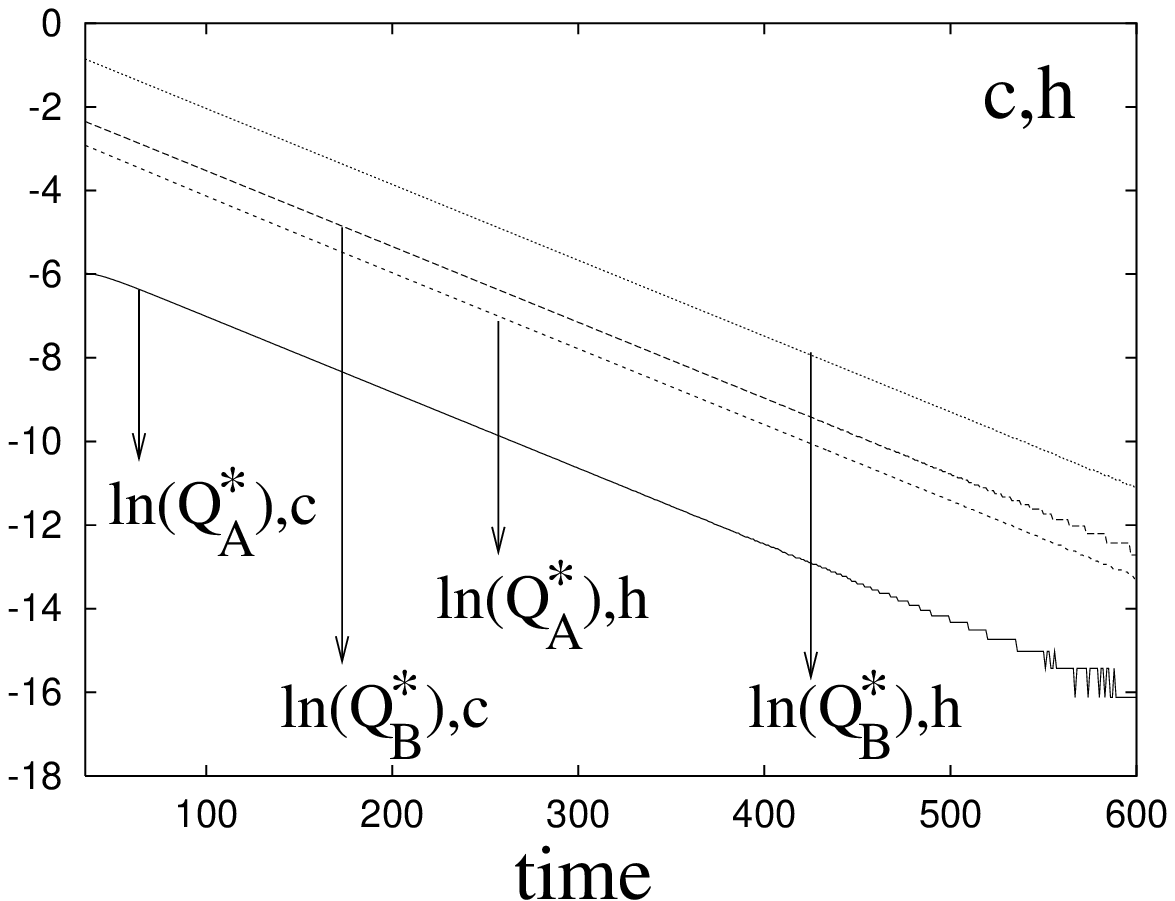, width=5.5cm} }
\subfigure {\epsfig {figure=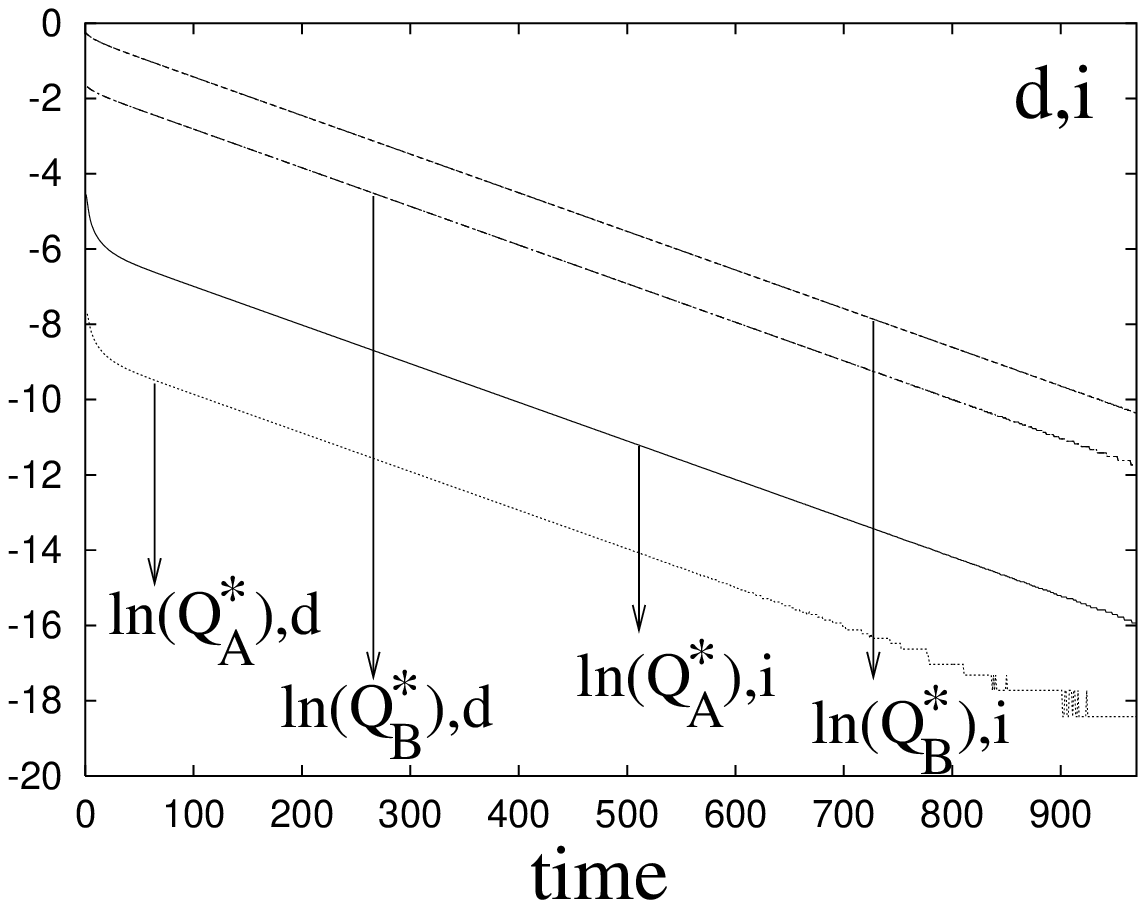, width=5.5cm} }
\subfigure {\epsfig {figure=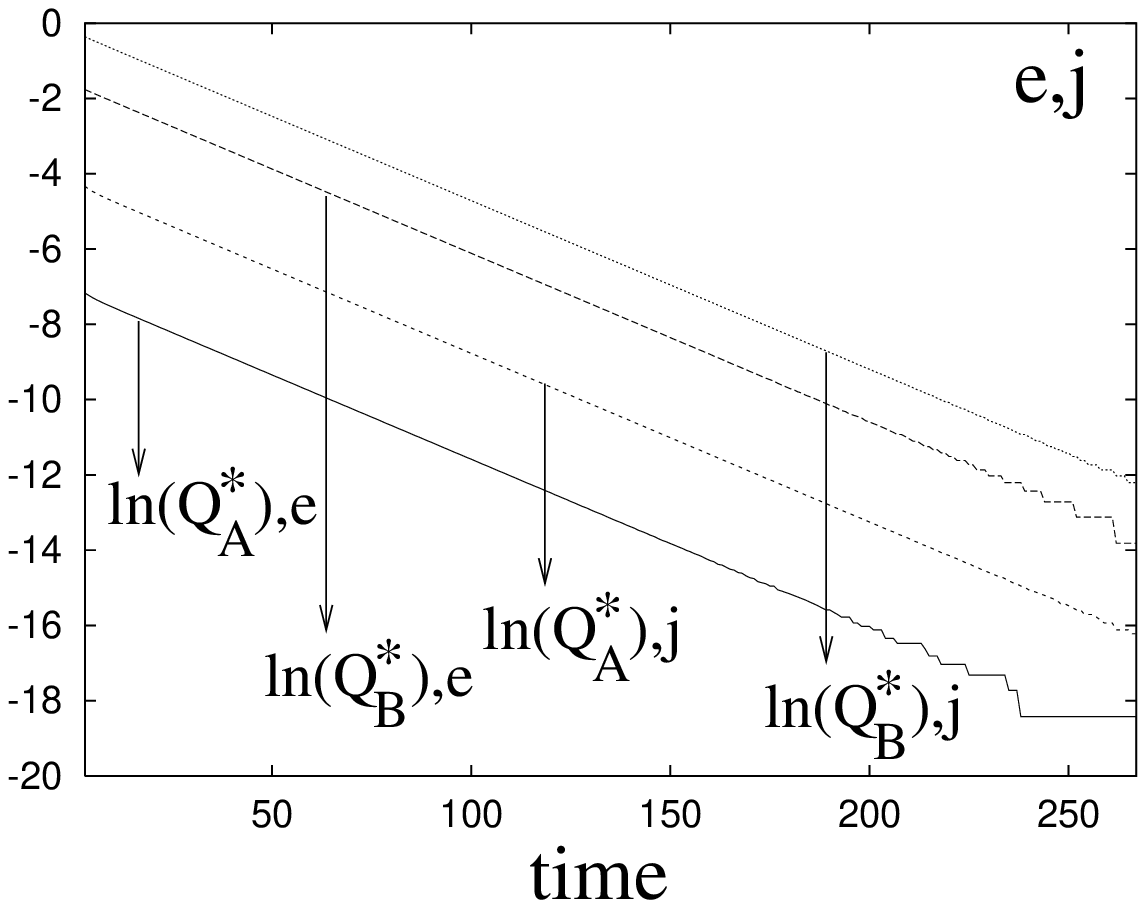, width=5.5cm} }
\caption { 
	   Time dependencies for $\ln({Q_{\rm A}}^{*})$ and $\ln({Q_{\rm
B}}^{*})$ for the cases a and f (slow diffusion-very slow reaction at low and high loadings), b and g
(slow diffusion-slow reaction at low and high loadings), c and h(fast
diffusion-slow reaction at high and low loadings), d and i (fast
diffusion-very fast reaction at low and high loadings), e and j(very fast
diffusion-fast reaction at low and high loadings) in TABLE I using MF. In
all the figures we have marked with * the absolute value of the difference between the current value and the steady-state value
of the parameter. 
         }
\label{ln_af_mf}
\end{figure*}

 We first look at the time dependence of the loading with $\rm A$ ($Q_{\rm A}$) and loading
with $\rm B$ ($Q_{\rm B}$).
 From the simulation results in figures~\ref{QA_QB_time_conv} and
~\ref{ln_af_mf} we see several regimes for the transients.
In the case diffusion is slow, the relaxation time is determined by
diffusion. When diffusion is fast and the reaction slow, then the relaxation
time is determined by reaction and when both are fast, relaxation time is
determined by adsorption/desorption.
 For all these cases the simulations results for the transients match the MF
results, except when we have low reaction rates and fast diffusion for both
low and high loading.

\begin{figure}
\centering
\subfigure{\epsfig{figure=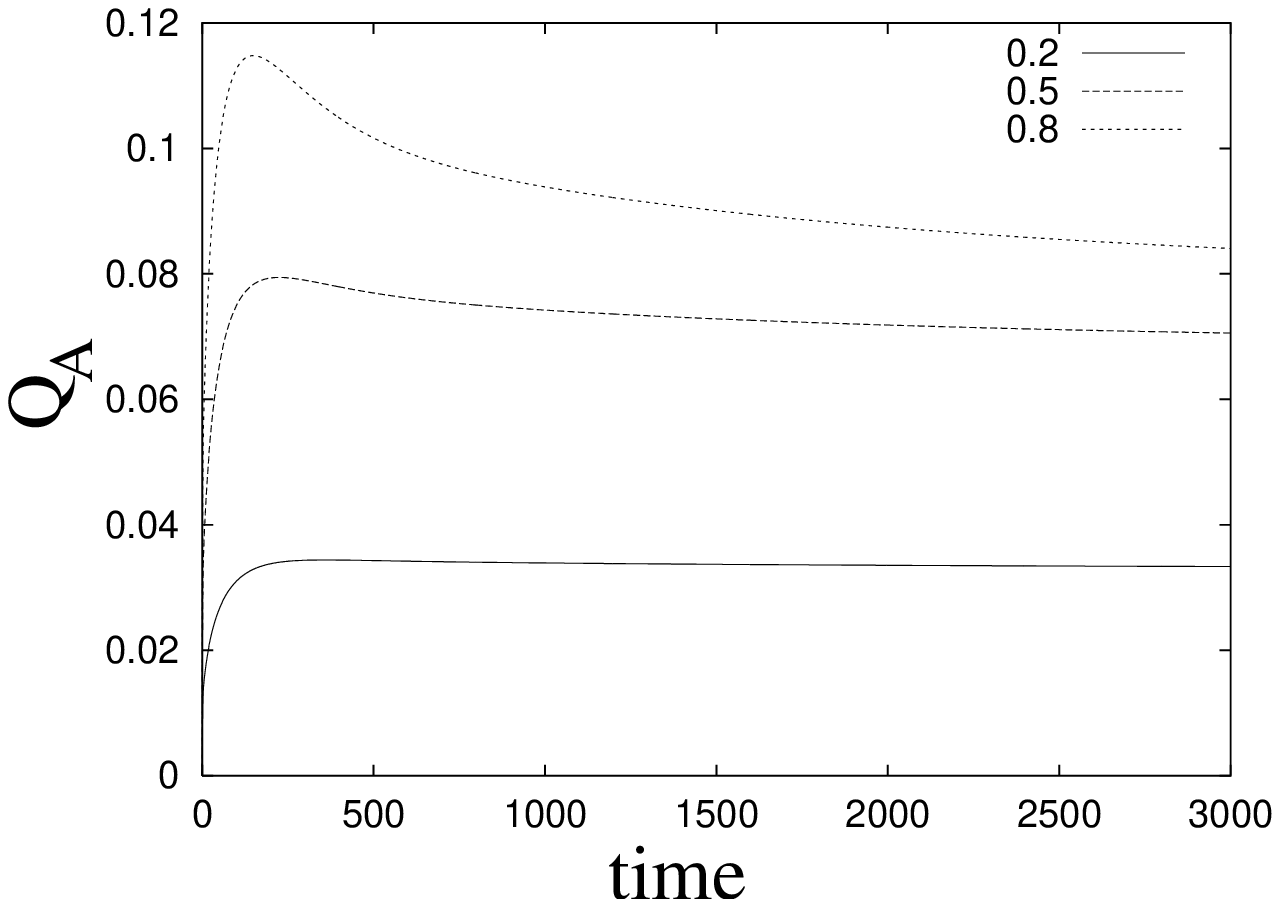,width=6.5cm} }
 \caption {
          MF results for the time dependence of $Q_{\rm A}$ for $Q$=0.2, 0.5, 0.8, $W_{\rm rx}$=0.01.
          }
\label{Aloading_time_totload}
\end{figure}

\begin {figure}
\centering
 \subfigure {\epsfig {figure=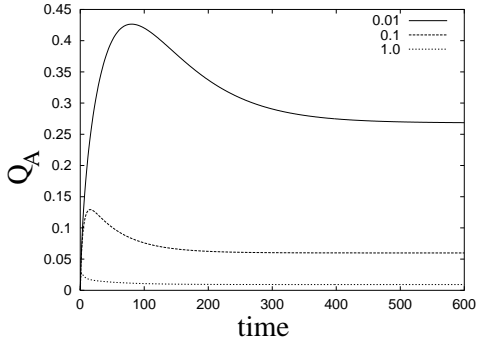,width=6.5cm} }
 \caption {
           MF results for the time dependence of $Q_A$ for $W_{\rm rx}$=0.01, 0.1, 1, when $W_{ads}$=0.8
          and $W_{\rm des}$=0.2.
          }
\label{Aloading_time_reac}
\end{figure}

\begin{figure}
\centering
\subfigure {\epsfig {figure=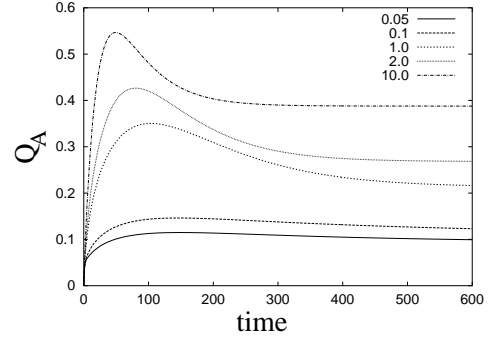,width=6.5cm} }
 \caption {
        MF results for the time dependence of $Q_{\rm A}$ for $W_{\rm diff}$=0.05, 0.1, 1, 2, 10, when
        $W_{\rm ads}$=0.8, $W_{\rm des}=0.2, W_{\rm rx}=0.01.$
          }
\label{Aloading_time_diff}
\end{figure}

  When diffusion is fast and reaction is slow, MF overestimates the amount
of $\rm A$'s in the pipe both for transients and for steady-state. 
DMC and MF results indicate an overshoot for $Q_{\rm A}$ both for high and low
loadings in the transient regime. The overshooting appears as a consequence 
of the difference between diffusion and reaction rates constants. 
Because the reaction is slow and diffusion fast, many $\rm A$'s start
accumulating into the system and they are only later converted into $\rm B$'s.
  The moment $t_{max}$ when the peak appears is determined by the ratio
between $W_{\rm diff}$ and $S$, but always shortly after the initial moment $t_0$
and it lasts only a short time.
   The height of the peak depends on the total loading
($W_{\rm ads}/W_{\rm des}$) (see figure~\ref{Aloading_time_totload}) and
on the ratio between $W_{\rm rx}$ and $W_{\rm diff}$(see figure
~\ref{Aloading_time_reac} and ~\ref{Aloading_time_diff}).
For the case of slow reaction and fast diffusion(f,g,h) in table II, the higher 
the peak, the lower the ratio between $W_{\rm rx}$ and $W_{\rm diff}$.
   In the table II we give the relative height of the peak ($\Delta H/{Q_{\rm A}}$) for different
$W_{\rm rx}$ and $W_{\rm diff}$, at high loading
($W_{\rm ads}=0.8, W_{\rm des}=0.2$), where $\Delta H$ is the height of
the peak.
\\
\begin {table}[h]
\vspace*{-0.5cm}
\begin {center}
\begin{tabular}{|l| p{0.9cm} p{0.9cm} p{1.0cm} p{1.0cm}|}
\hline \multicolumn{1}{|c}{case} & \multicolumn {1}{c}{$W_{rx}$} & \multicolumn {1}{c}{$W_{diff}$} & \multicolumn {2}{c|}{$\Delta H/Q_A$}\\
\hline   &          &            & MF & Sim\\
 a)& 0.01 & 2.0 & 1.5905 & 2.9521\\
 b)& 0.1 & 2.0 &  2.1590 & 2.693\\
 c)& 1.0 & 2.0 &  2.9911 & 2.9984\\
 d)& 0.01 & 0.05 & 1.4365 & 1.2761\\
 e)& 0.01 & 0.1 &  1.5070 & 1.4546\\
 f)& 0.01 & 1.0 &  1.6458 & 2.654\\
 g)& 0.01 & 2.0 &  1.5904 & 2.9384\\
 h)& 0.01 & 10.0 & 1.4095 & 3.141\\
 i)& 2.0 & 2.0 & 3.329 & 3.305 \\
\hline
\end{tabular}
\end{center}
\caption {Table containing the height of the peak for different
$W_{rx}$ and $W_{diff}$}
\end{table}

\begin {figure}[h]
\centering
\subfigure {\epsfig {figure=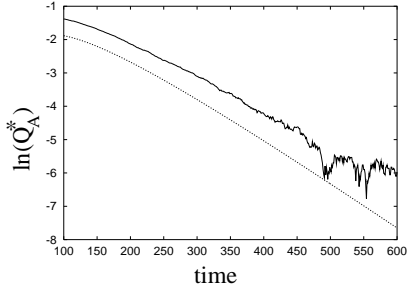, width=5.5cm} }
\caption { Analytical and simulation results
           for slow reaction systems, $W_{\rm diff}$=2, $W_{\rm rx}$=0.1, $W_{\rm ads}$=0.8,
           $W_{\rm des}$=0.2, when all the sites are reactive.
           The simulation (continous line) and analytical
           (dotted line) results for the $\ln({Q_{\rm A}}^{*})$ as a function on time.
           We have marked with * the difference between the current value and the
           steady-state value of $Q_{\rm A}$.
         }
\label{Sim_anal_Arelax}
\end{figure}

  MF overestimates the height of this peak comparing with simulation
results, but both DMC and MF results converge in the same way
to steady-state.
In figure~\ref{Sim_anal_Arelax} we show that MF results corresponds to DMC
results for transients for slow reaction systems.
 When difussion is slow and reaction is fast, MF predicts very well the DMC
results. We have also in this case an overshoot for $Q_{\rm A}$ in the transient regime.
 When we have slow diffusion and very slow reaction, the height of the
peak increases with both $W_{\rm rx}$ and $W_{\rm diff}$ (see table II
for d, e, f). Also in this case the MF corresponds to the DMC results.


  As MF ignores the spatial correlations between NN sites and MF
gives qualitatively good results comparing with DMC results, we
conclude that MF is a good enough approximation for the case when all the
sites are reactive. This is confirmed also by the comparison between MF and 
Pair Approximation. The Pair Approximation gives the same results as MF.

\subsubsection {\label{sec:lev4} Only some of the sites reactive }

  We want to see how the distribution of the reactive sites influence
the relaxation of $\rm A$ and $\rm B$ loadings. We compare the MF and the DMC
results for fast reaction and for slow reaction systems, for
different distributions of the reactive sites. We distinguish homogeneous
distribution of the reactive sites, marginal sites reactive, and middle sites reactive).

  We have previously seen that for fast reaction systems and all sites reactive,
the relaxation of $Q_{\rm A}$ and $Q_{\rm B}$ is the same as the relaxation of
the total loading $Q$. From~\cite{silvia} we know that the
results for the total loading, both for transients and for
steady-state, can be derived analytically from exact equations and
these results corresponds to the DMC results. When only the
marginal sites are reactive and reaction very fast, DMC and MF
results are similar with DMC and MF results for the cases when all
sites are reactive and when it is a homogeneous distribution of
the reactive sites in the pipe. When the reactive
sites are situated in the middle of the pipe, the loadings $Q,
Q_{\rm A}$,and $Q_{\rm B}$ relax slower to equilibrium than in the case
all the sites are reactive, because it takes more time for the $\rm A$
particles to reach the reactive sites.

  For slow reaction systems and all sites reactive, the relaxation of
$Q_{\rm A}$ and $Q_{\rm B}$ is slower than the relaxation of the total loading
$Q$. For this case we can not derive $Q_{\rm A}$ from exact equations.
The $\rm A$ loading is converging to equilibrium simultaneously with $\rm B$ loading.
 We compare thus DMC and MF results for different distributions of
the reactive sites, for different reaction rates. We will analyze
two cases, the first for slow reaction ($W_{\rm rx}=0.1$) and the
second for very slow reaction ($W_{\rm rx}=0.01$).

\begin {figure*}[ht]
\centering
\subfigure {\epsfig {figure=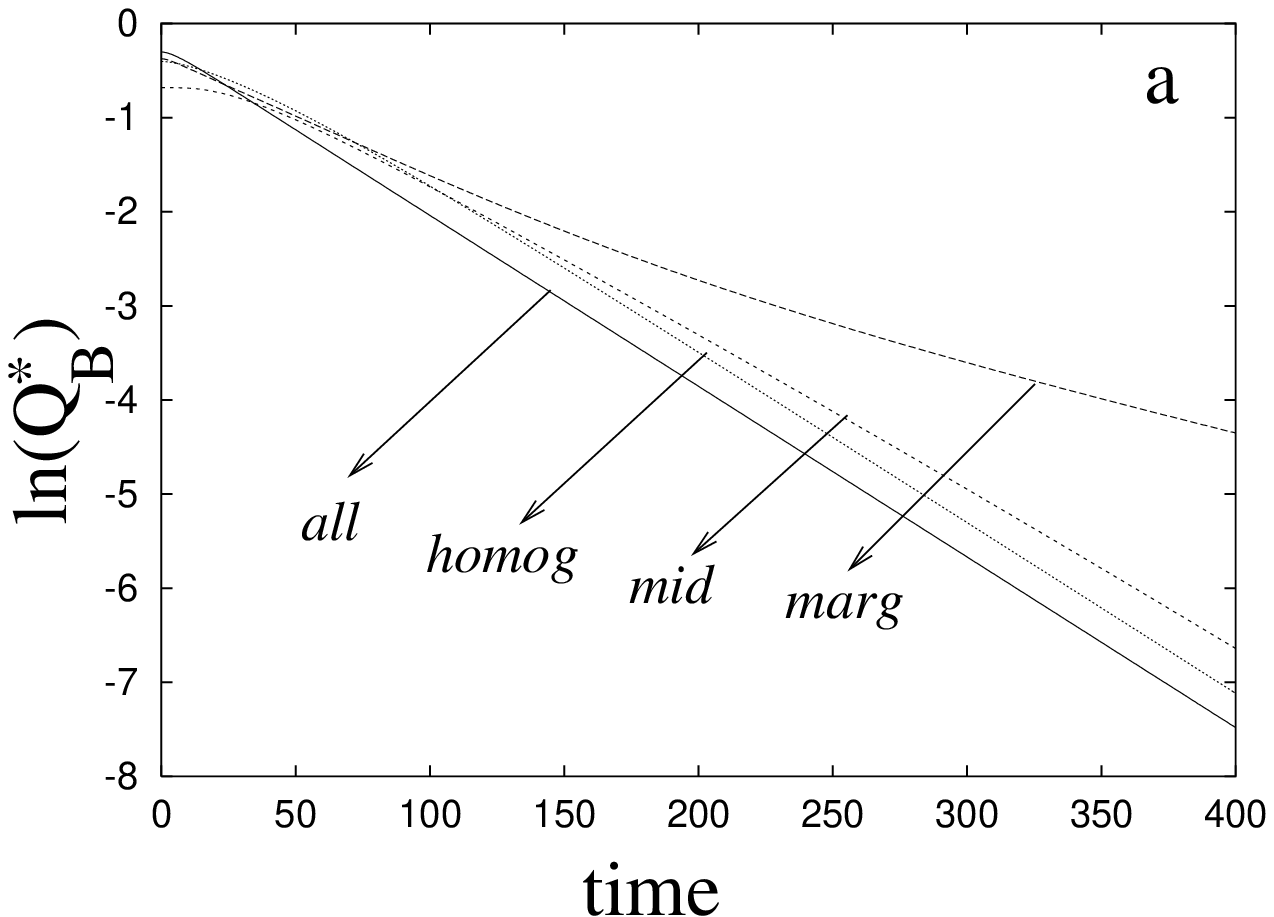, width=5.5cm} }
\subfigure {\epsfig {figure=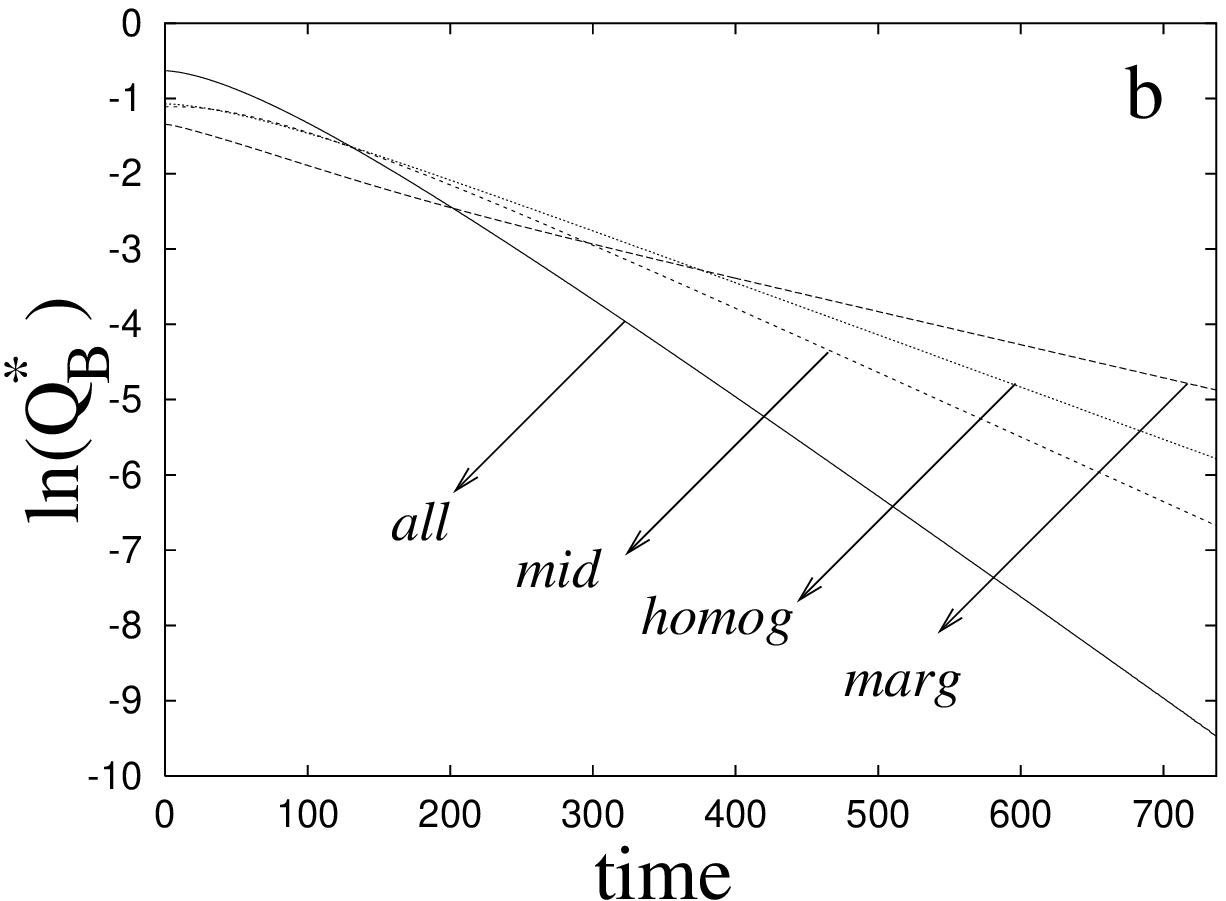, width=5.5cm} }
\caption {
          a)Time dependencies for $\ln({Q_{\rm B}}^{*})$ when $W_{\rm ads}=0.8$,
$W_{\rm des}=0.2, W_{\rm diff}=2, W_{\rm rx}=0.1$ using MF for marginal sites reactive,
homogeneous distribution, middle sites reactive, and all sites reactive.
          b)Time dependencies for $\ln({Q_{\rm B}^{*}})$ when $W_{\rm ads}=0.8$,
$W_{\rm des}=0.2, W_{\rm diff}=2, W_{\rm rx}=0.01$ using MF for the same distributions of
the reactive sites as in a.
          In a) and b) we have marked with * the absolute value of the difference
 between the current value and the steady-state value of the parameter.
         }
\label{tranzAB_sr_ssr_mf}
\end{figure*}

   In the first case ($W_{\rm rx}=0.1$), the $\rm B$ loading for homogenous
distribution and middle site reactive reaches equilibrium faster
than in the case of marginal sites reactive. This is happening
because for marginal sites reactive, the $\rm B$'s are formed near the
open ends and, consequently, they can easily desorb and the
equilibrium is reached later.
   When the marginal sites are reactive, the residence time of the $\rm B$'s
is small and the probability to find a $\rm B$ on a marginal site is high.
  The loading with $\rm B$'s is increasing more slowly to the steady-state value
because of the $\rm A$'s that are in the middle of the pipe.
The same behavior has the $\rm A$ loading converging to equilibrium simultaneously with
$\rm B$ loading.
   When the reactive sites are distributed 
in the middle of the pipe, because of blocking, the $\rm B$'s can not
reach easily the open ends, the probability to find a $\rm B$ on a marginal
site is small, and the residence time of the $\rm B$'s in the system is large.
The loading with $\rm B$'s is increasing fast to the steady-state value.
   From figure ~\ref{tranzAB_sr_ssr_mf}, we see that $Q_{\rm A}$ and $Q_{\rm B}$
for homogeneous and middle sites reactive have the same
relaxation.

  In the case of very slow reaction ($W_{\rm rx}=0.01$), when the marginal
sites are reactive, the $\rm B$ loading increases faster than in the
case of homogeneous and middle sites reactive, till a certain time
because $\rm A$ particles can reach faster the reactive sites. While
more $\rm B$'s are formed, the $\rm B$ loading is converging slower
because the $\rm B$'s can desorb relatively fast from the marginal
sites, the residence time of the formed $\rm B$'s is smaller than in
the case when the reactive sites are in the middle of the pipe or
homogeneously distributed (see figure~\ref{tranzAB_sr_ssr_mf}).
$Q_{\rm B}$ converges faster in case the middle sites are reactive than
in the case of homogeneous distribution of the reactive sites.

\begin {figure*}[th]
\centering
\subfigure {\epsfig {figure=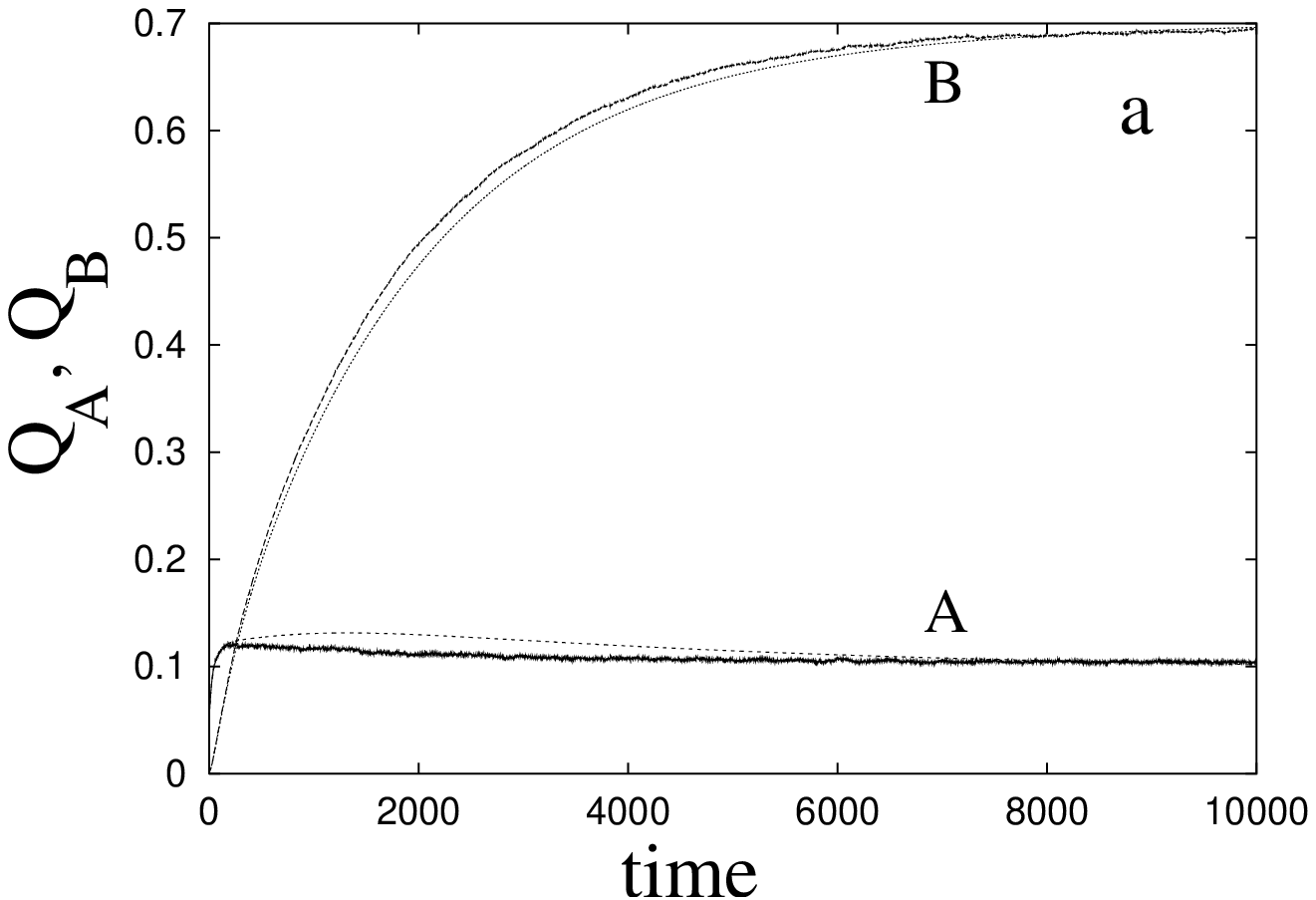, width=5.5cm} }
\subfigure {\epsfig {figure=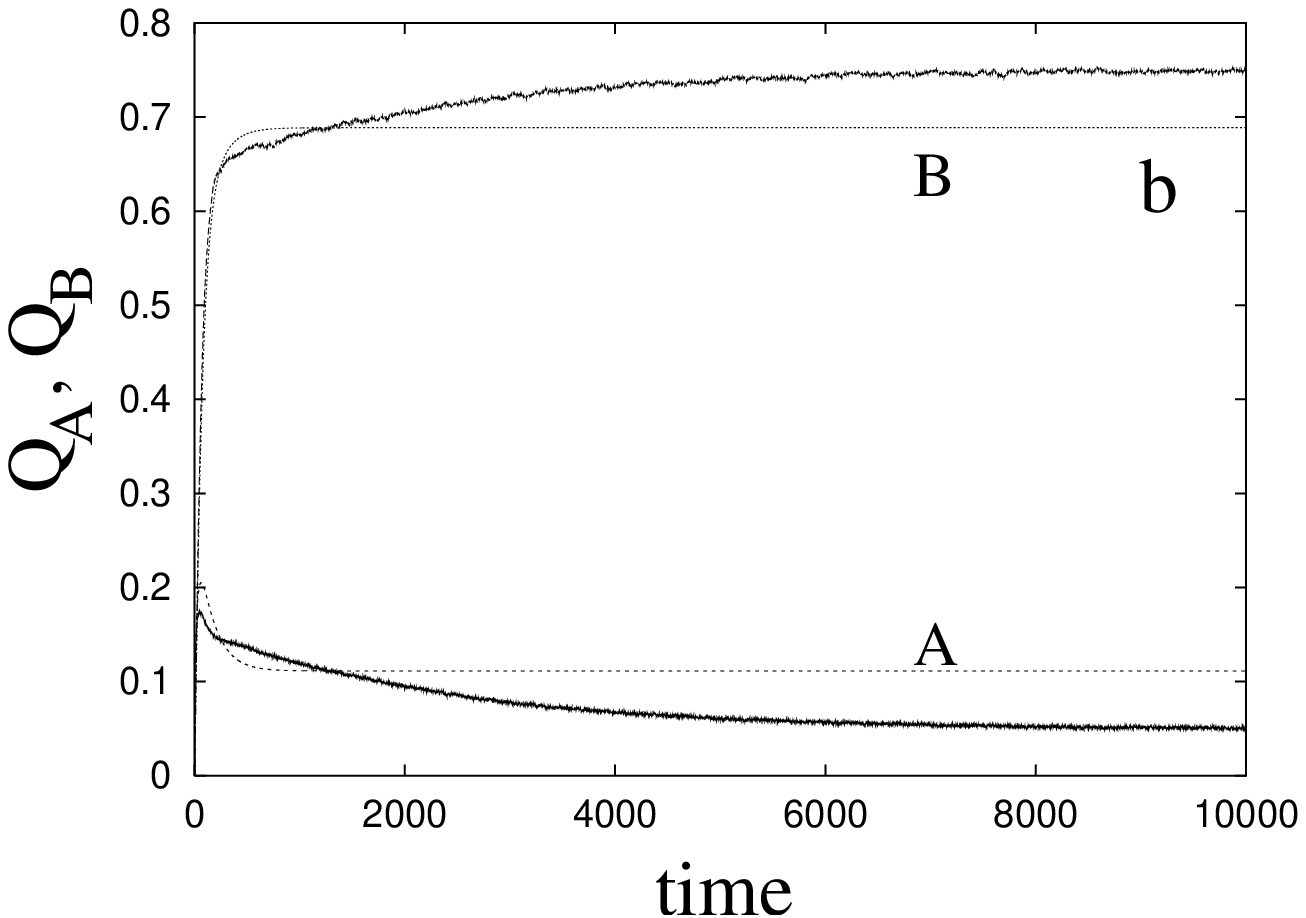, width=5.5cm} }
\caption {a) DMC simulation of time dependencies for $Q_{\rm A}$ and $Q_{\rm B}$ for
           $W_{\rm ads}$=0.8, $W_{\rm des}$=0.2, $W_{\rm diff}$=0.05,
           $W_{\rm rx}$=0.01, when blocks of 5 marginal sites
           are reactive ($S$=30). The straight lines corresponds to the MF
           results.
          b) DMC simulation of time dependencies for $Q_{\rm A}$ and $Q_{\rm B}$ for
           $W_{\rm ads}$=0.8, $W_{\rm des}$=0.2, $W_{\rm diff}$=2, $W_{\rm rx}$=0.1, when blocks of 5 marginal sites
           are reactive ($S$=30). The straight lines corresponds to the MF
           results.
         }
\label{QA_QB_time_marg}
\end{figure*}

\begin {figure*}[th]
\centering
\vspace*{2.5cm}
\subfigure {\epsfig {figure=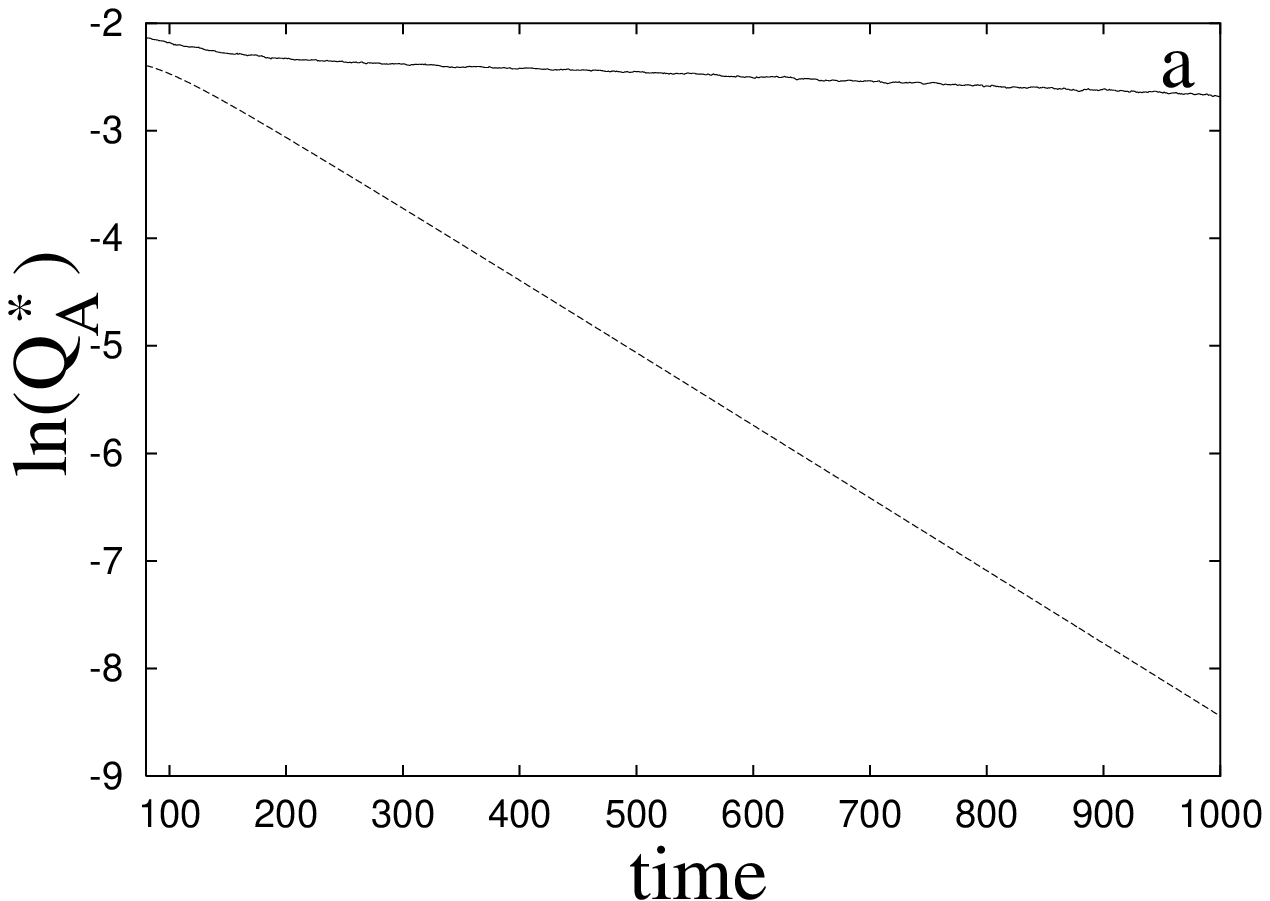, width=5.5cm} }
\subfigure {\epsfig {figure=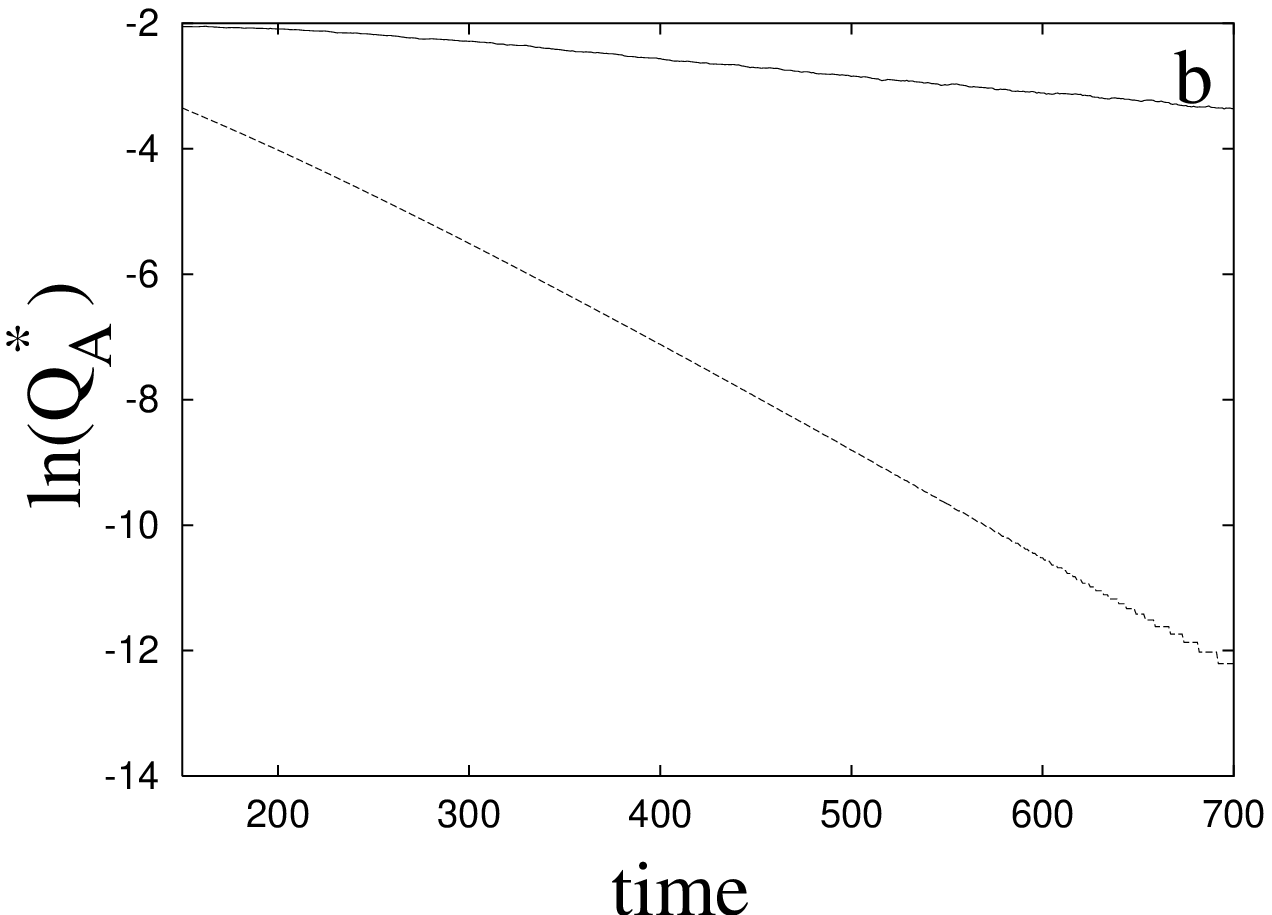, width=5.5cm} }
\subfigure {\epsfig {figure=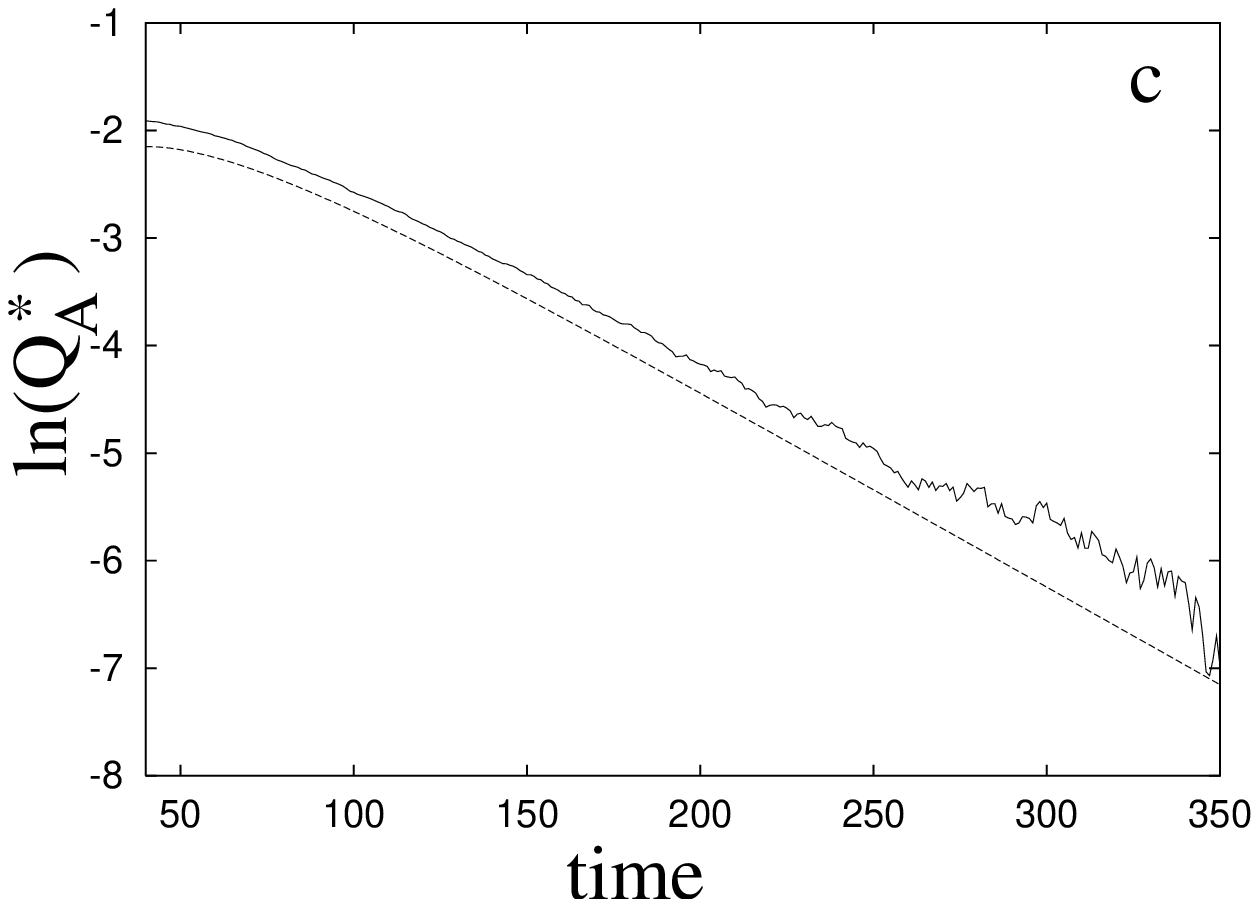, width=5.5cm} }
\caption {a) Time dependencies for $\ln({Q_{\rm A}}^{*})$ for a slow reaction
system ($W_{\rm ads}$=0.8, $W_{\rm des}$=0.2, $W_{\rm rx}$=0.1, $W_{\rm diff}=2$, $S$=30) when five
left and right marginal sites are reactive.
          b) Time dependencies for $\ln({Q_{\rm A}}^{*})$ for a slow reaction
system ($W_{\rm ads}$=0.8, $W_{\rm des}$=0.2, $W_{\rm rx}$=0.1, $W_{\rm diff}=2$, $S$=30) when ten
middle sites are reactive.
          c) Time dependencies for $\ln({Q_{\rm A}}^{*})$ for a slow reaction
system ($W_{\rm ads}$=0.8, $W_{\rm des}$=0.2, $W_{\rm rx}$=0.1, $W_{\rm diff}=2$, $S$=30) when ten
reactive sites are homogeneously distributed in the system. In a), b) and c)
the  continous line is for the DMC results and the dashed line is for the MF
results. We have marked with * the difference between the current value and the
steady-state value of the parameter.
                }
\label{slowrx_marg_mid_homog}
\end{figure*}

  We compare the time dependence of the $\rm A$ loading and of the $\rm B$
loading using DMC and MF for slow reaction systems and different rates of
conversion when diffusion is fast (see figure~\ref{QA_QB_time_marg}). 
 We see that differences appear in the transient region as well as in the 
steady-state for all the distributions of the reactive sites but very
prominent for marginal and middle sites reactive. In this case, for homogeneous
distribution of the reactive sites the differences between MF and DMC are
small (see figure~\ref{slowrx_marg_mid_homog}).

\begin {figure*}[th]
\centering
\subfigure {\epsfig {figure=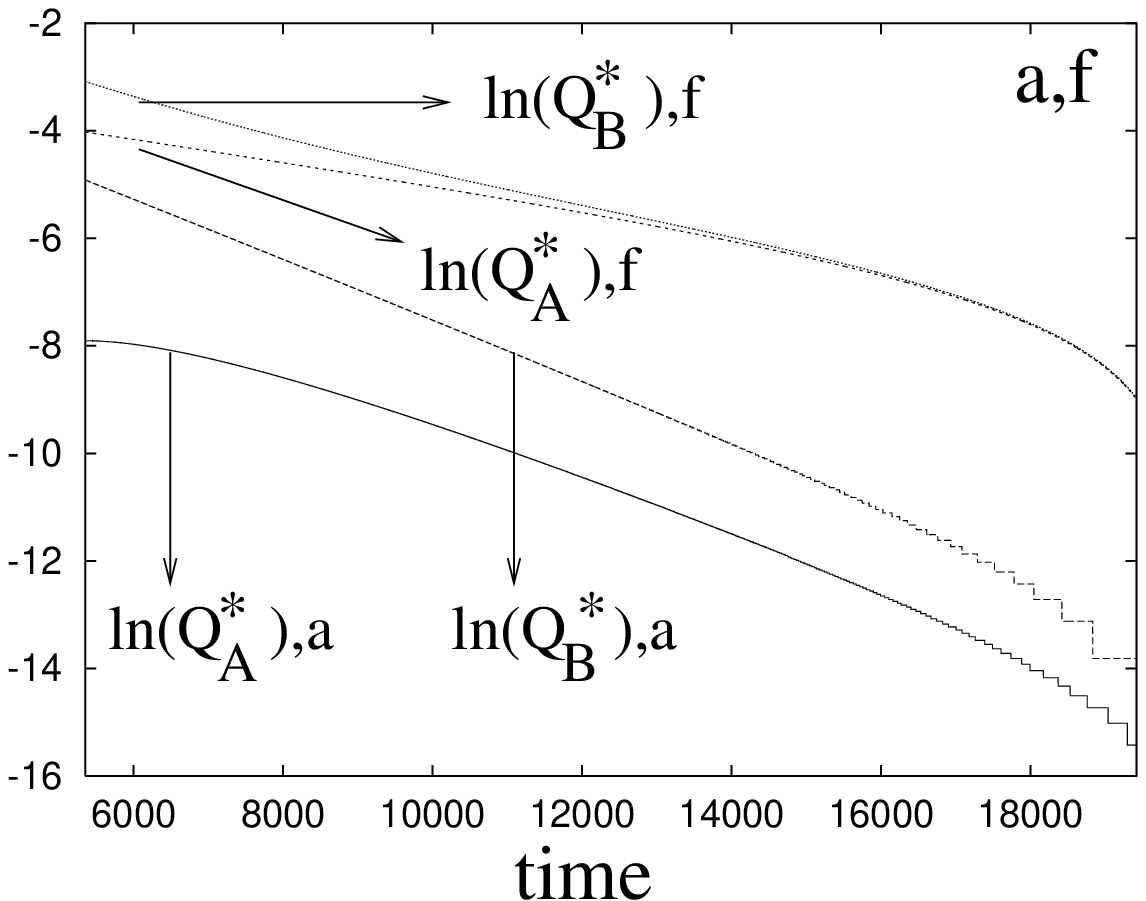, width=5.5cm} }
\subfigure {\epsfig {figure=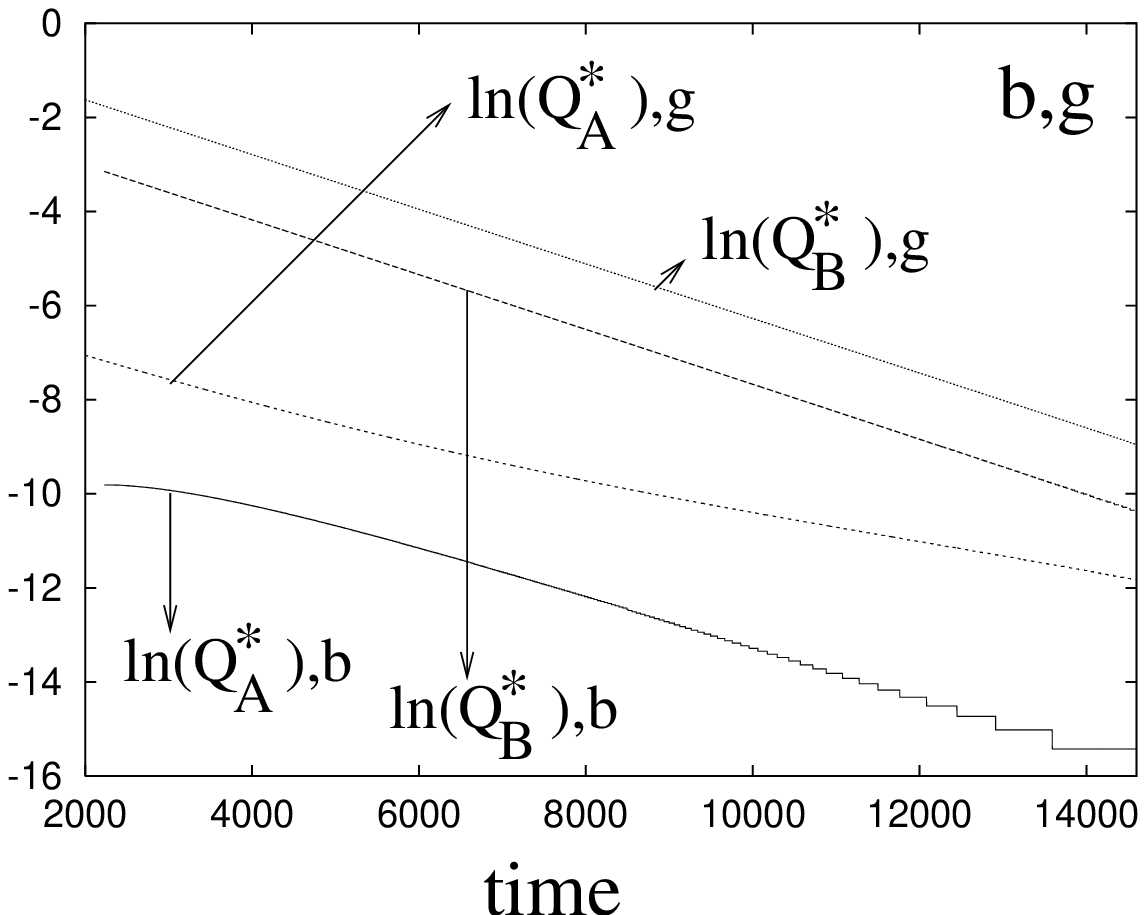, width=5.5cm} }
\subfigure {\epsfig {figure=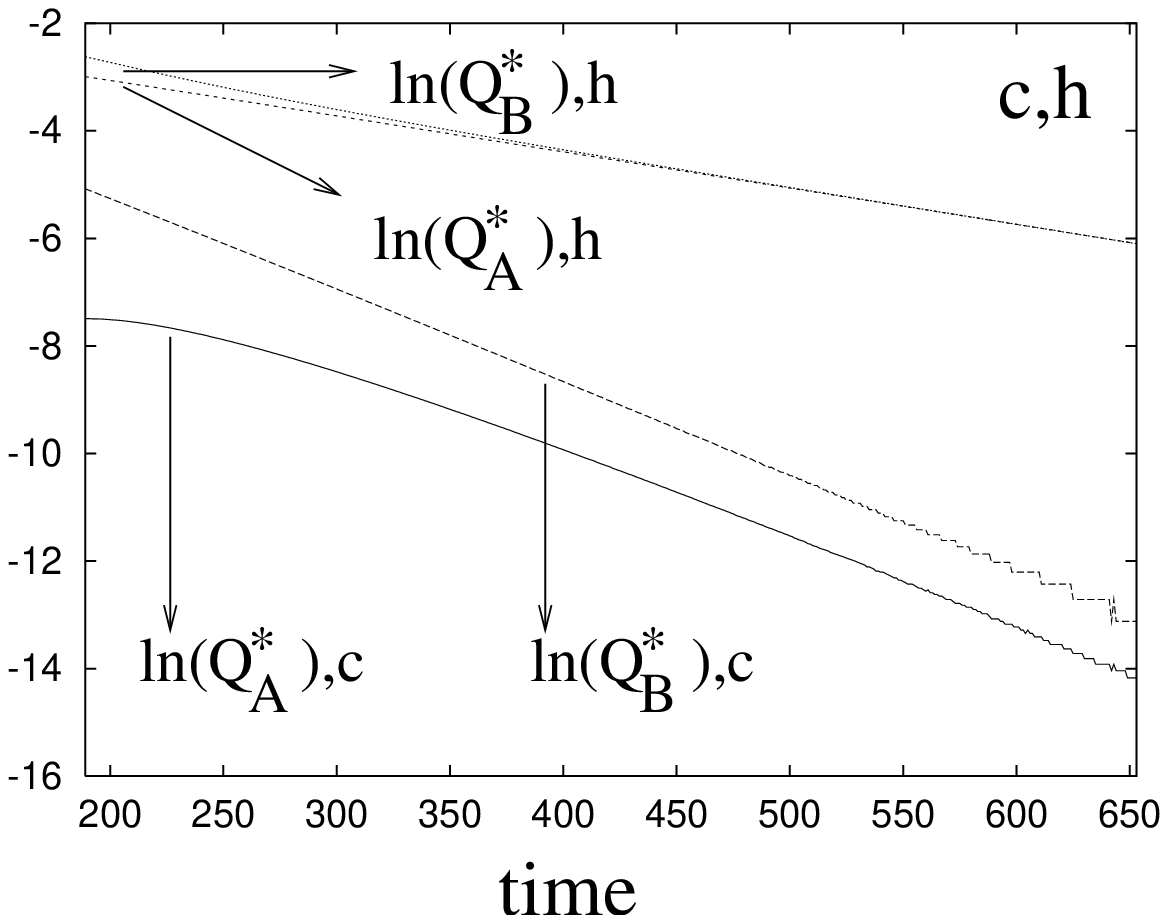, width=5.5cm} }
\subfigure {\epsfig {figure=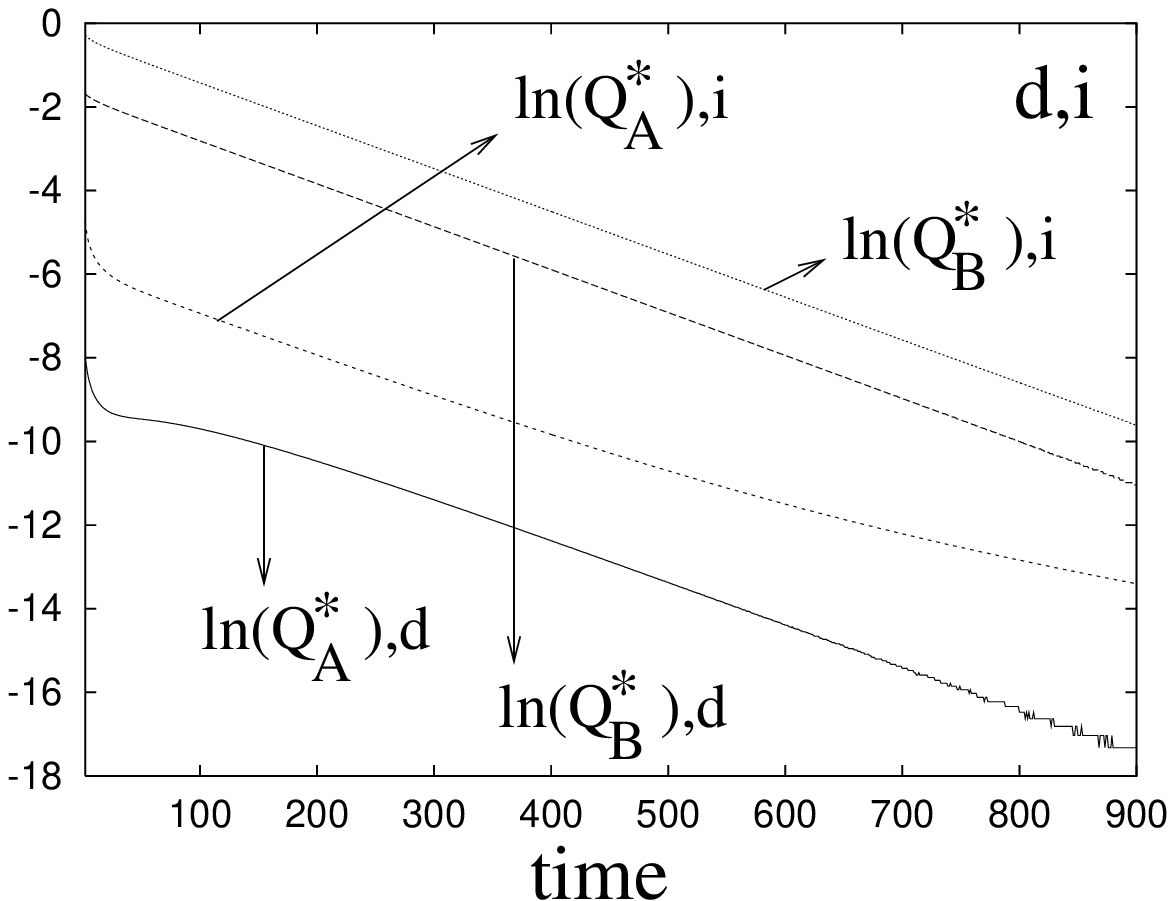, width=5.5cm} }
\subfigure {\epsfig {figure=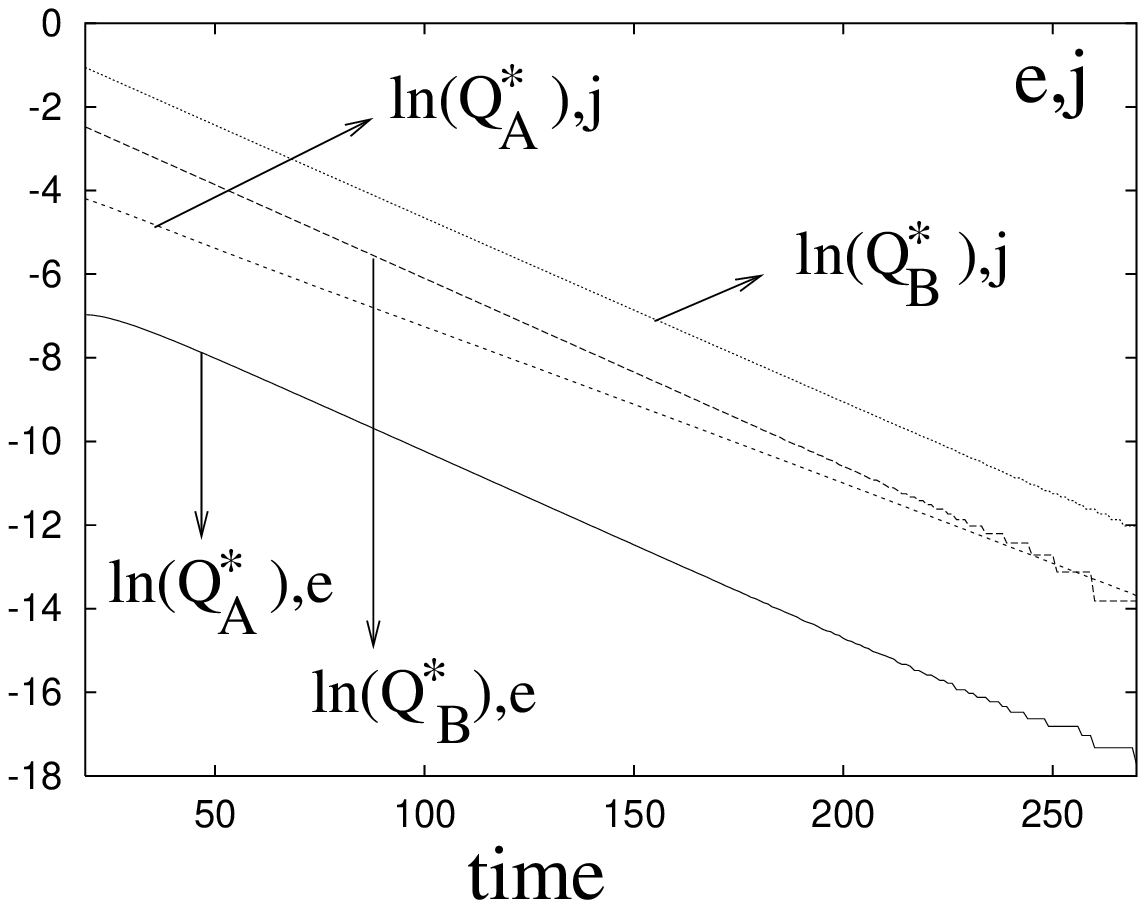, width=5.5cm} }
\caption { Time dependencies for $\ln({Q_{\rm A}}^{*})$ and $\ln({Q_{\rm B}}^{*})$ for the cases
            a and f(slow diffusion-very slow reaction at low and high
loadings), b and g(slow
diffusion-slow reaction at low and high loadings), c and h(fast diffusion-slow reaction
at low and high loadings), d and
i(fast diffusion-fast reaction at low and high loadings), e and j(very fast diffusion-fast reaction
at low and high loadings) in TABLE I using MF when 5 of
            the marginal sites are reactive. In all the figures we have
marked with * the absolute value of the difference between the current value and the steady-state
value of the parameter.
         }
\label{lnAB_time_marg}
\end{figure*}

 For the other cases (fast reaction-slow diffusion, fast reaction-fast
diffusion, slow reaction-slow diffusion), MF gives good results compared to
DMC for all the distributions of the reactive sites. In figure~\ref{lnAB_time_marg}
we compare the MF results for the transients for the case when five
marginal sites are reactive.

\begin {figure*}[th]
\centering
\subfigure {\epsfig {figure=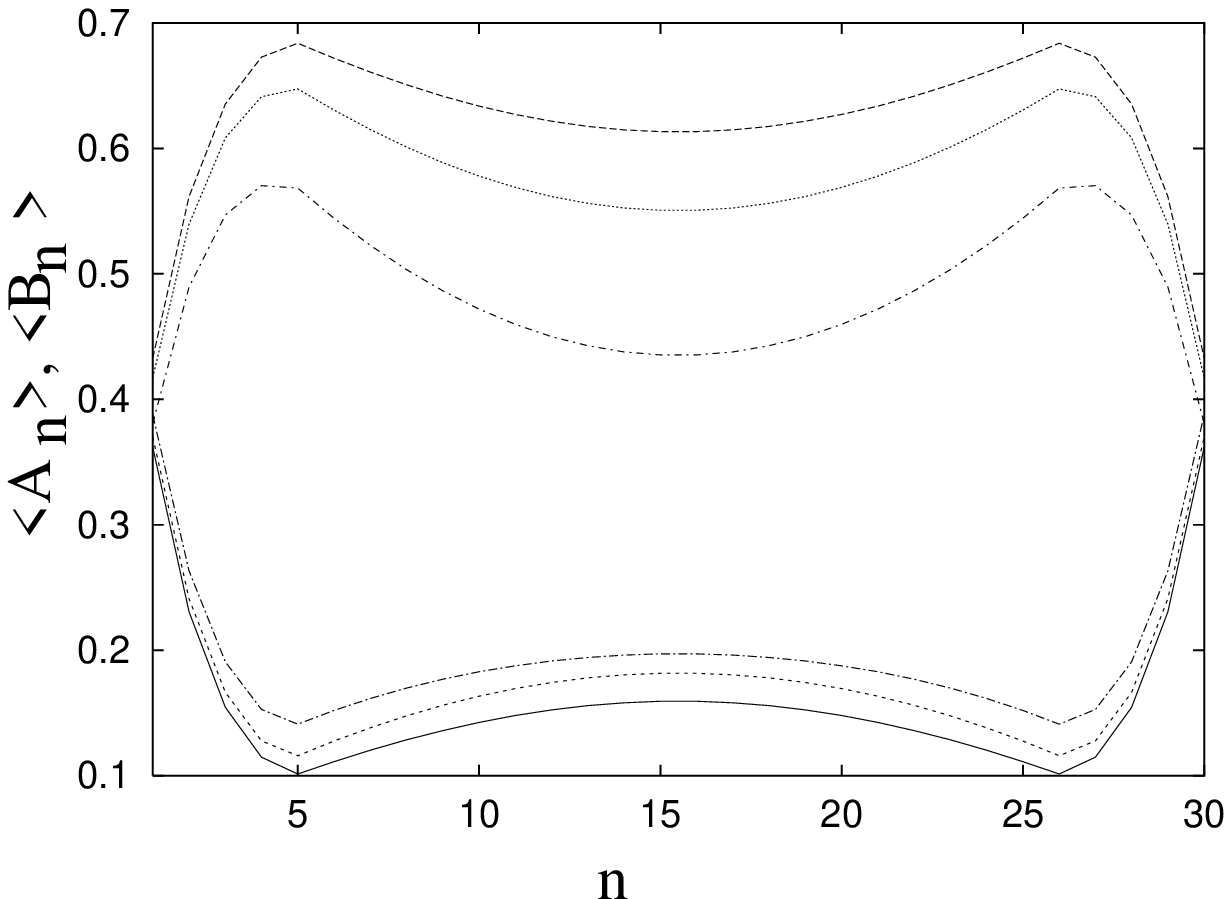, width=6.5cm} }
\caption {  MF profile occupancies ($\langle {\rm A_n} \rangle$ and $\langle {\rm B_n}
            \rangle$) for the case of 5 marginal sites reactive before steady-state is
         reached for a slow reaction system of length $S=30$ and parameters
         $W_{\rm ads}$=0.8 and $W_{\rm des}$=0.2, $W_{\rm diff}$=2, $W_{\rm rx}$=0.1.
         The lines at low occupancies correspunds to $\langle {\rm A_n} \rangle$ 
         profile occupancies after 200, 150 and 100 time units in this order
         from the bottom to the top.
         The lines at high occupancies correspunds to $\langle {\rm B_n} \rangle$ 
         profile occupancies after 200, 150 and 100 time units in this order
         from the top to the bottom.
           }
\label{tranzAB_profile_marg}
\end{figure*}

 We finally look at the site occupancy of the pipe. The MF profiles for
marginal sites reactive show in the transient regime accumulation of
the $\rm A$'s in the middle of the pipe for a slow reaction system and
fast diffusion. Because reaction is slow, the $\rm A$'s can pass without
reacting to the nonreactive sites.
 As the loading is increasing, the residence time of the particles increases,
and the probability to find an $\rm A$ in the middle of the pipe is
decreasing because of blocking (see figure~\ref{tranzAB_profile_marg}).

\begin {figure*}[bh]
\centering
\subfigure {\epsfig {figure=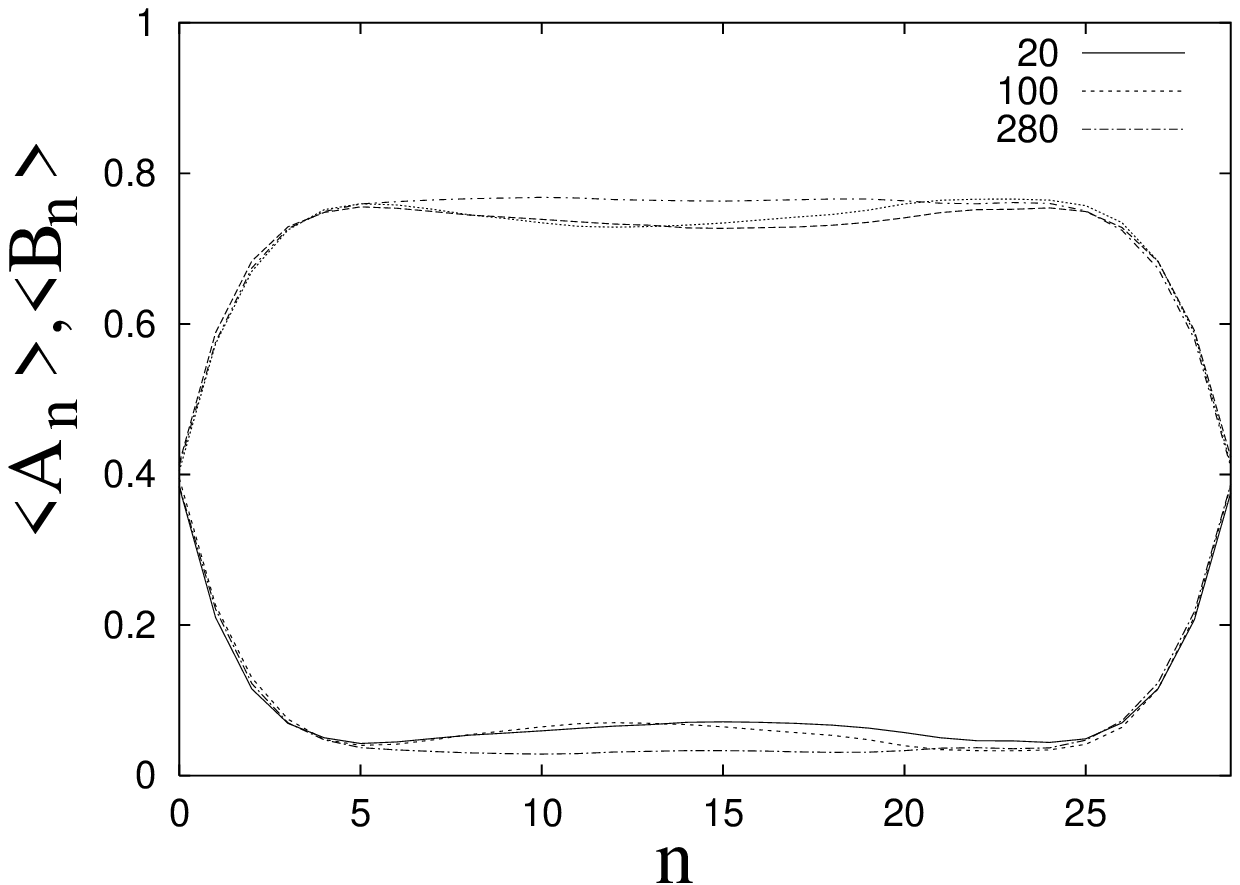, width=6.5cm} }
\caption {  DMC results for the occupancy profiles($\langle {\rm A_n} \rangle$ and $\langle {\rm
B_n} \rangle$) for slow reaction systems ($W_{\rm ads}$=0.8, $W_{\rm des}$=0.2,
$W_{\rm diff}$=2, $W_{\rm rx}$=0.1) in the case  5 marginal sites are reactive, before steady-state is reached.
 The lines at low occupancies corresponds to $\langle {\rm A_n} \rangle$ 
profile occupancies after 280, 100 and 20 time units in this order
from the bottom to the top.
  The lines at high occupancies correspunds to $\langle {\rm B_n} \rangle$ 
profile occupancies after 280, 100 and 20 time units in this order
from the top to the bottom.
         }
\label{tranzAB_carlos_profile_marg}
\end{figure*}

 Dynamic Monte Carlo simulations show the same behavior.
In figure~\ref{tranzAB_carlos_profile_marg} we see as well accumulation of
$\rm A$ particles in the middle of the pipe in the transient regime.
 As a consequence, until the equilibrium is reached the middle sites have different
contribution to the occupancy in the pipe depending on their
position. In the steady-state all the middle sites have the same
contribution to the occupancy profiles~\cite{silvia}.

\section { Summary }

  We have used DMC and analytical techniques to study the properties of
Single-File Systems in the transient regime.

 We have derived exact equations to solve the relaxation time of the whole
system ($t_{rel}$). We found that there are two regimes describing the dependence
on diffusion of the relaxation time $t_{rel}$. The first regime is for slow diffusion, when
$t_{rel}$ decreases fast with increasing diffusion, and the second for
fast diffusion, when $t_{rel}$ slowly decreases with diffusion to a
limiting value. We have analytically derived this limiting value of $t_{rel}$
for infinitely-fast diffusion.

 We have also studied the transients in the case with conversion.
MF results show that there are two different behaviors determined by 
conversion.
 For fast reaction systems, the relaxation time of the loading with $\rm A$'s 
($t_{relA}$) and $\rm B$'s ($t_{relB}$) is equal to the relaxation time of 
the total loading ($t_{rel}$). When $Q$ has reached equilibrium, $Q_{\rm A}$ 
and $Q_{\rm B}$ have also reached equilibrium.
 For slow reaction systems, the total loading $Q$ relaxes faster to equilibrium 
than the loading with $\rm A$'s ($Q_{\rm A}$) and $\rm B$'s ($Q_{\rm B}$).
 The regime between $Q$ reaching equilibrium and $Q_{\rm A}$ and $Q_{\rm B}$
reaching equilibrium we call the reaction limited regime. In the reaction
limited regime, MF shows that not only reaction, but also desorption has a 
strong influence influence on the transients. We find that the relaxation of 
$Q_{\rm A}$ ($t_{relA}$) as a function of desorption varies with reaction for
low desorption rates and converges to a limiting value for very high rates of
desorption.

 DMC results shows several regimes for the transients, in the case all the
sites are reactive.
 In the case diffusion is slow, the relaxation time is determined by
diffusion. When diffusion is fast and the reaction slow, then the relaxation
time is determined by reaction and when both are fast, relaxation time is
determined by adsorption/desorption.
 For all these cases the simulations results for the transients match the MF
results, except when we have low reaction rates and fast diffusion for both
low and high loading. In this case MF overestimates the amount
of $\rm A$'s in the pipe both for transients and for steady-state. 
DMC and MF results indicate also an overshoot for $Q_{\rm A}$ both for high and low
loadings in the transient regime, that appears as a consequence 
of the difference between diffusion and reaction rates constants.

 When only some of sites are reactive, for fast reaction systems, MF gives good
results compared to DMC for all the distributions of the reactive sites.
 When only the marginal sites are reactive and reaction very fast, DMC and MF
results are similar with DMC and MF results for the cases when all
sites are reactive and when it is a homogeneous distribution of
the reactive sites in the pipe. 
 When the reactive sites are situated in the middle of the pipe, the loadings $Q, Q_{\rm A}$, and $Q_{\rm B}$ relax slower 
to equilibrium than in the case all the sites are reactive, because it takes 
more time for the $\rm A$ particles to reach the reactive sites.

 For slow reactive systems, differences between DMC and MF results appear for 
transients for different distributions of the reactive 
sites when diffusion is fast, but are very prominent for marginal and middle sites reactive. 
 For homogeneous distribution of the reactive sites the differences between MF and DMC are
small. For slow reaction, we find that the $\rm B$ loading for homogenous
distribution and middle site reactive reaches equilibrium faster
than in the case of marginal sites reactive. For very slow reaction, $Q_{\rm B}$
increases faster at the beginning than in the case of homogeneous and middle 
sites reactive and becomes slower as the $\rm B$'s are formed. $Q_{\rm B}$
converges also faster in the case the middle sites are reactive than in the
case of homogeneous distribution of the reactive sites.

\section {Acknowledgments}

 The authors thank Prof.dr. R.A. van Santen for many stimulating
discussions.

\clearpage
\section {Appendix}
\subsection{Derivation of the equations for two-sites probabilities}

 In order to solve the rate equations of the system we use Cluster
Approximation. For simplicity we consider an approximation that considers
only the correlations between pairs of NN sites - Pair Approximation. 
 To the already derived rate equations for one-site probabilities we add
the two-site probability equations. We write write these equations 
in terms of three sites probabilities and we use then the decoupling scheme

\begin{equation}
\langle XYZ \rangle={{\langle XY \rangle \langle YZ \rangle}\over{\langle Y \rangle}}.
\end{equation}

 We give here the equations for the two-site probabilities for $\rm A$ and *
occupancy of these sites. The equations for the left marginal sites are

\begin{equation}
\begin{split}
{{d\langle A_1 A_2\rangle}\over dt}
&=W_{\rm diff}(\langle A_1 *_2 A_3 \rangle - \langle A_1
A_2 *_3 \rangle) - 2W_{\rm rx}\langle A_1 A_2\rangle \cr
&+ W_{\rm ads}\langle *_1 A_2 \rangle
- W_{\rm des} \langle A_1 A_2 \rangle\cr
{{d\langle A_1 *_2\rangle}\over dt}
&=W_{\rm diff}(\langle *_1 A_2\rangle + \langle A_1 A_2 *_3\rangle
+ \langle A_1 B_2 *_3 \rangle \cr
&- \langle A_1 *_2\rangle - \langle A_1 *_2 A_3\rangle -\langle A_1 *_2 B_3 \rangle) \cr
&- W_{\rm rx}\langle A_1 *_2 \rangle +W_{\rm ads}\langle
*_1*_2\rangle \cr
&-W_{\rm des}\langle A_1 *_2\rangle\cr
{{d\langle *_1 A_2\rangle}\over dt}
&= W_{\rm diff} (\langle A_1*_2\rangle + \langle *_1 *_2 A_3 \rangle
- \langle *_1 A_2 \rangle - \langle *_1 A_2 *_3\rangle) \cr
&- W_{\rm rx}\langle *_1 A_2\rangle -W_{\rm ads}\langle *_1 A_2\rangle \cr
& + W_{\rm des}(\langle A_1A_2 \rangle+\langle B_1 A_2\rangle)\cr
{{d\langle *_1 *_2\rangle}\over dt}
&=W_{\rm diff}(\langle *_1 A_2 *_3 \rangle + \langle *_1 B_2 *_3 \rangle \cr
&- \langle *_1 *_2 A_3 \rangle - \langle *_1 *_2 B_3 \rangle) \cr
&- W_{\rm ads} \langle *_1 *_2 \rangle + W_{\rm des}(\langle A_1 *_2\rangle + \langle B_1 *_2\rangle),
\end{split}
\end{equation}
where
$\langle A_1 *_2 A_3 \rangle={{ \langle A_1 *_2 \rangle \langle *_2 A_3 \rangle} \over{ \langle
*_2 \rangle}}$, etc.

Almost similar are the equations for the right marginal sites

\begin{equation}
\begin{split}
{{d\langle A_{S-1} A_S\rangle}\over dt}
&=W_{\rm diff}(\langle A_{S-2} *_{S-1} A_S \rangle - \langle *_{S-2}A_{S-1}A_S \rangle)\cr
&- 2W_{\rm rx}\langle A_{S-1} A_S\rangle + W_{\rm ads}\langle A_{S-1} *_S \rangle\cr
& - W_{\rm des} \langle A_{S-1} A_S \rangle\cr
{{d\langle A_{S-1} *_S\rangle}\over dt}
&=W_{\rm diff}(\langle A_{S-1}*_S *_{S+1}\rangle + \langle *_{S-1} A_S \rangle \cr
&- \langle A_{S-1} *_S \rangle - \langle *_{S-2}A_{S-1}*_S\rangle) \cr
& - W_{\rm rx}\langle A_{S-1} *_S \rangle - W_{\rm ads}\langle  A_{S-11}*_S\rangle \cr
&+ W_{\rm des}(\langle A_{S-1} A_S\rangle + \langle A_{S-1} B_S\rangle )\cr
{{d\langle *_{S-1} A_S\rangle}\over dt}
&=W_{\rm diff}(\langle A_{S-1} *_S\rangle + \langle *_{S-2} A_{S-1} A_S\rangle \cr
&+ \langle *_{S-2} B_{S-1} A_S \rangle  - \langle *_{S-1} A_{S}\rangle \cr
&- \langle A_{S-2} *_{S-1} A_S\rangle - \langle B_{S-2} *_{S-1} A_S \rangle) \cr
&- W_{\rm rx}\langle *_{S-1} A_S \rangle +W_{\rm ads}\langle
*_{S-1}*_S\rangle \cr
&-W_{\rm des}\langle *_{S-1} A_S\rangle\cr
{{d\langle *_{S-1} *_S\rangle}\over dt}
&=W_{\rm diff}(\langle *_{S-2} A_{S-1} *_S \rangle + \langle *_{S-2} B_{S-1} *_S \rangle\cr
& - \langle A_{S-2} *_{S-1} *_S \rangle - \langle B_{S-2} *_{S-1} *_S \rangle) \cr
&- W_{\rm ads} \langle *_{S-1} *_S \rangle \cr
&+ W_{\rm des}(\langle *_{S-1} A_S\rangle + \langle *_{S-1} B_S\rangle),
\end{split}
\end{equation}

 And, finally, the equations for the non-marginal sites

\begin{equation}
\begin{split}
{{d\langle A_n A_{n+1}\rangle}\over dt}
&=W_{\rm diff}(\langle A_n *_{n+1} A_{n+2} \rangle + \langle A_{n-1} *_n A_{n+1} \rangle \cr
&- \langle *_{n-1} A_n A_{n+1} \rangle - \langle A_{n} A_{n+1} *_{n+2} \rangle) \cr
&- 2W_{\rm rx}\langle A_n A_{n+1}\rangle \cr
{{d\langle A_n *_{n+1}\rangle}\over dt}
&=W_{\rm diff}(\langle *_n A_{n+1}\rangle + \langle A_{n-1} *_n *_{n+1}\rangle \cr
&+ \langle A_n A_{n+1} *_{n+2} \rangle + \langle A_n B_{n+1} *_{n+2}\rangle\cr
&- \langle A_n *_{n+1}\rangle - \langle A_n *_{n+1} A_{n+2}\rangle\cr
& - \langle A_n *_{n+1} B_{n+2} \rangle - \langle *_{n-1} A_n *_{n+1}\rangle) \cr
&- W_{\rm rx}\langle A_n *_{n+1} \rangle \cr
{{d\langle *_n A_{n+1}\rangle}\over dt}
&= W_{\rm diff} (\langle A_n *_{n+1}\rangle + \langle *_n *_{n+1} A_{n+2} \rangle\cr
&  + \langle *_{n-1}A_n A_{n+1} \rangle + \langle *_{n-1}B_n A_{n+1}\rangle \cr
& - \langle *_n A_{n+1} \rangle - \langle *_n A_{n+1} *_{n+2}\rangle \cr
&-\langle A_{n-1} *_n A_{n+1}\rangle - \langle B_{n-1}*_n A_{n+1}\rangle)\cr
& -W_{\rm rx}\langle *_n A_{n+1}\rangle \cr
{{d\langle *_n *_{n+1}\rangle}\over dt}
&=W_{\rm diff}(\langle *_n A_{n+1} *_{n+2} \rangle + \langle *_n B_{n+1} *_{n+2} \rangle \cr
&+ \langle *_{n-1}A_n*_{n+1} \rangle + \langle *_{n-1}B_n *_{n+1}\rangle \cr
&- \langle *_n *_{n+1} A_{n+2} \rangle - \langle *_n *_{n+1} B_{n+2} \rangle) \cr
&- \langle A_{n-1} *_{n} *_{n+2} \rangle - \langle B_{n-1} *_{n} *_{n+1} \rangle)
\end{split}
\end{equation}
We have the possibility to determine the one-site probabilities by
calculating the sum of the two-sites probabilities
$$\sum_Y{\langle XY \rangle}=\langle X \rangle.$$
\bibliographystyle{./apsrev}
\nocite{*}

\begin{thebibliography}{47}
\expandafter\ifx\csname natexlab\endcsname\relax\def\natexlab#1{#1}\fi
\expandafter\ifx\csname bibnamefont\endcsname\relax
  \def\bibnamefont#1{#1}\fi
\expandafter\ifx\csname bibfnamefont\endcsname\relax
  \def\bibfnamefont#1{#1}\fi
\expandafter\ifx\csname citenamefont\endcsname\relax
  \def\citenamefont#1{#1}\fi
\expandafter\ifx\csname url\endcsname\relax
  \def\url#1{\texttt{#1}}\fi
\expandafter\ifx\csname urlprefix\endcsname\relax\fi
\providecommand{\bibinfo}[2]{#2}
\providecommand{\eprint}[2][]{\url{#2}}

\bibitem[{\citenamefont{Baxter}(1982)}]{Baxter82}
\bibinfo{author}{\bibfnamefont{R.~J.} \bibnamefont{Baxter}},
  \emph{\bibinfo{title}{Exactly solved models in statistical mechanics}}
  (\bibinfo{publisher}{Academic Press, New York}, \bibinfo{year}{1982}).

\bibitem[{\citenamefont{Derrida and Evans}(1997)}]{DerridaEvans97}
\bibinfo{author}{\bibfnamefont{B.}~\bibnamefont{Derrida}} \bibnamefont{and}
  \bibinfo{author}{\bibfnamefont{M.}~\bibnamefont{Evans}},
  \bibinfo{journal}{Cambridge University Press, U.K.}  (\bibinfo{year}{1997}).

\bibitem[{\citenamefont{Alcaraz}(1994)}]{Alcaraz94}
\bibinfo{author}{\bibfnamefont{F.~C.} \bibnamefont{Alcaraz}},
  \bibinfo{journal}{Int. J. Mod. Phys} \textbf{\bibinfo{volume}{B 8}},
  \bibinfo{pages}{3349} (\bibinfo{year}{1994}).

\bibitem[{\citenamefont{Derrida et~al.}(1992)\citenamefont{Derrida, Domany, and
  Mukamel}}]{DDM92}
\bibinfo{author}{\bibfnamefont{B.}~\bibnamefont{Derrida}},
  \bibinfo{author}{\bibfnamefont{E.}~\bibnamefont{Domany}}, \bibnamefont{and}
  \bibinfo{author}{\bibfnamefont{D.}~\bibnamefont{Mukamel}},
  \bibinfo{journal}{J. Stat. Phys.} \textbf{\bibinfo{volume}{69}},
  \bibinfo{pages}{667} (\bibinfo{year}{1992}).

\bibitem[{\citenamefont{Schutz and Domany}(1993)}]{SchuetzDomany93}
\bibinfo{author}{\bibfnamefont{G.}~\bibnamefont{Schutz}} \bibnamefont{and}
  \bibinfo{author}{\bibfnamefont{E.}~\bibnamefont{Domany}},
  \bibinfo{journal}{J. Stat. Phys.} \textbf{\bibinfo{volume}{72}},
  \bibinfo{pages}{277} (\bibinfo{year}{1993}).

\bibitem[{\citenamefont{Derrida et~al.}(1993)\citenamefont{Derrida, Evans,
  Hakim, and Pasquier}}]{DEHP93}
\bibinfo{author}{\bibfnamefont{B.}~\bibnamefont{Derrida}},
  \bibinfo{author}{\bibfnamefont{M.~R.} \bibnamefont{Evans}},
  \bibinfo{author}{\bibfnamefont{V.}~\bibnamefont{Hakim}}, \bibnamefont{and}
  \bibinfo{author}{\bibfnamefont{V.}~\bibnamefont{Pasquier}},
  \bibinfo{journal}{J. Phys.} \textbf{\bibinfo{volume}{26}},
  \bibinfo{pages}{1493} (\bibinfo{year}{1993}).

\bibitem[{\citenamefont{J.G.Tsikoyannis and Wei}(1991)}]{tsikoyannis}
\bibinfo{author}{\bibnamefont{J.G.Tsikoyannis}} \bibnamefont{and}
  \bibinfo{author}{\bibfnamefont{J.}~\bibnamefont{Wei}},
  \bibinfo{journal}{Chem. Eng. Sci.} \textbf{\bibinfo{volume}{46}},
  \bibinfo{pages}{233} (\bibinfo{year}{1991}).

\bibitem[{\citenamefont{R{$\ddot{\rm o}$}denbeck
  et~al.}(1997)\citenamefont{R{$\ddot{\rm o}$}denbeck, K{$\ddot{\rm a}$}rger,
  and K.Hahn}}]{rodenbeck}
\bibinfo{author}{\bibfnamefont{C.}~\bibnamefont{R{$\ddot{\rm o}$}denbeck}},
  \bibinfo{author}{\bibfnamefont{J.}~\bibnamefont{K{$\ddot{\rm a}$}rger}},
  \bibnamefont{and} \bibinfo{author}{\bibnamefont{K.Hahn}},
  \bibinfo{journal}{Physical Review E} \textbf{\bibinfo{volume}{55}},
  \bibinfo{pages}{5697} (\bibinfo{year}{1997}).

\bibitem[{\citenamefont{Okino et~al.}(1999)\citenamefont{Okino, Snurr, Kung,
  Ochs, and Mavrovouniotis}}]{okino}
\bibinfo{author}{\bibfnamefont{M.~S.} \bibnamefont{Okino}},
  \bibinfo{author}{\bibfnamefont{R.~Q.} \bibnamefont{Snurr}},
  \bibinfo{author}{\bibfnamefont{H.~H.} \bibnamefont{Kung}},
  \bibinfo{author}{\bibfnamefont{J.~E.} \bibnamefont{Ochs}}, \bibnamefont{and}
  \bibinfo{author}{\bibfnamefont{M.~L.} \bibnamefont{Mavrovouniotis}},
  \bibinfo{journal}{J. Chem. Phys.} \textbf{\bibinfo{volume}{111}},
  \bibinfo{pages}{2210} (\bibinfo{year}{1999}).

\bibitem[{\citenamefont{Coppens et~al.}(1998)\citenamefont{Coppens, Bell, and
  Chakraborty}}]{coppens2}
\bibinfo{author}{\bibfnamefont{M.-O.} \bibnamefont{Coppens}},
  \bibinfo{author}{\bibfnamefont{A.}~\bibnamefont{Bell}}, \bibnamefont{and}
  \bibinfo{author}{\bibnamefont{Chakraborty}}, \bibinfo{journal}{Chem. Engng.
  Sci.} \textbf{\bibinfo{volume}{53}}, \bibinfo{pages}{2053}
  (\bibinfo{year}{1998}).

\bibitem[{\citenamefont{Coppens et~al.}(1999)\citenamefont{Coppens, Bell, and
  Chakraborty}}]{coppens1}
\bibinfo{author}{\bibfnamefont{M.-O.} \bibnamefont{Coppens}},
  \bibinfo{author}{\bibfnamefont{A.}~\bibnamefont{Bell}}, \bibnamefont{and}
  \bibinfo{author}{\bibnamefont{Chakraborty}}, \bibinfo{journal}{Chem. Engng.
  Sci.} \textbf{\bibinfo{volume}{54}}, \bibinfo{pages}{3455}
  (\bibinfo{year}{1999}).

\bibitem[{\citenamefont{Privman}(1997)}]{Privman97}
\bibinfo{author}{\bibfnamefont{V.}~\bibnamefont{Privman}},
  \emph{\bibinfo{title}{Nonequilibrium statistical mechanics in one dimension}}
  (\bibinfo{publisher}{Cambridge University Press, U. K.},
  \bibinfo{year}{1997}).

\bibitem[{\citenamefont{Marro and Dickman}(1998)}]{MarroDickman98}
\bibinfo{author}{\bibfnamefont{J.}~\bibnamefont{Marro}} \bibnamefont{and}
  \bibinfo{author}{\bibfnamefont{R.}~\bibnamefont{Dickman}},
  \emph{\bibinfo{title}{Nonequilibrium phase transitions in lattice models}}
  (\bibinfo{publisher}{Cambridge University Press, Cambridge},
  \bibinfo{year}{1998}).

\bibitem[{\citenamefont{Dickman and Jensen}(1991)}]{DickmanJensen91}
\bibinfo{author}{\bibfnamefont{R.}~\bibnamefont{Dickman}} \bibnamefont{and}
  \bibinfo{author}{\bibfnamefont{I.}~\bibnamefont{Jensen}},
  \bibinfo{journal}{Phys. Rev. Lett.} \textbf{\bibinfo{volume}{67}},
  \bibinfo{pages}{2391} (\bibinfo{year}{1991}).

\bibitem[{\citenamefont{Grinstein et~al.}(1989)\citenamefont{Grinstein, Lai,
  and Browne}}]{GLB89}
\bibinfo{author}{\bibfnamefont{G.}~\bibnamefont{Grinstein}},
  \bibinfo{author}{\bibfnamefont{Z.~W.} \bibnamefont{Lai}}, \bibnamefont{and}
  \bibinfo{author}{\bibfnamefont{D.~A.} \bibnamefont{Browne}},
  \bibinfo{journal}{Phys. Rev. A} \textbf{\bibinfo{volume}{40}},
  \bibinfo{pages}{4820} (\bibinfo{year}{1989}).

\bibitem[{\citenamefont{Dickman and Burschka}(1988)}]{DickmanBurschka88}
\bibinfo{author}{\bibfnamefont{R.}~\bibnamefont{Dickman}} \bibnamefont{and}
  \bibinfo{author}{\bibfnamefont{M.}~\bibnamefont{Burschka}},
  \bibinfo{journal}{Phys. Lett. A} \textbf{\bibinfo{volume}{127}},
  \bibinfo{pages}{132} (\bibinfo{year}{1988}).

\bibitem[{\citenamefont{S.V.Nedea et~al.}(2002)\citenamefont{S.V.Nedea,
  A.P.J.Jansen, J.J.Lukkien, and P.A.J.Hilbers}}]{silvia}
\bibinfo{author}{\bibnamefont{S.V.Nedea}},
  \bibinfo{author}{\bibnamefont{A.P.J.Jansen}},
  \bibinfo{author}{\bibnamefont{J.J.Lukkien}}, \bibnamefont{and}
  \bibinfo{author}{\bibnamefont{P.A.J.Hilbers}}, \bibinfo{journal}{Phys.Rev.E}
  (\bibinfo{year}{2002}).

\bibitem[{\citenamefont{Derrida and Evans}(1999)}]{DerridaEvans99}
\bibinfo{author}{\bibfnamefont{B.}~\bibnamefont{Derrida}} \bibnamefont{and}
  \bibinfo{author}{\bibfnamefont{M.~R.} \bibnamefont{Evans}},
  \bibinfo{journal}{J. Phys. A} \textbf{\bibinfo{volume}{32}},
  \bibinfo{pages}{4833} (\bibinfo{year}{1999}).

\bibitem[{\citenamefont{Ben-Naim and Krapivsky}(1994)}]{BenNaimKrapivsky94}
\bibinfo{author}{\bibnamefont{Ben-Naim}} \bibnamefont{and}
  \bibinfo{author}{\bibfnamefont{P.~L.} \bibnamefont{Krapivsky}},
  \bibinfo{journal}{J. Phys. A} \textbf{\bibinfo{volume}{27}},
  \bibinfo{pages}{481} (\bibinfo{year}{1994}).

\bibitem[{\citenamefont{Derrida et~al.}(1996)\citenamefont{Derrida, Hakim, and
  Zeitak}}]{DHZ96}
\bibinfo{author}{\bibfnamefont{B.}~\bibnamefont{Derrida}},
  \bibinfo{author}{\bibfnamefont{V.}~\bibnamefont{Hakim}}, \bibnamefont{and}
  \bibinfo{author}{\bibfnamefont{R.}~\bibnamefont{Zeitak}},
  \bibinfo{journal}{Phys. Rev. Lett.} \textbf{\bibinfo{volume}{77}},
  \bibinfo{pages}{2871} (\bibinfo{year}{1996}).

\bibitem[{\citenamefont{J.Mai et~al.}(1994{\natexlab{a}})\citenamefont{J.Mai,
  Kuzovkov, and von Niessen}}]{kuzovkov2}
\bibinfo{author}{\bibnamefont{J.Mai}},
  \bibinfo{author}{\bibfnamefont{V.}~\bibnamefont{Kuzovkov}}, \bibnamefont{and}
  \bibinfo{author}{\bibfnamefont{W.}~\bibnamefont{von Niessen}},
  \bibinfo{journal}{J.Chem.Phys.} \textbf{\bibinfo{volume}{100(11)}},
  \bibinfo{pages}{8522} (\bibinfo{year}{1994}{\natexlab{a}}).

\bibitem[{\citenamefont{von Smoluchowski}(1917)}]{smoluchowski}
\bibinfo{author}{\bibfnamefont{M.}~\bibnamefont{von Smoluchowski}},
  \bibinfo{journal}{Z. Phys. Chem.} \textbf{\bibinfo{volume}{92}},
  \bibinfo{pages}{129} (\bibinfo{year}{1917}).

\bibitem[{\citenamefont{Kampen}(1981)}]{kampen}
\bibinfo{author}{\bibfnamefont{V.}~\bibnamefont{Kampen}},
  \emph{\bibinfo{title}{Stochastic Processes in Physics and Chemistry}}
  (\bibinfo{publisher}{Elsevier Science Publishers B.V.},
  \bibinfo{year}{1981}).

\bibitem[{\citenamefont{Mamada and Takano}(1968)}]{mamda}
\bibinfo{author}{\bibfnamefont{H.}~\bibnamefont{Mamada}} \bibnamefont{and}
  \bibinfo{author}{\bibfnamefont{F.}~\bibnamefont{Takano}},
  \bibinfo{journal}{J. Phys. Soc. Japan 25} \textbf{\bibinfo{volume}{25}},
  \bibinfo{pages}{675} (\bibinfo{year}{1968}).

\bibitem[{\citenamefont{Lukkien et~al.}(1998)\citenamefont{Lukkien, Segers,
  P.A.J.Hilbers, R.J.Gelten, and A.P.J.Jansen}}]{lukkien}
\bibinfo{author}{\bibfnamefont{J.}~\bibnamefont{Lukkien}},
  \bibinfo{author}{\bibfnamefont{J.}~\bibnamefont{Segers}},
  \bibinfo{author}{\bibnamefont{P.A.J.Hilbers}},
  \bibinfo{author}{\bibnamefont{R.J.Gelten}}, \bibnamefont{and}
  \bibinfo{author}{\bibnamefont{A.P.J.Jansen}}, \bibinfo{journal}{Phys.Rev.E}
  \textbf{\bibinfo{volume}{58}}, \bibinfo{pages}{2598} (\bibinfo{year}{1998}).

\bibitem[{\citenamefont{R.J.Gelten
  et~al.}(1998{\natexlab{a}})\citenamefont{R.J.Gelten, A.P.J.Jansen, van
  Santen, J.J.Lukkien, and Hilbers}}]{gelten_all}
\bibinfo{author}{\bibnamefont{R.J.Gelten}},
  \bibinfo{author}{\bibnamefont{A.P.J.Jansen}},
  \bibinfo{author}{\bibfnamefont{R.}~\bibnamefont{van Santen}},
  \bibinfo{author}{\bibnamefont{J.J.Lukkien}}, \bibnamefont{and}
  \bibinfo{author}{\bibfnamefont{P.}~\bibnamefont{Hilbers}},
  \bibinfo{journal}{J.Chem.Phys.} \textbf{\bibinfo{volume}{108(14)}},
  \bibinfo{pages}{5921} (\bibinfo{year}{1998}{\natexlab{a}}).

\bibitem[{\citenamefont{R.J.Gelten et~al.}(1999)\citenamefont{R.J.Gelten, van
  Santen, and A.P.J.Jansen}}]{b_g_s_j}
\bibinfo{author}{\bibnamefont{R.J.Gelten}},
  \bibinfo{author}{\bibfnamefont{R.}~\bibnamefont{van Santen}},
  \bibnamefont{and} \bibinfo{author}{\bibnamefont{A.P.J.Jansen}},
  \emph{\bibinfo{title}{Dynamic Monte Carlo simulations of oscillatory
  heterogeneous catalytic reactions in P.B. Balbuena and J.M.Seminario}}
  (\bibinfo{publisher}{Elsevier, Amsterdam}, \bibinfo{year}{1999}).

\bibitem[{\citenamefont{R.J.Gelten
  et~al.}(1998{\natexlab{b}})\citenamefont{R.J.Gelten, van Santen, and
  A.P.J.Jansen}}]{gelten_santen_jansen}
\bibinfo{author}{\bibnamefont{R.J.Gelten}},
  \bibinfo{author}{\bibfnamefont{R.}~\bibnamefont{van Santen}},
  \bibnamefont{and} \bibinfo{author}{\bibnamefont{A.P.J.Jansen}},
  \bibinfo{journal}{Israel J.Chem.} \textbf{\bibinfo{volume}{38}},
  \bibinfo{pages}{415} (\bibinfo{year}{1998}{\natexlab{b}}).

\bibitem[{\citenamefont{A.P.J.Jansen}(1995)}]{jansen}
\bibinfo{author}{\bibnamefont{A.P.J.Jansen}}, \bibinfo{journal}{Comput. Phys.
  Comm.} \textbf{\bibinfo{volume}{86}}, \bibinfo{pages}{1}
  (\bibinfo{year}{1995}).

\bibitem[{\citenamefont{Segers}(1999)}]{segers}
\bibinfo{author}{\bibfnamefont{J.}~\bibnamefont{Segers}},
  \emph{\bibinfo{title}{Algorithms for the Simulation of Surface Processes}}
  (\bibinfo{publisher}{Ph.D. thesis, Eindhoven University of Technology},
  \bibinfo{year}{1999}).

\bibitem[{\citenamefont{Binder}(1986)}]{binder}
\bibinfo{author}{\bibfnamefont{K.}~\bibnamefont{Binder}},
  \emph{\bibinfo{title}{Monte Carlo methods in Statistical Physics}}
  (\bibinfo{publisher}{Springer, Berlin}, \bibinfo{year}{1986}).

\bibitem[{\citenamefont{Gillespie}(1976)}]{gillespie1}
\bibinfo{author}{\bibfnamefont{D.}~\bibnamefont{Gillespie}},
  \bibinfo{journal}{J.Comput.Phys.} \textbf{\bibinfo{volume}{22}},
  \bibinfo{pages}{403} (\bibinfo{year}{1976}).

\bibitem[{\citenamefont{Gillespie}(1977)}]{gillespie2}
\bibinfo{author}{\bibfnamefont{D.}~\bibnamefont{Gillespie}},
  \bibinfo{journal}{J.Phys.Chem.} \textbf{\bibinfo{volume}{81}},
  \bibinfo{pages}{2340} (\bibinfo{year}{1977}).

\bibitem[{\citenamefont{J.Mai et~al.}(1994{\natexlab{b}})\citenamefont{J.Mai,
  Kuzovkov, and von Niessen}}]{kuzovkov1}
\bibinfo{author}{\bibnamefont{J.Mai}},
  \bibinfo{author}{\bibfnamefont{V.}~\bibnamefont{Kuzovkov}}, \bibnamefont{and}
  \bibinfo{author}{\bibfnamefont{W.}~\bibnamefont{von Niessen}},
  \bibinfo{journal}{J.Chem.Phys.} \textbf{\bibinfo{volume}{100(8)}},
  \bibinfo{pages}{6073} (\bibinfo{year}{1994}{\natexlab{b}}).

\bibitem[{\citenamefont{Guttmann}(1989)}]{Guttmann89}
\bibinfo{author}{\bibfnamefont{A.~J.} \bibnamefont{Guttmann}},
  \emph{\bibinfo{title}{Asymptotic analysis of power-series expansions in Phase
  transitions and critical phenomena}} (\bibinfo{publisher}{Academic Press},
  \bibinfo{year}{1989}).

\bibitem[{\citenamefont{Amit}(1984)}]{Amit84}
\bibinfo{author}{\bibfnamefont{D.~J.} \bibnamefont{Amit}},
  \emph{\bibinfo{title}{Field theory, the renormalization group, and critical
  phenomen}} (\bibinfo{publisher}{World Scientific, Singapore},
  \bibinfo{year}{1984}).

\bibitem[{\citenamefont{Liggett}(1985)}]{Liggett85}
\bibinfo{author}{\bibfnamefont{T.~M.} \bibnamefont{Liggett}},
  \emph{\bibinfo{title}{Interacting particle systems}}
  (\bibinfo{publisher}{Springer, Berlin}, \bibinfo{year}{1985}).

\bibitem[{\citenamefont{Spohn}(1991)}]{Spohn91}
\bibinfo{author}{\bibfnamefont{H.}~\bibnamefont{Spohn}},
  \emph{\bibinfo{title}{Interacting particle systems}}
  (\bibinfo{publisher}{Springer, Berlin}, \bibinfo{year}{1991}).

\bibitem[{\citenamefont{McCoy and Wu}(1973)}]{McCoyWu73}
\bibinfo{author}{\bibfnamefont{B.~M.} \bibnamefont{McCoy}} \bibnamefont{and}
  \bibinfo{author}{\bibfnamefont{T.~T.} \bibnamefont{Wu}},
  \emph{\bibinfo{title}{The two-dimensional Ising model}}
  (\bibinfo{publisher}{Harvard University Press, Cambridge},
  \bibinfo{year}{1973}).

\bibitem[{\citenamefont{Polyakov}(1970)}]{Polyakov70}
\bibinfo{author}{\bibfnamefont{A.~M.} \bibnamefont{Polyakov}},
  \bibinfo{journal}{Sov. Phys. JETP Lett.} \textbf{\bibinfo{volume}{12}},
  \bibinfo{pages}{381} (\bibinfo{year}{1970}).

\bibitem[{\citenamefont{Cardy}(1987)}]{Cardy87}
\bibinfo{author}{\bibfnamefont{J.~L.} \bibnamefont{Cardy}},
  \emph{\bibinfo{title}{Phase transitions and critical phenomena}}
  (\bibinfo{publisher}{Academic Press, New York}, \bibinfo{year}{1987}).

\bibitem[{\citenamefont{Henkel}(1999)}]{Henkel99}
\bibinfo{author}{\bibfnamefont{M.}~\bibnamefont{Henkel}},
  \emph{\bibinfo{title}{Conformal Invariance and Critical Phenomena}}
  (\bibinfo{publisher}{Springer Verlag, Berlin}, \bibinfo{year}{1999}).

\bibitem[{\citenamefont{Schmittmann and Zia}(1995)}]{SchmittmannZia95}
\bibinfo{author}{\bibfnamefont{B.}~\bibnamefont{Schmittmann}} \bibnamefont{and}
  \bibinfo{author}{\bibfnamefont{R.~K.~P.} \bibnamefont{Zia}},
  \emph{\bibinfo{title}{Statistical mechanics of driven diffusive systems in
  Phase transitions and critical phenomen}} (\bibinfo{publisher}{Academic
  Press, New York}, \bibinfo{year}{1995}).

\bibitem[{\citenamefont{Alcaraz et~al.}(1994)\citenamefont{Alcaraz, M.~Droz,
  and Rittenberg}}]{ADHR94}
\bibinfo{author}{\bibfnamefont{F.~C.} \bibnamefont{Alcaraz}},
  \bibinfo{author}{\bibfnamefont{M.~H.} \bibnamefont{M.~Droz}},
  \bibnamefont{and}
  \bibinfo{author}{\bibfnamefont{V.}~\bibnamefont{Rittenberg}},
  \bibinfo{journal}{Ann. Phys. (N.Y.} \textbf{\bibinfo{volume}{230}},
  \bibinfo{pages}{250} (\bibinfo{year}{1994}).

\bibitem[{\citenamefont{Evans et~al.}(1999)\citenamefont{Evans, Rajewsky, and
  Speer}}]{ERS99}
\bibinfo{author}{\bibfnamefont{M.~R.} \bibnamefont{Evans}},
  \bibinfo{author}{\bibfnamefont{N.}~\bibnamefont{Rajewsky}}, \bibnamefont{and}
  \bibinfo{author}{\bibfnamefont{E.~R.} \bibnamefont{Speer}},
  \bibinfo{journal}{J. Stat. Phys} \textbf{\bibinfo{volume}{95}},
  \bibinfo{pages}{45} (\bibinfo{year}{1999}).

\bibitem[{\citenamefont{Gwa and Spohn}(1992)}]{GwaSpohn92}
\bibinfo{author}{\bibfnamefont{L.~H.} \bibnamefont{Gwa}} \bibnamefont{and}
  \bibinfo{author}{\bibfnamefont{H.}~\bibnamefont{Spohn}},
  \bibinfo{journal}{Phys. Rev.} \textbf{\bibinfo{volume}{A 46}},
  \bibinfo{pages}{844} (\bibinfo{year}{1992}).

\bibitem[{\citenamefont{Doering et~al.}(1991)\citenamefont{Doering, Burschka,
  and Horsthemke}}]{DBH91}
\bibinfo{author}{\bibfnamefont{C.~R.} \bibnamefont{Doering}},
  \bibinfo{author}{\bibfnamefont{M.~A.} \bibnamefont{Burschka}},
  \bibnamefont{and}
  \bibinfo{author}{\bibfnamefont{W.}~\bibnamefont{Horsthemke}},
  \bibinfo{journal}{J. Stat. Phys.} \textbf{\bibinfo{volume}{65}},
  \bibinfo{pages}{953} (\bibinfo{year}{1991}).

\end{thebibliography}

\end {document}